\documentclass[11pt,a4paper]{article}
\pdfoutput=1
\usepackage{jheppub}

\usepackage{placeins}
\usepackage{lmodern}
\usepackage[T1]{fontenc}
\usepackage[utf8]{inputenc}

\usepackage{eufrak} 

\usepackage[usenames,dvipsnames]{xcolor}
\usepackage{caption}
\usepackage{subcaption}

\def\){\right)}
\def\({\left( }
\def\]{\right] }
\def\[{\left[ }

\def\NO{\nonumber}

\newcommand{\be}{\begin{equation}}
\newcommand{\ee}{\end{equation}}

\def\bea{\begin{eqnarray}}
\def\eea{\end{eqnarray}}

\def\bal#1\eal{\begin{align}#1\end{align}}

\def\bald{\begin{aligned}}
\def\eald{\end{aligned}}

\def\bsub{\begin{subequations}}
\def\esub{\end{subequations}}

\def\beqx{\begin{displaymath}}
\def\eeqx{\end{displaymath}}

\newcommand{\bmat}{\left(\begin{array}}
\newcommand{\emat}{\end{array}\right)}

\def\a{\alpha}
\def\b{\beta}

\def\d{\delta}
\def\e{\epsilon}
\def\f{\phi}
\def\j{\psi}

{\ifmmode\def\l{\lambda}\fi}
\def\m{\mu}

\def\r{\rho}
\def\s{\sigma}
\def\t{\tau}
\def\x{\xi}

\def\G{\Gamma}

\def\L{\Lambda}

\def\P{\Pi}

\def\ve{\varepsilon}

    \def\vth{\vartheta}

\def\vf{\varphi}

\def\ca{{\cal A}}
\def\cb{{\cal B}}
\def\cc{{\cal C}}

\def\cf{{\cal F}}

\def\cj{{\cal J}}
\def\ck{{\cal K}}

\def\cm{{\cal M}}

\def\co{{\cal O}}

\def\ct{{\cal T}}

\def\cv{{\cal V}}
\def\cw{{\cal W}}

\def\bb#1{\ensuremath{\mathbb{#1}}} 

\def\bo{{\raise-.3ex\hbox{\large$\Box$}}}               
\def\pa{\partial}                                       
\def\face{{\raise.2ex\hbox{$\displaystyle \bigodot$}\mskip-2.2mu \llap {$\ddot
        \smile$}}}                                   
\def\>{\rangle}                                      
\def\<{\langle}                                      

\def\tx#1{\text{#1}}
\def\sbtx#1{{}_{\textsf{#1}}}                           
\newcommand{\sub}[1]{\phantom{}_{(#1)}\phantom{}}    
\def\leftrightarrowfill{$\mathsurround=0pt \mathord\leftarrow \mkern-6mu
        \cleaders\hbox{$\mkern-2mu \mathord- \mkern-2mu$}\hfill
        \mkern-6mu \mathord\rightarrow$}        
\def\dvec#1{\vbox{\ialign{##\crcr
        \leftrightarrowfill\crcr\noalign{\kern-1pt\nointerlineskip}
        $\hfil\displaystyle{#1}\hfil$\crcr}}}           
\def\diag{{\rm diag \,}}                                

\def\-{\hphantom{-}}

\newcommand\subpar[1]{
\medskip
\noindent\textit{#1}
}

\let\originalleft\left
\let\originalright\right
\renewcommand{\left}{\mathopen{}\mathclose\bgroup\originalleft}
\renewcommand{\right}{\aftergroup\egroup\originalright}

\newcommand{\eq}{\begin{equation}}
\newcommand{\eeq}{\end{equation}}

\newcommand\lp{\left(}
\newcommand\rp{\right)}
\newcommand\mc{\mathcal}
\newcommand\p{{\partial}}
\newcommand\dd{{\mathrm d}}

\newcommand\tr{{\,\mathrm{Tr}\,}}
\newcommand\ginv{{g^{-1}}{}}
\newcommand\g[1]{{g_{(#1)}}{}}
\newcommand\ggi[2]{{g_{(#1)}^{#2}}{}}
\newcommand\gij[1]{{g_{(#1)ij}}}
\newcommand\gi[1]{{g^{-1}_{(#1)}}{}}
\newcommand\dil[1]{{\varphi_{(#1)}}}
\newcommand\dilp[2]{{\varphi_{(#1)}^{#2}}} 
\newcommand\ax[1]{{\psi_I^{(#1)}}}
\newcommand\A[2]{{A^{(#1)}_{#2}}{}}

\newcommand\ga{{\gamma}}
\newcommand\ep{{\epsilon}}

\newcommand\qe{{q_{\textsf{e}}}} 
\newcommand\qm{{q_{\textsf{m}}}} 
\newcommand\qep[1]{{q_{\textsf{e}}^{#1}}} 
\newcommand\qmp[1]{{q_{\textsf{m}}^{#1}}} 

\usepackage{tikz}
\usetikzlibrary{patterns}

\tikzstyle{mybox} = [draw=white,  fill=white,
    rectangle, inner sep=2pt, inner ysep=2pt]

\graphicspath{{/home/work/Desktop/Horizon_Analysis/}}
\begin{document}

\title{Phases of planar AdS black holes with axionic charge}
\author[a]{Marco M. Caldarelli}
\author[a]{Ariana Christodoulou}
\author[b]{Ioannis Papadimitriou} 
\author[a]{Kostas Skenderis}

\affiliation[a]{
Mathematical Sciences and STAG Research Centre, University of Southampton,\\
Highfield, Southampton SO17 1BJ, United Kingdom}

\affiliation[b]{SISSA and INFN - Sezione di Trieste,
Via Bonomea 265, I 34136 Trieste, Italy}

\emailAdd{M.M.Caldarelli@soton.ac.uk}
\emailAdd{misc1g13@soton.ac.uk}
\emailAdd{Ioannis.Papadimitriou@sissa.it}
\emailAdd{K.Skenderis@soton.ac.uk} 

\abstract{Planar AdS black holes with axionic charge have finite DC conductivity due to momentum relaxation. We obtain a new family of exact asymptotically AdS$_4$ black branes with scalar hair, carrying magnetic and axion charge, and we study the thermodynamics and dynamic stability of these, as well as of a number of previously known electric and dyonic solutions with axion charge and scalar hair. The scalar hair for all solutions satisfy mixed boundary conditions, which lead to modified holographic Ward identities, conserved charges and free energy, relative to those following from the more standard Dirichlet boundary conditions. We show that properly accounting for the scalar boundary conditions leads to well defined first law and other thermodynamic relations. Finally, we compute the holographic quantum effective potential for the dual scalar operator and show that dynamical stability of the hairy black branes is equivalent to positivity of the energy density.}

\keywords{AdS/CFT, AdS/CMT, holographic renormalization, black holes, thermodynamics}

\preprint{SISSA 59/2016/FISI}

\maketitle

\section{Introduction}


Planar asymptotically anti-de Sitter (AdS) supergravity solutions with non trivial scalar profiles play an important role in the gauge/gravity duality since they describe holographic Renormalization Group (RG) flows between conformal fixed points. Finite temperature, i.e. black brane, AdS solutions supported by scalar fields have also attracted significant attention, especially in applications of holography to high energy physics and condensed matter systems.    

Prominent examples include non-conformal plasmas \cite{Benincasa:2005iv,Finazzo:2014cna,Gursoy:2015nza,Attems:2016ugt}, holographic superconductors \cite{Gubser:2008px,Hartnoll:2008vx}, where a charged scalar coupled to a U(1) gauge field condenses at low temperatures providing a holographic description of the superconducting phase transition, as well as non-relativistic RG flows with a neutral running dilaton exhibiting hyperscaling violation in the infrared \cite{Charmousis:2010zz,Huijse:2011ef,Iizuka:2011hg,Dong:2012se,Iizuka:2012iv,Gouteraux:2012yr}. A third type of solution where axion fields acquire a linear profile along the boundary directions \cite{Bardoux:2012aw} was put forward in \cite{Donos:2013eha, Andrade:2013gsa} as a mechanism of breaking translation invariance in the dual field theory, leading to finite DC conductivity. Linear axion backgrounds were considered earlier in \cite{Azeyanagi:2009pr}, and were proposed as a description of anisotropic holographic plasmas \cite{Mateos:2011ix} and fluids \cite{Jain:2014vka}. In fact, such backgrounds are a special case of the more general $Q$-lattices introduced in \cite{Donos:2013eha}, or the top-down  $\t$-lattices discussed recently in \cite{Donos:2016zpf}.  

Despite extensive work on AdS supergravity solutions with scalar fields and their physical importance in the context of holography, such backgrounds have been obtained mostly numerically. Moreover, their holographic dictionary, asymptotic conserved charges, and general thermodynamic properties remain somewhat opaque and are often incorrectly described. In this paper we discuss a number of exact black brane solutions that are simultaneously supported by two different types of scalars: a scalar field with a running profile $\phi$, as well as a number of axions $\psi_I$ with a linear profile in the field theory directions, but constant in the radial coordinate as in \cite{Andrade:2013gsa}. Both types of scalars are neutral with respect to a Maxwell field, that may carry electric or magnetic charge. We will call the scalar field with the running profile {\it dialton} for reasons to be explained below.

More specifically, we revisit the exact axionic black holes found in \cite{Bardoux:2012aw}, which do not have a running profile for the dialton, as well as those obtained in \cite{Bardoux:2012tr} and have a running dialton in addition to the non trivial axion background. A running dialton is also a feature present in the electrically charged black brane solutions found analytically in \cite{Gouteraux:2014hca}, which we also discuss. Finally, we obtain a new family of exact magnetically charged axionic black holes, which may be viewed as the magnetic version of those presented in \cite{Gouteraux:2014hca}.


The two different types of scalar fields generically turned on in these black brane solutions, namely the running dialton and the axion background, present two subtleties that we address in detail. Firstly, axions with a linear profile along the spatial boundary directions should be understood as 0-forms carrying magnetic charge proportional to the slope of the linear profile and not as regular scalars. This distinction is fundamental: axions with a linear profile along the spatial boundary directions are {\em primary hair}, while massive scalars or running dialtons are {\em secondary hair} \cite{Coleman:1991ku}. We show that linear axion backgrounds are exactly on the same footing as standard magnetically charged black holes, thus allowing for a straightforward understanding of their thermodynamics (see also \cite{Andrade:2013gsa, Park:2016slj}). Moreover, treating the axions as 0-forms leads to additional {\em global} Ward identities (see \eqref{globalWIs}) reflecting the fact that 0-forms cannot carry electric charge. Such Ward identities are not applicable to standard scalar operators. 

In holography, the linear axion background, $\psi_I \sim x^I$ corresponds to deforming the action of the dual QFT by $\int x^I  \mc O_{\psi_I}$. It is the presence of these couplings that explicitly break diffeomorphisms and introduces momentum dissipation. Note that the global Ward identity implies that the dual operators $\mc O_{\psi_I}$ are (locally) exact, so up to boundary terms one may remove the $x^I$ dependence by partial integration. This is the counterpart of the fact that the shift invariance in the bulk implies that there is no explicit $x$ dependence in the field equations. However, one cannot ignore the boundary terms as they  blow up at $x^I \to \pm \infty$. It is the presence of such boundary terms/global issues that it is ultimately responsible for momentum dissipation.

The second subtlety concerns the massive dialton field and the boundary conditions it satisfies. As we review in section \ref{model}, scalars with AdS mass in the window \eqref{window} admit Neumann or mixed boundary conditions, in addition to Dirichlet. Of course, bald black hole solutions are compatible with any boundary condition on the dialton field, but a solution with running dialton is compatible only with very specific boundary conditions, which are often unique. In particular, AdS$_4$ black holes with running dialtons typically satisfy mixed boundary conditions. Neumann or mixed boundary conditions on the scalars lead to a modification of the holographic stress tensor and on-shell action \cite{Papadimitriou:2007sj}, and hence of the associated conserved charges and free energy \cite{Papadimitriou:2005ii}. We show that these modifications are sufficient for the AdS black holes with running scalars satisfying Neumann or mixed boundary conditions to obey the standard thermodynamic relations, including the first law, without any `charges' associated with the running scalars, contrary to what has been claimed in a number of recent papers.
In particular, running scalars in AdS black holes are {\em secondary hair} -- not primary. 

In previous literature analogous scalar fields were often called dilatons. Strictly speaking, dilatons are associated with spontaneous breaking of scale invariance. In our case, scale invariance is broken explicitly by the axions\footnote{One may restore this invariance by appropriately scaling the parameters of the solutions and we will see that this is a indeed a property of the solutions we will discuss.} so it is not appropriate to call the scalar $\phi$ a dilaton. Holographically, the mixed boundary conditions obeyed by $\phi$ are associated with a multi-trace deformation of the dual QFT. Moreover, as we will see the condensate of the operator  dual to $\phi$ governs the different phases of the theory. We can thus use $\phi$ as a dial that can change the theory and/or move us across different phases and for this reason we will call it {\it dialton}. 

In the following table we summarize the above discussion for a generic scalar with AdS mass $m$ in $d+1$ dimensions, indicating when different boundary conditions are permitted, as well as if the scalar corresponds to primary or secondary hair and whether the first law gets modified. 
\begin{center}
	\begin{tabular}{c|c|c|c|c|c}
		& Dirichlet & Neumann/Mixed & Primary & Secondary & First law \\\hline
		$-\frac{d^2}{4} \leq m^2 \leq -\frac{d^2}{4}+1$ & $\checkmark$ & $\checkmark$ & $\times$ & $\checkmark$ & $\checkmark$\\\hline
		$-\frac{d^2}{4}+1 < m^2 <0$ & $\checkmark$ & $\times$ & $\times$ & $\checkmark$ & $\checkmark$\\\hline
		$m^2=0$, $0$-form charge$=0$   & $\checkmark$ & $\times$ & $\times$ & $\checkmark$ & $\checkmark$\\\hline
		$m^2=0$, $0$-form charge$\neq 0$   & $\checkmark$ & $\times$ & $\checkmark$ & $\times$ & modified\\\hline
	\end{tabular}	
\end{center}

Since the black hole solutions we discuss in this paper are known analytically, we are also able to compute analytically the off-shell holographic quantum effective potential of the scalar operator dual to the dialton field. This is given by the Legendre transform of the holographic generating functional with respect to the source of the scalar operator \cite{Papadimitriou:2007sj} and is related to the effective potential obtained from Designer Gravity \cite{Hertog:2004ns}. This computation allows us to show that dynamical stability of hairy planar AdS black holes with respect to scalar fluctuations is equivalent to positivity of the energy density.     
 
Finally, using our results for the thermodynamics of these exact black branes, we study their thermodynamic and dynamic stability, and the corresponding phase structure. Particularly interesting is the phase structure we observe for the electrically charged solutions of \cite{Gouteraux:2014hca} and their newly found magnetically charged versions, both of which exhibit a {\em zeroth} order phase transition at finite (respectively electric or magnetic) charge density, which becomes a standard second order phase transition at zero charge density. Zeroth order phase transitions have been predicted in the context of superfluidity and superconductivity and are related to the presence of metastable states \cite{MR2129341,Ibanez:2008xq}.\footnote{However, the situation described in \cite{MR2129341} is reversed in temperature relative to what we observe here. Namely, the metastable states in \cite{MR2129341} exist only below the critical temperature $T_c$, while here they only exist above $T_c$.} More recently such transitions have also been found to occur in higher dimensional black holes \cite{Altamirano:2013ane}. In the present context, we find that above a critical temperature and at non-zero charge density there are three hairy black holes, two with positive energy density --a large and a small black hole-- and one with negative. For the electric solutions the black hole with negative energy density is the largest of the three, while for magnetically charged solutions it is the smallest. As $T_c$ is approached from above, the two positive energy black holes converge and cease to exist below $T_c$. However, at non-zero charge density the negative energy solution at the critical temperature has a lower energy density than the other two solutions, which are therefore metastable. However, the larger of the two black holes with positive energy density has the smallest free energy and is therefore thermodynamically favored above $T_c$. Accordingly, the free energy is discontinuous at $T_c$, leading to a zeroth order phase transition. As the charge density is tuned to zero, however, the negative and positive energy solutions all converge as $T_c$ is approached from above, with their energy density approaching zero from below and above respectively, which leads to a regular second order phase transition at $T_c$. 

The paper is organized as follows. In section \ref{model} we present the general class of Einstein-Maxwell-Dialton-Axion models we are interested in, as well as the two specific models that admit the exact black brane solutions we discuss later on. In section \ref{dictionary} we summarize the results of holographic renormalization, which is carried out in more detail in appendix \ref{asexp}, and we demonstrate the effect of mixed scalar boundary conditions and axion charge on the holographic dictionary and, in particular, the holographic Ward identities. All new and previously known exact black brane solutions we study in this paper are presented in Section \ref{sec::solutions}. Section \ref{sec:thermo} contains a detailed analysis of the thermodynamics of planar AdS black holes, with particular emphasis on the role of mixed scalar boundary conditions and axion charge. These results allow us to study the phase structure  and thermodynamic stability of these black branes in section \ref{sec:phases}. Dynamic stability is also addressed in section \ref{sec:phases} using the calculation of the holographic quantum effective potential for the scalar operator in appendix \ref{effpot}. We conclude with few final remarks in section \ref{sec:conclusion}.

\section{The model}
\label{model}

We will consider theories of gravity in $d+1$ spacetime dimensions coupled to a {\em dialton} field $\phi$, a Maxwell field $A$ with field strength $F=\tx dA$, and $d-1$ free scalar fields $\psi_I$, with action
\begin{align}
S_{\textsf{bulk}}&=\int_{\mathcal M}\!d^{d+1}x\sqrt{-G}\left(
R-\frac12(\p\phi)^2-V(\phi)
-\frac12\sum_{I=1}^{d-1}(\p\psi_I)^2
-\frac14Z(\phi)F^2
\right).
\label{genAction}
\end{align}
We use units in which $16\pi G_\mathsf{N}=1$.
The spacetime metric $G_{\mu\nu}$ is used to raise/lower Greek indices $\mu, \nu,\ldots$ and the capital Latin indices $I,J,\ldots$ denote the flavor of the scalars $\psi_I$.
The function $V(\phi)$ defines the potential for the dialton, and it generates an effective cosmological constant $\Lambda=V(0)/2$ when the dialton vanishes.
For convenience, we will express it in terms of the radius $\ell$ of AdS, defined by
\eq
\Lambda=-d(d-1)/2\ell^2.
\eeq
The function $Z(\phi)$ determines the coupling of the dialton field to the Maxwell field and is normalized such that $Z(0)=1$.
The equations of motion following from the action \eqref{genAction} are\footnote
{ 
Our curvature conventions are as follows: $R_{\mu\nu\rho}{}^\sigma=\p_\nu\Gamma_{\mu\rho}^\sigma+\Gamma_{\nu\alpha}^\sigma\Gamma_{\mu\rho}^\alpha-(\mu\leftrightarrow\nu)$ and $R_{\mu\nu}=R^\rho{}_{\mu\rho\nu}$, which are those of Wald's book~\cite{Wald:1984rg}, but differ by an overall sign from those of \cite{Bianchi:2001de}.
We also omit the summation symbol over the axion flavor indices $I$ from now on, and use the Einstein convention for them too.
} 
\begin{align}
R_{\mu\nu}&=\frac12\p_\mu\psi_I\p_\nu\psi_I+\frac12\lp\p_\mu\phi\p_\nu\phi+\frac{2V(\phi)}{d-1}G_{\mu\nu}\rp
+\frac12Z(\phi)\lp F_{\mu\rho}F_\nu{}^\rho-\frac{F^2}{2(d-1)}G_{\mu\nu}\rp\,,\nonumber\\
\Box\psi_I&=0,\qquad
\Box\phi-V'(\phi)=\frac14Z'(\phi)F^2,\qquad
\nabla^\mu\lp Z(\phi)F_{\mu\nu}\rp=0.
\end{align}

The $d-1$ scalar fields $\psi_I$ enjoy a global $\textsf{ISO}(d-1)$ symmetry under which $\psi_I\mapsto\Lambda_{I}^J\psi_J+c_I$, where $c_I$ is a constant translation vector, and $\Lambda_{I}^J$ is a constant $\textsf{SO}(d-1)$ flavor rotation. The translation symmetry results from the fact that the scalars enter in the action through their field strength $F_I=\dd\psi_I$ only, and amounts to the gauge freedom of $0$-forms.
Taken together, these symmetries imply that the axions $\j_I$ are coordinates in a target Euclidean space $\bb R^{d-1}$. For planar solutions of the equations of motion this target space is isomorphic to the spatial part of the conformal boundary. Under this isomorphism the global $\textsf{ISO}(d-1)$ symmetry is mapped to the spatial isometries of the conformal boundary. Any solution of the equations of motion provides a particular {\em embedding} of the spatial part of the boundary to the target space $\bb R^{d-1}$. A generic embedding of the form $\j_I=p_I x^I+c_I$, for fixed constants $(p_I,c_I)$, breaks the global $\textsf{ISO}(d-1)$ symmetry completely, resulting in complete breaking of the spatial isometries of the conformal boundary. The solutions we are interested in correspond to special embeddings of the form $\j_I=p x^I$, which break only translations and preserve an $\textsf{SO}(d-1)$ symmetry. Breaking translation invariance in this way results in momentum dissipation in the dual filed theory \cite{Andrade:2013gsa}.
Since the axion flux is non-zero, i.e. $F_I=p\neq 0$, such solutions correspond to turning on a topological axion charge density background in the dual field theory, analogous to turning on a magnetic field on the boundary. These topological charges contribute to black hole thermodynamics and play a crucial role in understanding the phase diagram.

We will focus on two specific models for which planar black hole solutions with scalar hair are known analytically. Both models are special cases of \eqref{genAction} and have $d=3$. 
\paragraph{Theory I:}
The first model is obtained by setting $Z(\phi)=1$ and choosing
\eq
V(\phi)=2\Lambda\left[\cosh^4\lp\frac\phi{2\sqrt{3}}\rp-\alpha\sinh^4\lp\frac\phi{2\sqrt{3}}\rp\right],
\label{potential}\eeq
as potential for the dialton field. This potential can be obtained through field redefinition from AdS gravity with cosmological constant $\Lambda=-3/\ell^2$, conformally coupled to a scalar field with a conformal self interaction coupling $\a$ \cite{deHaro:2006ymc,Papadimitriou:2007sj}. For $\a=1$ this potential was discussed earlier in \cite{Martinez:2004nb}, where a black hole solution with a horizon of constant negative curvature was obtained, and in \cite{Papadimitriou:2006dr}, where this potential was embedded in the $U(1)^4$ truncation of maximally supersymmetric gauge supergravity in four dimensions. In fact, for $\a=1$, taking $Z(\f)$ to be a specific exponential function of the scalar $\f$ and setting the axions $\j_I$ to zero, the full action \eqref{genAction} can be embedded in the $U(1)^4$ truncation of gauged supergravity. No embedding is known for $\a \neq 1$, or for $Z(\f)=1$. In this article we will focus on solutions with non-trivial axion profiles, and so our action \eqref{genAction} should be treated as a bottom up model. In the presence of the axion fields, this theory and its black brane solutions have been studied in \cite{Bardoux:2012tr}.     

Defining the constants 
\eq\label{coeffs}
V_k\equiv\lim_{r\rightarrow0}V^{(k)}(\phi),\qquad
Z_k\equiv\lim_{r\rightarrow0}Z^{(k)}(\phi),
\eeq
where $V^{(k)}(\phi)$ and $Z^{(k)}(\phi)$ denote respectively the $k$-th derivative of the functions $V(\f)$ and $Z(\f)$, for the potential \eqref{potential} all odd coefficients vanish since $V(\f)$ is a manifestly even function, while the first four even coefficients are 
\eq
V_0=-\frac{6}{\ell^2},\quad
V_2=-\frac2{\ell^2},\quad
V_4=-\frac{5-3\alpha}{3\ell^2},
\quad
V_6=-\frac{17-15\alpha}{9\ell^2}.
\label{vcoeffsI}\eeq
In particular, the AdS mass of the scalar $\phi$ is given by 
\eq\label{AdSmass}
m_\phi^2=V''(0)=-\frac2{\ell^2},
\eeq
and so it lies in the window 
\eq
-\frac{d^2}4\leq\ell^2m_\phi^2\leq -\frac{d^2}4+1,
\label{window}
\eeq
where both modes of the scalar $\phi$ are normalizable. The lower bound is the usual Breitenlohner-Freedman bound that ensures stability of the AdS vacuum under scalar perturbations \cite{Breitenlohner:1982bm}. The upper bound is not necessary for stability, but if satisfied then both scalar modes are normalizable, which allows for non-Dirichlet boundary conditions for the scalar \cite{Balasubramanian:1998sn,Klebanov:1999tb}. Recalling the relation between the AdS mass of a scalar field and the conformal dimension of the dual operator, $\ell^2 m_\f^2=\Delta(\Delta-d)$, one sees that there are two generically distinct conformal dimensions for a given mass, namely $\Delta_-\leq d/2$ and $\Delta_+\geq d/2$. In the standard quantization, where Dirichlet boundary conditions are imposed on the scalar, the dual operator has dimension $\Delta_+$. When the scalar mass lies in the window \eqref{window}, however, Neumann and mixed boundary conditions are also admissible, leading to a dual operator of dimension $\Delta_-$. The upper bound in \eqref{window}, which ensures that both modes are normalizable, is equivalent to the condition that $\Delta_-$ is above the unitarity bound, i.e. $\Delta_-\geq (d-2)/2$. For the scalar mass \eqref{AdSmass}, $\Delta_+=2$ and $\Delta_-=1$. The boundary conditions we are going to impose on the scalar are dictated by the black hole solutions we are interested in.

\paragraph{Theory II:}
Another family of theories we will be interested in is defined by
\be
V(\phi)=\sigma_1 e^{\frac{(d-2)(d-1)\delta^2-2}{2(d-1)\delta}\phi}+\sigma_2e^{\frac{2\phi}{\delta(1-d)}}+\sigma_3e^{(d-2)\delta\phi},\qquad 
Z(\phi)=e^{-(d-2)\delta\phi},
\label{genPot}
\ee
where
\begin{align}
\sigma_0&=-\frac{d(d-1)}{\ell^2}=2\Lambda,
&\sigma_1&=\sigma_0\frac{8(d-2)(d-1)^2\delta^2}{d(2+(d-2)(d-1)\delta^2)^2},
\nonumber\\
\sigma_2&=\sigma_0\frac{(d-2)^2(d-1)\delta^2(d(d-1)\delta^2-2)}{d(2+(d-2)(d-1)\delta^2)^2},
&\sigma_3&=-2\sigma_0\frac{(d-2)^2(d-1)\delta^2-2d}{d(2+(d-2)(d-1)\delta^2)^2},
\end{align}
and $\delta$ is a free parameter in the potential.
This action, and a corresponding family of analytic black brane solutions, were presented in appendix~C of~\cite{Gouteraux:2014hca}. We will be mostly interested in the $d=3$ case, and for simplicity we trade the parameter $\delta$ for a new parameter $\xi$ defined by
\eq
\delta=\sqrt{\frac{2-\xi}\xi},\qquad
0<\xi<2.
\label{delta}\eeq
In terms of $\x$, the potential and gauge kinetic coupling of the four-dimensional theory simplify to
\begin{align}
V(\phi)&=-\frac1{\ell^2}e^{-\sqrt{\frac{\xi}{2-\xi}}\phi}\lp
(2-\xi)(3-2\xi)+4\xi(2-\xi)e^{\frac\phi{\sqrt{\xi(2-\xi)}}}+\xi(2\xi-1)e^{\frac{2\phi}{\sqrt{\xi(2-\xi)}}}
\rp,
\nonumber
\\
Z(\phi)&=e^{-\sqrt{\frac{2-\xi}{\xi}}\phi}.
\label{pot2}
\end{align}
For this potential the coefficients \eqref{coeffs} read
\eq
V_0=-\frac6{\ell^2},\quad
V_2=-\frac2{\ell^2},\quad
V_4=\frac{4-6\xi(2-\xi)}{\ell^2\xi(2-\xi)},
\quad
V_5=-\frac{4(\xi-1)(2\xi-3)(2\xi-1)}{\ell^2 [\xi(2-\xi)]^{3/2}}.
\label{vcoeffsII}\eeq
Notice that the potential~\eqref{pot2} with $\xi=1/2$ (or equivalently with $\xi=3/2$) coincides with the potential~\eqref{potential} with $\alpha=1$. In all other cases they do not match. However, the coefficient $V_2$ is the same for the two potentials independently of the value of $\x$, and so the AdS masses of the scalar $\f$, and hence the conformal dimensions of the dual operators, are the same in the two theories. An interesting feature of the potential \eqref{pot2} is that it can be written globally in terms of a superpotential as   
\be
U'^2(\f)-\frac34U^2(\f)=2V(\f),
\ee
where
\be
U(\f)=-\frac{2}{\ell}\((2-\x)e^{-\frac{\x/2}{\sqrt{\x(2-\x)}}\f}+\x e^{\frac{(1-\x/2)}{\sqrt{\x(2-\x)}}\f}\).
\ee
Moreover, for $\x=1/2$ (or equivalently with $\xi=3/2$) and $\x=1$, this potential can be embedded in the $U(1)^4$ truncation of maximally supersymmetric gauge supergravity, including the gauge field with $Z(\f)$ as given in \eqref{pot2}. 


The limits $\xi \to 0$ and $\xi \to 2$ require separate treatment and correspond to special cases of Theory I.  In order to consider these limiting cases one must first rescale the fields appropriately to ensure that the potential and couplings remain finite. In particular, to study the $\xi \to 0$ limit we set $\xi =\epsilon$ and redefine the dialton and the gauge field as
\eq
\phi \to \tilde{\phi} = \frac{\phi}{\sqrt{\epsilon}}, \qquad A \to \tilde{A} = \frac{A}{\sqrt{\epsilon}}.
\eeq
In the limit $\epsilon \to 0$, the potential $V(\phi)$ becomes $-6/\ell^2 = 2\Lambda$ and the coupling of the dialton to the Maxwell field $Z(\phi)\to e^{-2\tilde{\phi}}$. Rewriting the action (\ref{genAction}) in terms of the rescaled fields and noting that the kinetic terms for the dialton and gauge field acquire a factor of $\epsilon$, in the limit $\epsilon \to 0$ , we obtain
\eq
S_{\textsf{bulk}}=\int_{\mathcal M}\!d^{4}x\sqrt{-G}\left(
R-2\Lambda
-\frac12\sum_{I=1}^{2}(\p\psi_I)^2
\right). \label{act_nodil_noq}
\eeq
This is a consistent truncation of Theory I, obtained by setting $\f$ and the gauge field to zero. 

Similarly, the limit $\xi \to 2$ is obtained by setting $\xi = 2-\epsilon$ and redefining the dialton as above. In this case one need not rescale the gauge field, which therefore survives the limit. Letting $\epsilon \to 0$ we have $V(\phi)\to 2 \Lambda$ and $Z(\phi) \to 1$ and the action becomes
\eq
S_{\textsf{bulk}}=\int_{\mathcal M}\!d^{4}x\sqrt{-G}\left(
R-2\Lambda
-\frac12\sum_{I=1}^{2}(\p\psi_I)^2
-\frac14 F^2
\right). \label{action_nodil}
\eeq
Once again, this is a consistent truncation of Theory I corresponding to setting $\f=0$, but keeping the gauge field.
Since both limits $\xi \to 0$ and $\xi \to 2$ result in truncations of Theory I we will not consider these cases further, focusing instead in the cases $0<\x<2$.


\section{Scalar boundary conditions and the holographic dictionary}
\label{dictionary}

Before discussing the specific black hole solutions of Theories I and II we are interested in, it is instructive to consider the holographic dictionary for these theories, and how this depends on the boundary conditions imposed on the dialton field.
The solutions we are interested in correspond to mixed boundary conditions for the scalar $\f$, which leads to a modification of the holographic dictionary and of the conserved charges relative to the usual Dirichlet problem. As we shall see, this modification is critical for correctly describing the thermodynamics of these solutions.  

\subsection{Holographic dictionary for the Dirichlet problem}
The holographic dictionary is directly related to the boundary conditions imposed on the bulk fields and consequently to the variational problem and the corresponding boundary terms. In order to formulate the variational problem we introduce a radial cutoff $\e$ and define the regularized action 
\be\label{Sreg}
S\sbtx{reg}=S\sbtx{bulk}+S\sbtx{GH},
\ee
where $S\sbtx{bulk}$ is given by \eqref{genAction} and $S_{\textsf{GH}}$ is the standard Gibbons-Hawking term 
\be\label{GH}
S_{\textsf{GH}}=\int_{z=\e} d^dx \sqrt{-\gamma}\;2K,
\ee
where $\gamma_{ij}$ is the induced metric on the radial cutoff and $K$ stands for the trace of the extrinsic curvature $K_{ij}=-\frac{1}{2\ell}z\pa_z\gamma_{ij}$.
The regulator $\e$ here refers to the radial coordinate defined in \eqref{FGmetric} in appendix \ref{asexp}, so that $\e\to 0$ as the cutoff is removed. A well posed variational problem on the conformal boundary requires that additional local and covariant boundary terms, $S\sbtx{ct}$, are introduced so that $S_{\textsf{reg}}+S_{\textsf{ct}}$ is asymptotically independent of the radial cutoff $\e$, and hence the limit  
\eq\label{Sren}
S_{\textsf{ren}}=\lim_{\ep\rightarrow0}\lp S_{\textsf{reg}}+S_{\textsf{ct}}\rp,
\eeq
exists. The counterterms $S\sbtx{ct}$ can be computed in a number of ways. For the action \eqref{genAction} with $d=3$ and for the scalar mass \eqref{AdSmass}, corresponding to $\Delta_+=2$ and $\Delta_-=1$, the counterterms are a special case of those computed using the radial Hamiltonian formalism in \cite{Papadimitriou:2011qb} and \cite{Lindgren:2015lia} and take the form\footnote{Notice that we cannot integrate by parts the axionic term to put it in the familiar $\psi_I\Box_\gamma\psi_I$ form, since we will consider configurations where the axions do not falloff to zero as $x^i$ goes to infinity on the boundary.} 
\eq\label{counterterms}
S_{\textsf{ct}}=-\int_{z=\ep}\!d^3x\sqrt{-\ga}\left(
\frac4{\ell}+\ell R[\ga]+\frac1{2\ell}\phi^2-\frac\ell2\ga^{ij}\p_i\psi_I\p_j\psi_I
\right).
\eeq
In appendix \ref{asexp} we provide an alternative derivation of these counterterms by solving asymptotically the second order field equations, following \cite{Henningson:1998gx, Henningson:1998ey, deHaro:2000vlm, Bianchi:2001kw} (see also the review \cite{Skenderis:2002wp}).

The renormalized action $S_{\textsf{ren}}$ is identified holographically with the generating function in the dual theory corresponding to Dirichlet boundary conditions on the bulk fields. In particular, identifying the leading modes in the asymptotic expansions \eqref{FG}, namely $\g0$, $\dil0$, $\ax0$, and $A^{(0)}$, with the sources of the dual operators and $S_{\textsf{ren}}$ with the generating function, allows us to {\em define} the 1-point functions of the dual operators through the variation \cite{deHaro:2000vlm},
\be
\delta S_{\textsf{ren}}=\int\!d^3x\sqrt{-\g0}\(\frac{1}2\< \ct^{ij}\>\sbtx{D}\d\gij0+\< \cj^i\>\sbtx{D}\d A_i^{(0)}+\<\mc O_{\psi_I}\>\sbtx{D}\d\ax0+\<\mc O_{\Delta_+}\>\sbtx{D}\d\dil0\).
\label{deltaS}
\ee
Imposing Dirichlet boundary conditions on the fields corresponds to keeping the leading modes $\g0$, $\dil0$, $\ax0$, and $A^{(0)}$ fixed on the boundary,\footnote{Since the bulk fields do not induce boundary fields on a conformal boundary, but rather a conformal class of boundary fields, the only meaningful Dirichlet boundary conditions keep the conformal class fixed and not the conformal representative \cite{Papadimitriou:2005ii}. Namely, one can only demand that $\delta\g0=2\d\s(x)$, $\delta A_i^{(0)}=0$, $\delta\ax0=0$, and $\delta\dil0=-(d-\Delta_+)\d\s(x)\dil0$, where $\d\s(x)$ is an arbitrary infinitesimal scalar function on the boundary. The variation \eqref{deltaS} still vanishes under these generalized Dirichlet boundary conditions due to the trace Ward identity \eqref{D-trace-WI}, provided the conformal anomaly is zero.} which leads to a well posed variational problem since the boundary variation \eqref{deltaS} vanishes.

The definition of the Dirichlet 1-point functions through the variation \eqref{deltaS} is general, but the expressions for these 1-point functions in terms of the coefficients of the Fefferman-Graham expansions depend on the details of the theory. For the action \eqref{genAction} in $d=3$ and scalar mass given by \eqref{AdSmass}, the asymptotic analysis in appendix \ref{asexp} determines 
\begin{align}\label{D-1pt-fns}
\langle\mc O_{\psi_I}\rangle\sbtx{D}&=3\ell^2\ax3,&
\quad
\langle\mc O_{\Delta_+}\rangle\sbtx{D}&=\ell^2\dil1,\NO\\
\langle \cj^i\rangle\sbtx{D}&=Z_0A_{(1)}^i,&
\quad
\langle \ct^{ij}\rangle\sbtx{D}&=\ell^2\lp3g_{(3)}^{ij}+\dil0\dil1g_{(0)}^{ij}\rp,
\end{align}
where recall that $Z_0$ is defined in \eqref{coeffs} and we assume through this article that $Z_0>0$.
\paragraph{Local Ward identities} The Ward identities can be derived in general through a Noether procedure using the invariance (up to anomalies) of $S\sbtx{ren}$ under boundary diffeomorphisms, U(1) gauge transformations and Weyl rescaling of the sources.
Such a derivation does not reply on the expressions \eqref{D-1pt-fns} for the 1-point functions in terms of the coefficients of the Fefferman-Graham expansions. However, for any specific model, the Ward identities can be verified explicitly using the asymptotic analysis in appendix \ref{asexp}. In particular, the relations~\eqref{divg3} and~\eqref{divA} imply that the 1-point functions \eqref{D-1pt-fns} satisfy the identities 
\begin{align}\label{D-WIs}
&-D_{(0)}^j\langle \ct_{ij}\rangle\sbtx{D}+\langle\mc O_{\Delta_+}\rangle\sbtx{D}\p_i\dil0+\langle\mc O_{\psi_I}\rangle\sbtx{D}\p_i\ax0
+\langle \cj^j\rangle\sbtx{D} F_{ij}^{(0)}=0,\NO
\\
&D_{(0)i}\langle \cj^i\rangle\sbtx{D}=0,
\end{align}
reflecting respectively boundary diffeomorphism and U(1) gauge invariance. Moreover, 
\eqref{trg3} leads to the trace Ward identity
\eq\label{D-trace-WI}
-\langle \ct^i{}_i\rangle\sbtx{D}
+(d-\Delta_+)\langle\mc O_{\Delta_+}\rangle\sbtx{D}\dil0=\ca(\vf_{(0)}),
\eeq
where $\ca(\vf_{(0)})$ is the conformal anomaly. For $d=3$ the only possible contribution to the anomaly is due to the scalar operator dual to $\f$, but for the particular potentials we consider in Theories I and II this contribution is zero and so the conformal anomaly vanishes.

\paragraph{Global Ward identities for axions} As we saw in section \ref{model}, both the bulk Lagrangian \eqref{genAction} and the counterterms \eqref{counterterms} are invariant under the global $\textsf{ISO}(d-1)$ transformation $\j_I\to\L^J_I\j_J+c_I$, and hence so is the renormalized action $S\sbtx{ren}$. Applying the Noether procedure to these global symmetries leads to additional {\em global} Ward identities, or {\em selection rules}, which cannot be derived from the local constraints of the bulk dynamics. In particular, an infinitesimal transformation of this form in \eqref{deltaS} leads to the integral constraints  
\be\label{globalWIs}
\int_{\pa\cm}\hspace{-0.1cm} d^dx\sqrt{-\g0}\;\<\mc O_{\psi_I}\>=0,\qquad \int_{\pa\cm}\hspace{-0.1cm} d^dx\sqrt{-\g0}\;\j^{(0)}_{[I}\langle\mc O_{\psi_{J]}}\rangle=0,
\ee
where the axion indices are antisymmetrized in the second identity. 

The second identity is a special feature of our model \eqref{genAction} while the first identity is a fundamental property of axion fields -- it corresponds to gauge transformations for $0$-forms. Contrary to higher rank forms, the gauge freedom of $0$-forms does not correspond to a shift by local exact form, but rather a shift by a constant. For $0$-forms, therefore, the analogue of the local current conservation is the first global constraint in \eqref{globalWIs}.

\subsection{Mixed boundary conditions and multi-trace deformations}

Since the AdS mass of the scalar field $\f$ in both Theories I and II is in the window \eqref{window}, both scalar modes are normalizable and one can impose more general boundary conditions on $\f$, corresponding to keeping a generic function $J(\dil0,\dil1)$ fixed on the boundary. As we reviewed above, Dirichlet boundary conditions correspond to choosing the source as $J\sbtx{D}=\dil0$, in which case the dual scalar operator has conformal dimension $\Delta_+$. Choosing instead $J\sbtx{N}=-\ell^2\dil1$ as the source corresponds to Neumann boundary conditions for the scalar and accordingly the dual operator has dimension $\Delta_-$. The choice 
\begin{equation}
J_\cf=-\ell^2\dil1-\cf'(\dil0),
\end{equation}
 where $\cf(\dil0)$ is a polynomial of degree $n$ with $2\leq n \leq d/\Delta_-$, corresponds to a multi-trace deformation of the Neumann theory by $\cf(\co_{\Delta_-})$ \cite{ Witten:2001ua, Berkooz:2002ug, Mueck:2002gm,Papadimitriou:2007sj}.   

The choice of source $J(\dil0,\dil1)$ that we want to keep fixed on the boundary dictates the additional {\em finite} boundary term that we must add to the renormalized on-shell action for the Dirichlet problem, $S\sbtx{ren}$. For $J_\cf=-\ell^2\dil1-\cf'(\dil0)$ the required additional term is 
\eq\label{Sf}
S_\cf
=\int_{r=\ep}\hspace{-0.2cm}d^3x\sqrt{-\g0}\lp
J_\cf\dil0+\cf(\dil0)\rp, 
\eeq
since 
\begin{align}
\delta_\phi (S_{\textsf{reg}}+S_{\textsf{ct}}+S_\cf)&=\int_{r=\ep}\!d^3x\sqrt{-\g0}\lp
-\ell^2\dil0\delta\dil1-\dil0\cf''(\dil0)\delta\dil0
\rp+\mc O(\ep)
\nonumber\\
&=\int_{r=\ep}\!d^3x\sqrt{-\g0}\,
\dil0\,\delta J_\cf+\mc O(\ep).
\end{align}
This variation also allows us to read off the corresponding scalar 1-point function in the theory where $J_\cf$ is the source, namely\footnote{Correlation functions with subscript ${}\sbtx{D}$ or ${}\sbtx{N}$ denote respectively observables computed in the theories defined by Dirichlet and Neumann boundary conditions for the scalar $\f$. Correlation functions without any suffix denote observables in the theory where $J_\cf=-\ell^2\dil1-\cf'(\dil0)$ is the source of the dual scalar operator, which is the theory relevant for the black hole solutions we are interested in. The function $\cf$ is uniquely determined by these black hole solutions.}
\eq
\langle\mc O_{\Delta_-}\rangle=\dil0.
\eeq

Besides the 1-point function of the scalar operator dual to $\f$, changing the boundary conditions also affects the expression for the renormalized stress tensor since the extra boundary term \eqref{Sf} depends on the boundary metric $g_{(0)}$ \cite{Papadimitriou:2007sj}. In particular, a general variation of 
\be\label{Sren-mixed}
S'_{\textsf{ren}}=\lim_{\ep\rightarrow0}\lp S_{\textsf{reg}}+S_{\textsf{ct}}+S_\cf\rp,
\ee
takes the form
\be
{\delta S}'_{\textsf{ren}}=\int\!d^3x\sqrt{-\g0}\(\frac{1}2\< \ct^{ij}\>\d\gij0+\< \cj^i\>\d A_i^{(0)}+\<\mc O_{\psi_I}\>\d\ax0+\<\mc O_{\Delta_-}\>\d J_\cf\),
\label{deltaS-f}
\ee
where now 
\begin{align}
\langle\mc O_{\psi_I}\rangle&=3\ell^2\ax3,&
\quad
\langle\mc O_{\Delta_-}\rangle&=\dil0,\NO\\
\langle \cj^i\rangle&=Z_0A_{(1)}^i,&
\quad
\langle \ct^{ij}\rangle&=3\ell^2g_{(3)}^{ij}+\lp \cf(\dil0)-\dil0\cf'(\dil0)\rp g_{(0)}^{ij}.\label{M-1pt-fns}
\end{align}
Note that the 1-point functions for the current $\cj^i$ and the scalar operators $\co_{\j_I}$ are unchanged relative to \eqref{D-1pt-fns}, but the stress tensor and the operator dual to the scalar $\f$ have been modified.  

\paragraph{Local Ward identities} 
Using the relation between the 1-point functions \eqref{D-1pt-fns} and \eqref{M-1pt-fns}, as well as the expression for the new scalar source $J_\cf$, it is straightforward to rewrite the Ward identities \eqref{D-WIs} and \eqref{D-trace-WI} in terms of the variables relevant for the new boundary conditions. This leads to the new Ward identities
\bal\label{M-WIs}
&-D_{(0)}^j\langle \ct_{ij}\rangle+\langle\mc O_{\Delta_-}\rangle\p_iJ_\cf+\langle\mc O_{\psi_I}\rangle\p_i\ax0
+\langle \cj^j\rangle F_{ij}^{(0)}=0,\NO\\
&D_{(0)i}\langle \cj^i\rangle=0, 
\eal
and
\eq\label{M-trace-WI}
-\langle \ct^i{}_i\rangle+(d-\Delta_-)\dil0J_\cf+d\cf(\dil0)-\Delta_-\dil0\cf'(\dil0)=0,
\eeq
where we have assumed that the conformal anomaly $\ca$ vanishes in \eqref{D-trace-WI}, as is the case for the examples we are interested in here. 
Note that the trace of the stress tensor gets a new contribution due to the multi-trace deformation $\cf(\co_{\Delta_-})$. As expected, this contribution vanishes if and only if the multitrace deformation is marginal, i.e. $\cf(\vf_{(0)})\propto \vf_{(0)}^{d/\Delta_-}$. The black hole solutions we are going to discuss turn out to satisfy precisely such marginal triple-trace boundary conditions with $d=3$ and $\Delta_-=1$, which we parameterize in terms of a marginal coupling $\vth$ as  
\eq\label{bc}
\cf(\dil0)=\frac13\vartheta\dilp03.
\eeq
We are going to specify the value of the coupling $\vth$ for the solutions of Theories I and II later on.

\section{Planar black holes and their properties}
\label{sec::solutions}

In this section we present a number of black hole solutions of both Theories I and II and study their properties. Using the holographic dictionary discussed in section \ref{dictionary}, we identify the boundary conditions that they are compatible with and interpret these solutions holographically. The solutions we discuss are not new, except for a new magnetic black hole solution for Theory II.

\subsection{Exact planar solutions of Theory I}

\paragraph{Bald solution}
When $\phi=0$ the theory reduces to AdS gravity coupled to a gauge field $F$ and the two free scalars $\psi_i$. Electrically charged planar black solutions of this theory, with an axion profile breaking translational invariance, were found in \cite{Bardoux:2012aw} and take the form
\begin{align}
&ds^2=-f(r)dt^2+f^{-1}(r)dr^2+r^2\( dx^2+dy^2\),
\label{bald}\NO\\
&f(r)=-\frac{p^2}2-\frac{m}{r}+\frac{q^2}{4r^2}+\frac{r^2}{\ell^2},
\qquad \psi_1=px,\qquad \psi_2=py,\nonumber\\
&A=\mu\lp1-\frac{r_0}{r}\rp\dd t,
\qquad F=\dd A=-\frac{q^2}{r^2}\dd t\wedge\dd r,
\qquad\text{with $q=\mu r_0$.}
\end{align}
The parameters $m$, $p$, and $q$ are related to the mass, axion charge, and $U(1)$ charge densities respectively. The event horizon is located at $r_0$, the largest zero of $f(r)$, and we have chosen a gauge in which $A$ vanishes on the horizon; $\mu$ is then interpreted as the chemical potential in the dual field theory. This metric was used in \cite{Andrade:2013gsa} as a simple holographic model of momentum relaxation. It should be noted that this metric can be easily extended to include a magnetic field, by taking $q^2=q_{\textsf{e}}^2+q_{\textsf{m}}^2$ in the above metric, and the gauge field to be of the form
\eq
A=\mu\lp1-\frac{r_0}{r}\rp\dd t+q_{\textsf{m}}x\,\dd y,
\qquad F=\dd A=\frac{q_{\textsf{e}}}{r^2}\,\dd t\wedge\dd r+q_{\textsf{m}}\,\dd x\wedge\dd y,
\qquad\text{with $q_{\textsf{e}}=\mu r_0$.}
\label{dyonic}\eeq

\paragraph{Electric hairy solution}
The electrically charged hairy black brane solution found in~\cite{Bardoux:2012tr} can be transformed into a solution of Theory~I by rewriting it in the Einstein frame, where it becomes 
\begin{align}
&ds^2=\Omega(r)\lp-f(r)dt^2+f^{-1}(r)dr^2+r^2 (dx^2+dy^2)\rp,\NO\\
&\Omega(r)=1-\frac{v^2}{(r+\sqrt\alpha\, v)^2},\qquad
f(r)=\frac{r^2}{\ell^2}-\frac{p^2}2\lp1+\frac{\sqrt\alpha\,v}r\rp^2,
\nonumber
\\
&\phi(r)=2\sqrt{3}\tanh^{-1}\!\lp\frac{v}{r+\sqrt\alpha\,v}\rp,
\qquad \psi_1=px,\qquad \psi_2=py,
\nonumber
\\
&A=\mu\lp1-\frac{r_0}{r}\rp\dd t,
\qquad F=\dd A=\frac q{r^2}\dd t\wedge\dd r,
\qquad\text{with $q=\mu r_0$.}
\label{hairySol}
\end{align}
As for the bald solution \eqref{bald}, the axion profile breaks translational invariance in the spatial boundary directions, but in this case the dialton $\f$ also has a non-trivial profile. The parameters $p$, and $q$ are again related respectively to the axion and electric charge densities, but importantly, $v$ (originally denoted by $m$ in \cite{Bardoux:2012tr}) is in fact not an independent parameter. The equations of motion require that it is related to the charge densities $p$, $q$ and the parameter $\a$ in the scalar potential as $q^2=2p^2v^2(1-\alpha)$, which in turn implies that the solution \eqref{hairySol} only exists for $0\leq\alpha\leq1$. As we shall see later on, $v$ is related to the vacuum expectation value ({\em vev}) of the scalar operator dual to $\f$. The event horizon of these solutions is located at $r_0$, the largest zero of $f(r)$, and we again choose a gauge in which $A$ vanishes on the horizon so that $\mu$ corresponds to the chemical potential in the dual field theory.

\paragraph{Dyonic hairy solution}
In analogy with the bald solution, the hairy solution \eqref{hairySol} can be easily extended to include a constant magnetic field since we are in four dimensions. The main difference is that for the dyonic solution the parameter $v$ is related to the  
the electromagnetic duality invariant quantity $q_{\textsf{e}}^2+q_{\textsf m}^2$ instead of the electric charge alone, i.e. 
\eq
q_{\textsf{e}}^2+q_{\textsf m}^2=2p^2v^2(1-\alpha).
\label{constraint}\eeq
The metric, dialton, and axion fields are the same as in the purely electric solution~\eqref{hairySol}, but the electromagnetic field picks up an extra magnetic contribution, as given in equation~\eqref{dyonic}.
The resulting planar black hole is dyonic, and carries both an electric charge density $q_{\textsf{e}}$ and a magnetic charge density $q_{\textsf m}$. 
When the magnetic field is turned on, the parameter space changes slightly, but the blackening function $f(r)$, the conformal factor $\Omega(r)$, and the dialton profile remain unchanged, and therefore the geometric and thermodynamic properties of these dyonic branes can be derived in a straightforward way from those of the electrically charges branes.

\subsection{Exact planar solutions of Theory II}

\paragraph{Bald solutions}
In Theory II with $0<\x<2$, the gauge field $A$ sources the dialton via the coupling $Z(\phi)F^2$. Hence, bald solutions of this theory are necessarily neutral 
and take the form
\begin{align}
ds^2&=-f(r)dt^2+\frac{dr^2}{f(r)}+r^2d\vec x^2\,,
&\psi_I&=px^I,\nonumber\\
f(r)&=\frac{r^2}{\ell^2}-\frac{r_0^3}{r\ell^2}-\frac{p^2}{2}\lp1-\frac{r_0}{r}\rp\,,
&\phi&=0,\qquad A=0.
\label{baldII}\end{align}
This solution depends on the two parameters $r_0$ and $p$, and, since for $\f=0$ and $A=0$ Theory~I and Theory~II coincide, is the same as the bald solution~\eqref{bald} of Theory~I, with appropriate identification of the parameters. 
The only case where Theory~II admits charged bald solutions is in the limit $\x\to2$ so that $Z(\f)\to 1$. However, as we saw in \eqref{action_nodil}, this leads to a truncation of Theory~I and so we need not consider this case further.

\paragraph{Electric hairy solution}
A family of electrically charged hairy black hole solutions of Theory~II was presented in \cite{Gouteraux:2014hca} and reads\footnote{This fixes a typo in equation~(C.5) of reference~\cite{Gouteraux:2014hca}.}
\begin{align}
ds^2&=-f(r)h(r)^{\frac{-4}{2+(d-2)(d-1)\delta^2}}dt^2+h(r)^{\frac4{(d-2)(2+(d-2)(d-1)\delta^2)}}
\lp f^{-1}(r)dr^2+r^2d\vec x^2\rp,\nonumber\\
f(r)&=\frac{r^2}{\ell^2}\lp h(r)^{\frac{4(d-1)}{(d-2)(2+(d-2)(d-1)\delta^2)}}-\frac{r_h^d}{r^d}
h(r_h)^{\frac{4(d-1)}{(d-2)(2+(d-2)(d-1)\delta^2)}}\rp-\frac{p^2}{2(d-2)}\lp1-\frac{r_h^{d-2}}{r^{d-2}}\rp,
\nonumber\\
\phi&=\frac{-2(d-1)\delta}{2+(d-2)(d-1)\delta^2}\log h(r),\qquad
h(r)=1+\frac{v\sbtx{e}}{r^{d-2}},\qquad \psi_I=px^I,\nonumber\\
A&=\frac{\sqrt{4(d-1)v\sbtx{e}\((d-2)\frac{r_h^{d+2}}{\ell^2}h(r_h)^{\frac{2(2-(d-2)^2(d-1)\delta^2)}{(d-2)(2+(d-1)(d-2)\delta^2)}}-\frac{p^2r_h^d}{2h(r_h)}\)}}{(d-2)r_h^{d-1}h(r)\sqrt{2+(d-2)(d-1)\delta^2}}\lp1-\frac{r_h^{d-2}}{r^{d-2}}\rp\,\dd t.
\label{genSol}
\end{align}
This family of solutions is valid for arbitrary boundary dimension $d$ and depends on the three parameters $r_h$, $p$, and $v\sbtx{e}$ (called $Q$ in \cite{Gouteraux:2014hca}). However, this parameterization of the solution treats the electric charge density as a dependent parameter, expressed in terms of the radius of the horizon $r_h$ and the parameter $v\sbtx{e}$, which, as we will see, is proportional to the {\em vev} of the scalar operator dual to $\f$. This not only obscures the limiting process of taking the charge density to zero, but also does not reflect the change of the sign of the gauge potential when the charge density changes sign.  

Focusing on the four-dimensional case, i.e. $d=3$, from now on, we will therefore adopt an alternative parameterization of the solution \eqref{genSol}, by introducing explicitly the charge density $q\sbtx{e}$ as an independent parameter, in addition to replacing $\delta$ with $\x$ as in~\eqref{delta}. With these modifications the solution takes the form  
\begin{align}
	ds^2&=-f(r)h^{-\xi}(r)dt^2+h^{\xi}(r)
	\lp f^{-1}(r)dr^2+r^2d\vec x^2\rp,\nonumber\\
	f(r)&=\frac{r^2}{\ell^2}h^{2\xi}(r)-\frac{p^2}2h(r)-\frac{q\sbtx{e}^2}{2\x v\sbtx{e}r},
	\NO\\
	\phi&=-\sqrt{\xi(2-\xi)}\log h(r)\,,\qquad h(r)=1+\frac{v\sbtx{e}}{r}\,,
	\nonumber\\
	A&=q\sbtx{e}\(\frac{1}{r_0 h(r_0)}-\frac{1}{rh(r)}\)
	\,\dd t\,,\qquad \hskip-0.2cm \psi_I=px^I.
\label{bsolxi}\end{align}
Here $r_0$ is the largest zero of $f(r)$ and we are in a gauge in which $A$ vanishes on the horizon.
Notice that the expression for the blackening factor $f(r)$ implies that as long as $q\sbtx{e}\neq 0$ we must necessarily have $v\sbtx{e}\neq 0$, and hence the scalar field must have a non-trivial profile. In summary, $q\sbtx{e}\neq 0$ implies that $v\sbtx{e}\neq 0$, but the converse is not true.  

\paragraph{New magnetic hairy solution} We have been able to find in addition an analytical black brane solution that is purely magnetically charged and takes the form  
\bal\label{mag-sol-II}
ds^2&=-f(r)h^{-(2-\x)}(r)dt^2+h^{2-\x}(r)\(\frac{dr^2}{f(r)}+r^2d\vec{x}^2\),\NO\\
f(r)&=h(r)\(\frac{r^2}{\ell^2}+\frac{(3-2\x)v\sbtx{m} r}{\ell^2}\(1+\frac{(1-\x)v\sbtx{m}}{r}\)-\frac{p^2}{2}+\frac{q\sbtx{m}^2}{2\x v\sbtx{m} r}\),\NO\\
\f&=-\sqrt{\x(2-\x)}\;\log h(r),\qquad h(r)=1+\frac{v\sbtx{m}}{r},\NO\\
A&= q_{\textsf{m}}\,x\dd y,\qquad \j_I=px^I.
\eal
In addition to the axion charge density $p$, this solution is parameterized by the  magnetic charge density $q\sbtx{m}$ as well as the independent parameter $v\sbtx{m}$, which as we will see later, is again related to the {\em vev} of the scalar operator dual to $\f$.

\subsection{Black hole properties: horizons and extremality}
\label{ss::horizons}

\paragraph{Bald black branes}
Let us consider first the bald dyonic solution~\eqref{bald}-\eqref{dyonic} of Theory~I. This analysis also covers the bald solutions~\eqref{baldII} of Theory~II, since the latter can be embedded in the former as the neutral $\qe=\qm=0$ subfamily.
For these bald solutions the mass parameter can be expressed in terms of $(r_0,p,\mu,\qm)$ by solving $f(r_0)=0$ for $m$, namely
\eq
m=\frac{r_0^3}{\ell^2}\lp1+\frac{\ell^2}{4r_0^2}\lp\mu^2-2p^2\rp+\frac{\ell^2}{4r_0^4}\qmp2\rp.
\eeq
The temperature of the black brane is then given by
\eq
T=\frac{f'(r_0)}{4\pi}=\frac{3r_0}{4\pi\ell^2}\lp1-\frac{\ell^2}{12r_0^2}\lp\mu^2+2p^2\rp-\frac{\ell^2}{12r_0^4}\qmp2\rp.
\label{Tbald}\eeq
The solution becomes extremal when $r_0$ is a double zero of $f(r)$, or equivalently, when its temperature $T$ vanishes. The location of the extremal horizon is at 
\eq
r_{0,\mathsf{extr}}^2=\frac{\ell^2}{24}\lp\mu^2+2p^2\rp
\lp1+\sqrt{1+\frac{48\qmp2}{\ell^2(\mu^2+2p^2)^2}}\rp.
\eeq
The entropy density 
-- defined as $s=a_{\textsf{h}}/4G_{\textsf{N}}$, with $a_{\textsf{h}}=r_0^2$ the area density of the horizon and using our convention that $16\pi G_{\textsf{N}}=1$ -- is given by
\eq s=\frac{r_0^2}{4G_\mathsf{N}}=4\pi r_0^2,\eeq
and remains finite in the extremal $T=0$ case. The near horizon geometry can be obtained by  defining $r=r_0+\epsilon\rho$, $t=\tau/\epsilon$, and taking the $\epsilon\rightarrow0$ limit, resulting in the AdS$_2\times\mathbb{R}^2$ geometry
\eq
ds^2_{\mathsf{IR}}=-\frac{\rho^2}{\ell^2_{\mathsf{IR}}}d\tau^2+\frac{\ell^2_{\mathsf{IR}}}{\rho^2}d\rho^2+r_0^2\lp dx^2+dy^2\rp,
\eeq
where the radius of the AdS$_2$ factor is given by
\eq
\ell^2_{\mathsf{IR}}=\frac2{f''(r_{0,\mathsf{extr}})}
=\frac{2r_{0,\mathsf{extr}}^4}{(p^2+\mu^2)r_{0,\mathsf{extr}}^2+\qmp2}.
\eeq
When $\qm=0$, this expression reduces to \cite{Andrade:2013gsa} 
\eq
\ell^2_{\mathsf{IR}}=\frac{\ell^2}6\frac{\mu^2+2p^2}{\mu^2+p^2}.
\eeq
The full geometry, therefore, interpolates between AdS$_4$ (in Poincar\'e coordinates) in the UV and a near horizon AdS$_2\times\mathbb{R}^2$ geometry in the IR.

\subpar{Symmetry enhancement:}
As shown by Davison and Gout\'eraux \cite{Davison:2014lua}, for particular values of the parameters the bald solution becomes exactly conformal to AdS$_2\times\mathbb R^2$, and so it possesses an enhanced $SL(2,\mathbb{R})\times SL(2,\mathbb{R})$ symmetry. In that case one can solve the linearized perturbation equations exactly in terms of hypergeometric functions.\footnote{
Notice that such a symmetry is also enjoyed by the scalar wave equation in the nonextremal Kerr black hole background, in the low frequency limit \cite{Castro:2010fd} (see \cite{Bertini:2011ga} for the Schwarschild black hole case). This \textit{hidden conformal symmetry} is not derived from an underlying symmetry of the spacetime itself, but is rather related to the fact that black hole scattering amplitudes are given in terms of hypergeometric functions, which are well-known to form representations of the conformal group $SL(2,\mathbb{R})$. What is notable in the bald black hole case is that this symmetry becomes an exact symmetry of the linearized gravitational perturbation equations for those values of the parameters.}
This happens precisely when the form of the lapse function simplifies to
\eq
f(r)=-\frac{p^2}2+\frac{r^2}{\ell^2},
\label{enhancedf}\eeq
i.e.~when $m$ and $q$ both vanish. The crucial point is that the metric becomes \textit{conformal} to a patch of AdS$_2\times\mathbb R^2$, which also happens for  the hairy solution~\eqref{hairySol} when $\alpha=0$ (besides the case $v=0$ that coincides with the bald solution). It is however unlikely that in the case of the hairy solution with $\a=0$ the coupled linearized perturbation equations for the fields $g_{\mu\nu}$, $A_\m$, $\phi$, and $\psi_I$ remain exactly solvable as a consequence of the enhanced symmetry. This is left for further exploration.
\paragraph{Hairy black branes of Theory~I}
We first need to find the range of parameters for which the solution supports a regular event horizon. Without loss of generality, we take $p\geq0$. Then $f(r_0)=0$ can be solved for its largest real root, determining the location of the horizon to be
\eq
r_0=\left\{
\begin{array}{lll}
\displaystyle \frac{\ell p}{2\sqrt{2}}\lp1+\sqrt{1+4\sqrt{2\alpha}\frac{v}{\ell p}}\rp
& \text{for } v\geq-\frac{\ell p}{4\sqrt{2\alpha}}
& \text{(case A)},\\
&\\
\displaystyle -\frac{\ell p}{2\sqrt{2}}\lp1-\sqrt{1-4\sqrt{2\alpha}\frac{v}{\ell p}}\rp
& \text{for } v<-\frac{\ell p}{4\sqrt{2\alpha}}
& \text{(case B)}.
\end{array}
\right.
\eeq
However, not all of these are genuine black branes. If the conformal factor $\Omega(r)$ vanishes outside the horizon, the geometry suffers from a naked singularity. As long as the largest zero of $\Omega(r)$, namely $r_\Omega=|v|-\sqrt{\alpha}\,v$, is smaller than $r_0$, there is an event horizon hiding the singularity. This condition translates to\footnote{We will see below, however, that only solutions with $v>0$ are dynamically stable.}
\eq
-\frac{\ell p}{\sqrt{2}(1+\sqrt{\alpha})^2}<v<\frac{\ell p}{\sqrt2(1-\sqrt{\alpha})^2},
\label{regularity}\eeq
and restricts us to a subset of case~A only; all case~B geometries display a naked singularity. In conclusion, for parameters $(p,v;\alpha)$ satisfying the relation \eqref{regularity} we get a regular black brane with
\eq\label{r0}
r_0=\frac{\ell p}{2\sqrt{2}}\lp1+\sqrt{1+4\sqrt{2\alpha}\frac{v}{\ell p}}\rp.
\eeq
Alternatively, we can invert this relation, and express all quantities in terms of the parameters $(p,r_0;\alpha)$ using
\eq
v=\sqrt{\frac2\alpha}\frac{r_0^2}{\ell p}-\frac{r_0}{\sqrt{\alpha}}.
\eeq
The regularity condition~\eqref{regularity} then becomes
\eq
\frac{\ell p}{\sqrt{2}(1+\sqrt{\alpha})}<r_0<\frac{\ell p}{\sqrt2(1-\sqrt{\alpha})}.
\label{r0range}\eeq
The stability condition $v>0$ (see section \ref{sec:phases}) then further restricts 
\eq
\frac{\ell p}{\sqrt2}<r_0<\frac{\ell p}{\sqrt2(1-\sqrt{\alpha})}.
\label{r0range-stable}\eeq 

\subpar{Absence of extremal horizons:}
The largest zero of $f(r)$ becomes a double zero located at $r_0=\ell p/2\sqrt{2}$ when $v=-\ell p/4\sqrt{2\a}$, and so extremal solutions would be dynamically unstable, if they existed. 
However, when $v=-\ell p/4\sqrt{2\a}$ the conformal factor $\Omega(r)$ vanishes at
\eq\label{rOmega}
r_\Omega=\frac{1+\sqrt\alpha}{4\sqrt{2\alpha}}\ell p >r_0,
\eeq
and, hence, extremal solutions are singular: there is no extremal limit of the hairy black brane.\footnote{When $\alpha=1$ we have $r_\Omega=r_0$ and so $\Omega$ vanishes on the horizon. However, the entropy vanishes too in that case.}

\subpar{Temperature and entropy density:}
The temperature of the hairy branes is not affected by the presence of the conformal factor $\Omega(r)$ and takes the form
\eq
T=\frac{f'(r_0)}{4\pi}=\frac1{\pi\ell^2}\lp r_0-\frac{p\ell}{2\sqrt2}\rp.
\label{Thairy}\eeq
Given that for regular hairy black branes the range of $r_0$ is limited by~\eqref{r0range-stable}, we see that the temperature of the hairy black branes is bounded both from below and above according to
\eq
\frac{p}{2\sqrt2\pi\ell}<T<\frac{1+\sqrt\alpha}{1-\sqrt\alpha}\,\frac{p}{2\sqrt2\pi\ell}.
\label{Tbound}\eeq
Hence, the temperature of the brane can never vanish, as expected since the hairy solution \eqref{hairySol} becomes singular in the extremal limit.

The fact that the temperature is a fixed function of the charge densities, as follows by combining the expressions \eqref{Thairy}, \eqref{r0} and \eqref{constraint}, as well as that this temperature has a lower non-zero bound, are some of the puzzling features of the hairy solutions \eqref{hairySol} of Theory I. A related property, reminiscent of extremal solutions, is that the entropy density of these black branes is also uniquely determined in terms of the charge densities and is given by
\eq 
s=\frac{r_0^2\,\Omega(r_0)}{4G_\mathsf{N}} = 4\pi r_0^2\,\Omega(r_0).
\label{sHairyI}
\eeq 
However, as we have seen, the solution cannot be extremal. We believe that the reason behind these unusual properties is that in fact the analytic solution \eqref{hairySol} is a single member of a continuous family of solutions, where the temperature, or the mass, is a free parameter. It would be interesting to find these solutions.

\paragraph{Hairy black branes of Theory~II}
Starting with the electric solution \eqref{bsolxi}, we again need to know the range of parameters for which these solutions describe regular black branes, and to find the location of their event horizon.
The solution has singularities at both $r=0$ and $r=-v\sbtx{e}$ (where $h(r)$ vanishes). To be regular, the solution must thus have an event horizon at a location $r_0$ such that $r_0>0$ 
and $r_0>-v\sbtx{e}$.
Horizons correspond to the zeros of the function $f(r)$ and can be determined in terms of the electric and axion charge densities, respectively $p$ and $q\sbtx{e}$, as well as the parameter $v\sbtx{e}$, by solving the (generically transcendental) equation 
\be\label{horizon-eq}
\frac{r^2_0}{\ell^2}h^{2\xi}(r_0)-\frac{p^2}2h(r_0)-\frac{q\sbtx{e}^2}{2\x v\sbtx{e}r_0}=0, \qquad h(r_0)=1+\frac{v\sbtx{e}}{r_0}.
\ee 
For $\x=1,\, 1/2,\, 3/2$, this equation is cubic in $r_0$ and the roots can be found explicitly, although they are still highly involved expressions.

The temperature of these black branes can be computed as usual by requiring that there is no conical singularity in the Euclidean section of the metric, giving
\begin{eqnarray}
T&=&\frac1{4\pi}
\frac{d}{dr}\left.\lp\frac{f(r;q\sbtx{e},v\sbtx{e},p)}{h^\xi(r)}\rp\right|_{r=r_0} \nonumber \\
&=& \frac{1}{4 \pi \ell^2h^\x(r_0)} \left( 2 r_0 h^{2 \xi}(r_0) - v\sbtx{e} \xi h^{2 \xi-1}(r_0)+ \frac{p^2 \ell^2 v\sbtx{e} (1-\xi)}{2 r_0^2}  + \frac{q\sbtx{e}^2 \ell^2 (r +(1-\xi)v\sbtx{e})}{2 \xi v\sbtx{e}r^3 h(r_0)}\right).
\label{ThairyII}\end{eqnarray}
Moreover, the entropy density is given by the area density of the event horizon,
\eq
s=\frac{r_0^2\,h^\xi(r_0)}{4G_\mathsf{N}}=4\pi r_0^2\, h^\xi(r_0).
\label{sHairyII}
\eeq

In contrast to the hairy solutions of Theory~I, these solutions admit extremal limits, corresponding to the cases where the temperature~\eqref{ThairyII} vanishes.
Combining this condition with the defining equation $f(r_0)=0$ for the horizon determines the location of the {\em extremal} horizon to be
\be\label{r-extr-e}
r_0^{\tt ex}=\frac{\x(2\x-5) p^2v\sbtx{e}^2-3 q\sbtx{e}^2+\text{sgn}(v\sbtx{e})\sqrt{9q\sbtx{e}^4+\x^2(2\x-1)^2p^4v\sbtx{e}^4+2\x(2\x+3)p^2q\sbtx{e}^2v\sbtx{e}^2}}{4\x p^2 v\sbtx{e}}.
\ee
Requiring in addition that $f(r_0^{\tt ex})=0$ gives the extremality condition, which can be expressed in the form $v\sbtx{e}^{\tt ex}(q\sbtx{e},p)$. It is not possible to obtain this extremality condition analytically for generic $\x$, but it can be done for specific values.\footnote{The extremality condition can also be determined by requiring that the  discriminant of the polynomial $f(r)$ vanishes, in which case the horizon becomes a multiple root. } The simplest case is
\bal
&\x=1,\qquad v\sbtx{e}=-\sqrt{\frac{\ell\(9\ell p^2 q\sbtx{e}^2-\ell^3p^6+(\ell^2p^4-6q\sbtx{e}^2)^{3/2}\)}{\ell^2 p^4-8 q\sbtx{e}^2}},
\eal
in which case the extremal horizon simplifies to (see \eqref{r0-e})
\be\label{r-extr-e-xi=1}
r_0^{\tt ex}=-v\sbtx{e}+\frac{(6q\sbtx{e}^2-\ell^2 p^4)v\sbtx{e}}{9q\sbtx{e}^2+p^2v\sbtx{e}^2}.
\ee
However, as we shall see, the energy density for this solution, given in equation \eqref{energyDensityIIe}, is negative and so it is dynamically unstable (see section \ref{sec:phases}).

The magnetically charged solutions \eqref{mag-sol-II} can be studied similarly. The location $r_0$ of the horizon is determined by the equation
\be\label{horizon-eq-m}
\frac{r^2_0}{\ell^2}+\frac{(3-2\x)v\sbtx{m} r_0}{\ell^2}\(1+\frac{(1-\x)v\sbtx{m}}{r_0}\)-\frac{p^2}{2}+\frac{q\sbtx{m}^2}{2\x v\sbtx{m} r_0}=0,
\ee
which in this case is cubic for arbitrary $\x$. The generic expression for $r_0$, however, is still too lengthy to usefully reproduce here. The same applies to the temperature and entropy density, which are given by 
\begin{align}\label{TsTheoryIIm}
		T&=\frac1{4\pi}
		\frac{d}{dr}\left.\lp\frac{f(r;q\sbtx{m},v\sbtx{m},p)}{h^{2-\xi}(r)}\rp\right|_{r=r_0}
		=\frac{h^{\xi-1}(r_0)}{8\pi\ell^2r_0}
		\lp
		-\ell^2p^2+6r_0^2h^2(r_0)-8v\sbtx{m}r_0\xi+2v\sbtx{m}^2\xi(2\xi-5)
		\rp,
		\nonumber\\
		s&=\frac{r_0^2\,h^{2-\xi}(r_0)}{4G_\mathsf{N}}=4\pi r_0^2\, h^{2-\xi}(r_0).
\end{align}
As for the electrically charged solutions, extremal solutions can be found, even analytically for specific values of the parameters. The simplest case is again
\bal
&\x=1, \qquad v\sbtx{m}=\sqrt{\frac{\ell\(9\ell p^2 q\sbtx{m}^2-\ell^3p^6+(\ell^2p^4-6q\sbtx{m}^2)^{3/2}\)}{\ell^2 p^4-8 q\sbtx{m}^2}},
\eal
with the extremal horizon given by (see \eqref{r0-m})
\be
r_0^{\tt ex}=\frac{9\ell^2q\sbtx{m}^2+\ell^2p^2v\sbtx{m}^2}{2v\sbtx{m}\(3\ell^2p^2+2v\sbtx{m}^2\)}. 
\ee
Again, the energy density \eqref{energyDensityIIm} is negative for this solution, and so it is also dynamically unstable.

\paragraph{Scaling symmetry}
Finally, all the families of solutions presented above enjoy a scaling symmetry: they are left invariant by the scaling $(t, \vec x, r)\rightarrow(\lambda t,\lambda\vec x, \lambda^{-1}r)$ of the coordinates, when accompanied by the rescaling $(p,v\sbtx{e/m},q\sbtx{e/m})\rightarrow(\lambda^{-1}p,\lambda^{-1}v\sbtx{e/m},\lambda^{-2} q\sbtx{e/m})$ of the parameters. Under such a rescaling, the temperature and entropy transform according to $T\rightarrow\lambda^{-1}T$ and $s\rightarrow\lambda^{-2}s$. This invariance will simplify the study of the phases of these black branes, since it allows us to scale away one of the parameters such as the axion charge density $p$, as long as it is non vanishing.

\section{Black hole thermodynamics}
\label{sec:thermo}

We now turn to the thermodynamics of the black holes presented in section \ref{sec::solutions}. Here we will define only some of the thermodynamic variables from first principles, such as the temperature, entropy density and free energy, but other variables --in particular the thermodynamic potentials conjugate to the magnetic and axionic charge densities-- will be obtained through general thermodynamic relations. This means that the analysis below does not provide an independent confirmation of thermodynamic relations such as the first law. A first principles definition of all the thermodynamic variables for planar black holes with axionic charge, and correspondingly a general derivation of the first law without relying on specific solutions, will appear elsewhere \cite{thermo2}.  

The renormalized Euclidean generating function, when evaluated on black hole solutions, gives the Gibbs free energy, or the grand canonical potential, where all intensive variables are kept fixed \cite{Gibbons:1976ue}. Since the black holes we are interested in correspond to imposing mixed boundary conditions on the scalars specified by the parameter $\vth$ in \eqref{bc}, the generating function is given by $S'_{\textsf{ren}}$, defined in \eqref{Sren-mixed}. The grand canonical potential, $\cw$, is therefore related to the Euclidean generating function $S_\textsf{ren}'^E$ as   
\eq
S_\textsf{ren}'^E=-S_\textsf{ren}'=\b\mc W(T,\mc V,\mu,\mc B,\Pi),
\label{Wdef}\eeq
and is a function of the temperature $T=1/\b$, spatial volume $\cv$, chemical potential $\mu$, magnetic field $\mc B$, and axionic strength $\Pi$, which for an isotropic collection of axionic fields $\psi_I$ is defined as
\eq
\Pi=\frac1\ell\sqrt{\frac1{d-1}\sum_I(\p\psi_I)^2}\,.
\label{defpi}\eeq

An important simplification for spatially homogeneous systems, like the planar black holes we are interested in here, is that the renormalized on-shell action is simply proportional to the (formally infinite) spatial volume $\mc V$, while the corresponding free energy density
\eq\label{free-energy-density}
w(T,\mu,\mc B,\Pi)=\mc W(T,\cv,\mu,\mc B,\Pi)/\mc V,
\eeq
is independent of $\cv$. Variations of the free energy, therefore, satisfy
\eq
dw=-s\,\dd T-\rho\,\dd\mu-\mc M\,\dd\mc B-\varpi\,\dd\Pi,
\label{dw}\eeq
where we have defined the entropy density $s$, the charge density~$\rho$, the magnetization $\mc M$ and the axionic magnetization $\varpi$ as the conjugate variable to $T$, $\mu$, $\mc B$, and $\Pi$ respectively,
\eq
\lp\frac{\p{w}}{\p T}\rp_{\mu,\mc B,\Pi}=-s,\quad
\lp\frac{\p{w}}{\p\mu}\rp_{T,\mc B,\Pi}=-\rho,\quad
\lp\frac{\p{w}}{\p\mc B}\rp_{T,\mu,\Pi}=-\mc M,\quad
\lp\frac{\p{w}}{\p\Pi}\rp_{T,\mu,\mc B}=-\varpi.
\label{conjugate}\eeq
Performing a Legendre transformation with respect to the temperature and the chemical potential, we obtain a description of the system in terms of its internal energy density $\varepsilon$,
\eq
\varepsilon(s,\rho,\mc B,\Pi)=w+Ts+\mu\rho,
\label{energyDensity}\eeq
whose variation expresses the first law of thermodynamics for an infinitesimal volume,
\eq\label{first-law-density}
\dd\varepsilon = T\,\dd s+\mu\,\dd\rho-\varpi\,\dd\Pi-\mc M\,\dd\mc B.
\eeq

The total energy
\eq
\mc E(S,\mc V,Q_{\mathsf e},\mc B,\Pi)=\mc V\,\varepsilon\lp\frac{S}{\mc V},\frac{Q_{\mathsf e}}{\mc V},\mc B,\Pi\rp,
\eeq
depends naturally on the total entropy $S=s\mc V$, charge $Q_{\mathsf e}=\rho\mc V$, and volume $\mc V$ of the system. Allowing for volume variations, we obtain the first law of thermodynamics in its usual form,
\eq\label{first-law}
\dd\mc E=T\,\dd S-\mc P\,\dd \cv+\mu\,\dd Q_{\mathsf e}-(\varpi\mc V)\dd\Pi-(\mc M\mc V)\,\dd\mc B,
\eeq
where the pressure of the system has been defined as the conjugate variable to the volume, i.e.
\eq
\mc P=-\lp\frac{\p\mc E}{\p\mc V}\rp_{S,Q_{\mathsf e},\mc B,\Pi}\,.
\label{pressure}\eeq
Combining \eqref{first-law} with~\eqref{dw} and~\eqref{energyDensity}, we find that the pressure is related to the free energy as
\eq
w=-\mc P,\label{wP}
\eeq
and satisfies the Gibbs-Duhem relation
\eq\label{GD}
\varepsilon+\mc P=Ts+\mu\rho.
\eeq

This general thermodynamic analysis is valid for any planar black hole. To apply it to specific systems one needs in addition an equation of state, relating the thermodynamic variables. This is strongly constrained by symmetries, such as conformal invariance. As we shall see, the boundary conditions that the dialton satisfies in the black holes of Theories I and II discussed above do not explicitly break conformal invariance and hence, conformal invariance is only broken explicitly by the background magnetic field and axion charge. A simple scaling argument allows us to generalize the equation of state for conformal theories to theories where conformal symmetry is explicitly broken by magnetic and axionic charges, and we explicitly confirm that all planar black holes discussed in section \ref{sec::solutions} satisfy such an equation of state. Interestingly, the constraint~\eqref{constraint} between the parameters of the hairy black hole of Theory~I endows the system with a linear structure evoking the structure of supersymmetric black holes.

\subsection{Conformal thermodynamics in the presence of magnetic and axionic charge}
In a $d$-dimensional theory with no explicit breaking of conformal symmetry the stress tensor is traceless in any state. This implies that in a state of thermal equilibrium the system is governed by the equation of state $\mc P=\varepsilon/(d-1)$. As we now show, in the presence of a magnetic field $\mc B$, a chemical potential $\mu$ for the electric charge, and an isotropic collection of axionic fields with charge density $\Pi$ as defined by~\eqref{defpi}, this equation of state gets modified.
These are all intensive quantities, and we can thus rewrite the grand-canonical potential, using~\eqref{Wdef} and~\eqref{energyDensity}, as
\eq
{\mc W}(T,\mc V,\mu,\mc B,\Pi)=(\varepsilon-Ts-\mu\rho)\mc V.
\label{Phidef}\eeq
Conformal invariance and extensivity restrict thus the form of the state function ${\mc W}$ to
\eq\label{Wscaling}
{\mc W}(T,\mc V,\mu,\mc B,\Pi)=-\mc V T^d\,h\lp\frac{\mu}{T},\frac{\mc B}{T^2},\frac{\Pi}{T}\rp,
\eeq
where the function $h$ depends only on the dimensionless ratios $\mu/T$, $\mc B/T^2$ and $\Pi/T$. We should stress that in writing this relation we assume that there are no 
dimensionful couplings, either single- or multi-trace, for the scalar operator dual to $\f$. This assumption is justified for the planar black holes we consider here, but in general dimensionful scalar couplings must be included in the scaling argument (see e.g. \cite{Attems:2016ugt}). As a consequence of \eqref{Wscaling}, ${\mc W}$ possesses the scaling property
\eq
{\mc W}(\lambda T,\lambda^{1-d}\mc V,\lambda\mu,\lambda^2\mc B,\lambda\Pi)=\lambda{\mc W}(T,\mc V,\mu,\mc B,\Pi).
\eeq
Differentiating this relation with respect to $\lambda$ and setting $\lambda=1$ we obtain
\eq
-sT-(1-d)\mc P-\rho\mu-2\mc M\mc B-\varpi\Pi={\mc W}/{\mc V},
\label{stateW}\eeq
where we have used the conjugate variables introduced in~\eqref{conjugate} and~\eqref{pressure}. Equivalently,
\eq
\frac{\p{\mc W}}{\p T}=-s\mc V,\quad
\frac{\p{\mc W}}{\p\mc V}=-\mc P,\quad
\frac{\p{\mc W}}{\p\mu}=-\rho\mc V,\quad
\frac{\p{\mc W}}{\p\mc B}=-\mc M\mc V,\quad
\frac{\p{\mc W}}{\p\Pi}=-\varpi\mc V.\quad
\eeq
Combining equation~\eqref{stateW} with the defining relation~\eqref{Phidef} finally gives the equation of state 
\eq
\varepsilon=(d-1)\mc P-2\mc M\mc B-\varpi\Pi.
\label{eqState}\eeq
We will see that all black holes under consideration indeed have an equation of state of this form.

\subsection{Thermodynamics of the bald dyonic black branes of Theory~I}

Let us start with the bald, dyonic black hole solution given by equations~\eqref{bald}-\eqref{dyonic}. For that metric, the Fefferman-Graham radial coordinate $z$ is related to $r$ by
\eq\label{rz-bald-I}
r=\frac{\ell^2}z+\frac18p^2z+\frac m{6\ell^2}z^2-\frac{\qep2+\qmp2}{32\ell^4}z^3+\mc O(z^4),
\eeq
and the Fefferman-Graham expansions of the fields are (see appendix \ref{asexp}) 
\begin{align}
g_{ij}&=
\lp
\begin{array}{ccc}
-1 & 0 & 0 \\
0 & \ell^2 & 0 \\
0 & 0 & \ell^2
\end{array}
\rp
+z^2
\lp
\begin{array}{ccc}
p^2/4\ell^2 & 0 & 0 \\
0 & p^2/4 & 0 \\
0 & 0 & p^2/4
\end{array}
\rp
+z^3
\lp
\begin{array}{ccc}
2m/3\ell^4 & 0 & 0 \\
0 & m/3\ell^2 & 0 \\
0 & 0 & m/3\ell^2
\end{array}
\rp+\mc O(z^4),
\nonumber\\
\psi_I&=px^I,
\qquad
A_i=\lp\frac{\qe}{r_0},\,0,\,\qm x\rp+z \lp-\frac{\qe}{\ell^2},\,0,\,0\rp+\mc O(z^3).
\end{align}

In order to evaluate the bulk integral of the on-shell action we integrate over the radial coordinate $r$ from the horizon at $r_0$ to a UV cutoff $\bar r$ in the $r$ coordinate. At the end of the calculation we express $\bar r(\e)$ in terms a cutoff $\e$ in the $z$ coordinate using the asymptotic expansion \eqref{rz-bald-I}. This gives 
\eq
S_{\textsf{bulk}}=
\int_{r_0}^{\bar r(\e)}\!\!\!d r\!\int\!d^{3}x\,
\sqrt{-G}\left(V(\phi)-\frac14F^2\right)
=-\frac{2\ell^4}{\ep^3}-\frac{3\ell^2p^2}{4\ep}+
\lp\frac{2r_0^3}{\ell^2}-m+\frac{\qep2-\qmp2}{2r_0}\rp+\mc O(\ep).
\eeq

The bald black hole solutions are clearly compatible with any boundary condition imposed on the dialton $\f$. However, we want to consider these black holes as solutions of the same theory that admits the hairy solutions and hence we should impose the same boundary conditions on $\f$ as those the hairy solutions satisfy. It follows that the renormalized action is given by \eqref{Sren-mixed} instead of \eqref{Sren}, even though the two numerically coincide for bald solutions. The Euclidean renormalized action, therefore, takes the form 
\eq
S_{\textsf{ren}}'^E=-S_{\textsf{ren}}'=\b\mc V\lp m-\frac{2r_0^3}{\ell^2}-\frac{\qep2-\qmp2}{2r_0}\rp.
\label{SrenBaldI}\eeq
The corresponding one-point functions for the stress tensor and the conserved electric current are given by \eqref{M-1pt-fns} and, for the bald solutions, take the values
\eq\label{1ptBaldI}
\langle \ct_{ij}\rangle=
\lp
\begin{array}{ccc}
2m/\ell^2 & 0 & 0 \\
0 & m & 0 \\
0 & 0 & m
\end{array}
\rp,
\qquad
\langle \cj^{i}\rangle=\lp\frac{\qe}{\ell^2},\,0,\,0\rp.
\eeq
Note that this stress tensor is traceless, as it should. Moreover, the one-point functions of the scalar operators vanish identically, i.e. $\langle\mc O_{\Delta_-}\rangle=\langle\mc O_{\psi_I}\rangle=0$.

%
These expressions for the renormalized on-shell action and one-point functions allow us to evaluate all thermodynamic variables and to confirm the general identities derived above. Firstly, the Gibbs free energy $\mc W$ is immediately obtained from \eqref{Wdef} and \eqref{SrenBaldI}. Recalling the expressions for the temperature and entropy density of these black holes obtained in section~\ref{ss::horizons}, namely
\eq
T=\frac{f'(r_0)}{4\pi}
=\frac{3r_0}{4\pi\ell^2}-\frac{1}{16\pi r_0}\lp2p^2+\frac{\qep2+\qmp2}{r_0^2}\rp,\qquad
s=4\pi r_0^2,
\eeq
and defining the energy density, chemical potential, and charge density respectively as
\eq
\varepsilon=\ell^2\langle \ct^{tt}\rangle=2m,\qquad
\mu=\lim_{r\rightarrow\infty}A_t=\frac{\qe}{r_0},\qquad
\rho=\ell^2\langle \cj^t\rangle=\qe,
\eeq
we can then confirm that the relation~\eqref{energyDensity} is indeed satisfied. As a consistency check, one can also check that\footnote
{
This can be done parameterizing $w=w(r_0,\qe,p)$, $T=T(r_0,\qe,p)$, and $\mu=\mu(r_0,\qe,p)$, in terms of $r_0$, $\qe$, and $p$, and using the implicit function theorem, leading for example to
\eq
\lp\frac{\p w}{\p T}\rp_{\mu,p}=\frac
{\left.\frac{\p w}{\p r_0}\right|_{\qe,p}-\left.\frac{\p w}{\p\qe}\right|_{r_0,p}\frac{{\p\mu/\p r_0}|_{\qe,p}}{{\p\mu/\p \qe}|_{r_0,p}}}
{\left.\frac{\p T}{\p r_0}\right|_{\qe,p}-\left.\frac{\p T}{\p\qe}\right|_{r_0,p}\frac{{\p\mu/\p r_0}|_{\qe,p}}{{\p\mu/\p \qe}|_{r_0,p}}},
\qquad\textrm{etc.}
\eeq
}
\eq
s=-\lp\frac{\p w}{\p T}\rp_{\mu,\mc B,\Pi},
\qquad
\rho=-\lp\frac{\p w}{\p \mu}\rp_{T,\mc B,\Pi}.
\eeq
Moreover, given that the total energy, electric charge and entropy are obtained from the corresponding densities by multiplying by the (regularized version of the infinite) spatial volume $\cv$, i.e. $\mc E=\varepsilon\mc V$, $Q_{\textsf e}=\rho\mc V$, and $S=s\mc V$, the thermodynamic identity~\eqref{pressure} determines the pressure to be  
\eq\label{PBaldI}
\mc P=-\lp\frac{\p\mc E}{\p\mc V}\rp_{S,Q_{\textsf e},\mc B,\Pi}\!
=\langle \ct_{xx}\rangle-\frac{\qmp2}{r_0}+p^2r_0.
\eeq
From \eqref{SrenBaldI} and \eqref{1ptBaldI} then follows that $\mc P=-w$, in agreement with the general result \eqref{wP}. Finally, \eqref{PBaldI} and \eqref{energyDensity} give the Gibbs-Duhem relation \eqref{GD}.

The thermodynamic relations we checked so far for the bald solutions of Theory I do not involve the magnetic and axion charges, or their conjugate potentials. However, these variables are required in order to verify the first law \eqref{first-law-density} and the equation of state \eqref{eqState}.
As we argued above, for homogeneous systems the Gibbs free energy density \eqref{free-energy-density} is a function of the magnetic and axionic charge {\em densities}, respectively $\mc B$ and $\Pi$ (defined in~\eqref{defpi}), namely
\eq
\mc B=\frac1{\ell^3}F_{(0)xy}=\frac\qm{\ell^3},\qquad
\Pi=\frac{|p|}{\ell},
\eeq
and not of the corresponding total charges.\footnote
{
\label{foot:pressure}
This property of homogeneous systems is directly related to the fact that $\mc P\neq\langle \ct_{xx}\rangle$ for such systems. If, instead, one kept fixed the total charges $\Pi\mc V$ and $\mc B\mc V$, one would have found the pressure to be equal to $\langle \ct_{xx}\rangle$ \cite{Andrade:2013gsa}.
} 
The conjugate thermodynamic potentials, namely the magnetization $\mc M$ and the axionic charge potential $\varpi$, are then defined as in \eqref{conjugate} and take the values
\eq
\mc M=-\lp\frac{\p w}{\p \cb}\rp_{T,\mu,\Pi}\!=-\ell^3\frac{\qm}{r_0},\qquad
\varpi=-\lp\frac{\p w}{\p \Pi}\rp_{T,\mu,\mc B}\!=2\ell pr_0.
\eeq
Using these results it is straightforward to confirm that both the equation of state~\eqref{eqState} and the first law~\eqref{first-law-density} are satisfied. The extra contribution to the pressure~\eqref{PBaldI} is thus due to the pressure $\mc B\mc M+\frac12\Pi\varpi$ exerted by the magnetization.

\subsection{Thermodynamics of the hairy dyonic black branes of Theory~I}

Next we turn to the hairy dyonic black hole solution~\eqref{hairySol} with the gauge field given in~\eqref{dyonic}. In studying the thermodynamics of these solutions it is crucial to recall that the parameter $v$ is not independent and it is determined in terms of the rest of the parameters of the solution via the condition~\eqref{constraint}. As we shall see below, $v$ corresponds to the {\em vev} of the scalar operator $\co_{\Delta_-}$ dual to the dialton $\f$.

The relation between the Fefferman-Graham radial coordinate $z$ and $r$ for the metric~\eqref{hairySol} is
\be
r=\frac{\ell^2}z+\frac18\lp p^2-\frac{2v^2}{\ell^2}\rp z+
\frac{v\sqrt\alpha}{6\ell^2}\lp p^2+\frac{2v^2}{\ell^2}\rp z^2 +\frac{(1+\alpha)v^2}{16\ell^4}\lp p^2-\frac{2(1+3\alpha)v^2}{(1+\alpha)\ell^2}\rp z^3+\mc O(z^4),
\ee
while the Fefferman-Graham expansions now take the form (see appendix \ref{asexp})
\begin{align}
g_{ij}&=
\lp
\begin{array}{ccc}
-1 & 0 & 0 \\
0 & \ell^2 & 0 \\
0 & 0 & \ell^2
\end{array}
\rp
+z^2
\lp
\begin{array}{ccc}
\frac{p^2}{4\ell^2}-\frac{3v^2}{2\ell^4} & 0 & 0 \\
0 & \frac{p^2}4-\frac{3v^2}{2\ell^2} & 0 \\
0 & 0 & \frac{p^2}4-\frac{3v^2}{2\ell^2}
\end{array}
\rp\nonumber\\
&\quad
+z^3
\lp
\begin{array}{ccc}
\sqrt\alpha\frac{2v}{3\ell^4}\lp p^2-\frac{4v^2}{\ell^2}\rp & 0 & 0 \\
0 & \sqrt\alpha\frac{v}{3\ell^2}\lp p^2+\frac{8v^2}{\ell^2}\rp & 0 \\
0 & 0 & \sqrt\alpha\frac{m}{3\ell^2}\lp p^2+\frac{8v^2}{\ell^2}\rp
\end{array}
\rp+\mc O(z^4),
\nonumber\\
\phi&=2\sqrt3\frac{v}{\ell^2}z-2\sqrt{3\alpha}\frac{v^2}{\ell^4}z^2+\mc O(z^3),
\qquad\qquad
\psi_I=px^I,\NO\\
A_i&=\lp\frac{\qe}{r_0},\,0,\,\qm x\rp+z \lp-\frac{\qe}{\ell^2},\,0,\,0\rp+\mc O(z^3).\label{FG-HI}
\end{align}
From the asymptotic expansion for the scalar $\f$ we can immediately deduce the boundary conditions that these black holes are compatible with. Comparing the relation $\dil1=-\sqrt\alpha\dilp02/{2\sqrt3}$ between the two modes in \eqref{FG-HI}
with the condition that the single trace source for the dual scalar operator vanishes, i.e. $J_\cf=-\ell^2\dil1-\cf'(\dil0)=0$, determines that the multi-trace deformation function $\cf(\vf_{(0)})$ is of the form \eqref{bc} with
\eq\label{th-I}
\vartheta\sbtx{I}=\frac{\ell^2\sqrt\alpha}{2\sqrt3}.
\eeq

Introducing again a UV cutoff at $z=\ep$, the integral of the bulk part of the on-shell action is
\begin{align}
S_{\textsf{bulk}}&=
\int_{r_0}^{\bar r(\e)}\!\!\!dr\!\int\!d^{3}x\,
\sqrt{-G}\left(V(\phi)-\frac14F^2\right)
=-\frac{2\ell^4}{\ep^3}-\frac{3\ell^2}{4\ep}\lp p^2-\frac{2v^2}{\ell^2}\rp\NO\\
&\hphantom{=}
+\lp\frac{2r_0^3}{\ell^2}-v\sqrt\alpha\lp p^2+\frac{2v^2}{\ell^2}\rp+\frac{\qep2-\qmp2}{2r_0}
+\frac{2v^4\alpha}{\ell^2}\frac{3r_0^2+3vr_0\sqrt\alpha+v^2\alpha}{(r_0+v\sqrt\alpha)^3}\rp
+\mc O(\ep).
\end{align}
Taking into account the value \eqref{th-I} for the parameter $\vth$ that determines the scalar boundary conditions, we find that the renormalized on-shell action~\eqref{Sren-mixed} is 
\eq\label{Sren-hairy-I}
S_{\textsf{ren}}'^E=-S_{\textsf{ren}}'=-\b\mc V\lp\frac{2r_0^3}{\ell^2}+\frac{\qep2-\qmp2}{2r_0}-p^2v\sqrt{\alpha}
-\frac{2v^3}{\ell^2}\frac{r_0^3\sqrt\alpha}{(r_0+v\sqrt\alpha)^3}\rp.
\eeq
Finally, inserting the expansions \eqref{FG-HI} in the expressions \eqref{M-1pt-fns} for the renormalized one-point functions we get
\begin{align}\label{vevs-I}
\langle \ct_{ij}\rangle&=
\lp
\begin{array}{ccc}
2p^2v\sqrt\alpha/\ell^2 & 0 & 0 \\
0 & p^2v\sqrt\alpha & 0 \\
0 & 0 & p^2v\sqrt\alpha
\end{array}
\rp,\quad
\langle \cj^{i}\rangle =\lp\frac{\qe}{\ell^2},\,0,\,0\rp,\quad
\langle\mc O_{\Delta_-}\rangle=\dil0=2\sqrt3\frac{v}{\ell^2},
\end{align}
while again the expectation value of the scalar operator dual to the axions vanishes, i.e. $\langle\mc O_{\psi_I}\rangle=0$. Notice that the stress tensor is indeed traceless, in agreement with~\eqref{M-trace-WI} for the boundary condition~\eqref{bc}. Moreover, these holographic relations allows us to identify the parameter $v$ with the expectation value of the scalar operator $\co_{\Delta_-}$.

%
Using the above expressions for the renormalized Euclidean action and one-point functions we can determine the remaining thermodynamic variables for these black holes and verify the general thermodynamic relations derived at the beginning of this section. We already obtained the temperature and entropy density for these black holes in equations~\eqref{Thairy} and~\eqref{sHairyI}, respectively.
The energy density is defined as for the bald solutions and is given by
\eq
\varepsilon=\ell^2\langle \ct^{tt}\rangle=2vp^2\sqrt\alpha.
\label{epDyon}
\eeq
Moreover, the chemical potential, the electric charge density, as well as the magnetic and axion charge densities are identical to those of the bald solution, namely 
\eq
\mu=\lim_{r\rightarrow\infty}A_t=\frac{\qe}{r_0},\qquad
\rho=\ell^2\langle \cj^t\rangle=\qe,\qquad \mc B=\frac1{\ell^3}F_{(0)xy}=\frac\qm{\ell^3},\qquad
\Pi=\frac{|p|}{\ell}.
\eeq
An important property of the hairy solutions of Theory I is that, as a direct consequence of the condition \eqref{constraint}, the variables $(T,\mu,\mc B,\Pi)$ are {\em not} independent and satisfy the constraint  
\eq
\mc B^2+\lp\pi T+\frac{\Pi}{2\sqrt2}\rp^2\left[
\frac{\mu^2}{\ell^2}
-\frac4\alpha(1-\alpha)\lp\pi T-\frac{\Pi}{2\sqrt2}\rp^2
\right]=0.
\label{constraint2}\eeq
Since all the variables $T,\mu,\mc B,\Pi$ are a priori external tunable parameters, we conclude that these black holes exist only when these external parameters lie on the constraint submanifold defined by \eqref{constraint2}.  

The Gibbs free energy is obtained from the renormalized Euclidean on-shell action through the definition \eqref{Wdef} and can be expressed in the form 
\eq
w(T,\mu,\mc B,\Pi)=
-\sqrt2\,\ell^4\Pi
\left[
\lp\pi T+\frac{\Pi}{2\sqrt2}\rp^2
+\frac{\mu^2}{4\ell^2(1-\alpha)}
\right]
+\ell^4\mc B^2\,\frac{\pi T-\alpha\lp\pi T+\frac{\Pi}{2\sqrt2}\rp}
{(1-\alpha)\lp\pi T+\frac{\Pi}{2\sqrt2}\rp^2},
\label{wHairyIb}
\eeq
where again all the variables $T,\mu,\mc B,\Pi$  lie on the constraint submanifold \eqref{constraint2}. Using the above expressions for the energy density, temperature, entropy, electric charge density and chemical potential, one can verify that the free energy density satisfies the thermodynamic relation \eqref{energyDensity}, provided the constraint \eqref{constraint2} is taken into account. For zero magnetic field \eqref{wHairyIb} simplifies to 
\eq
w=-\frac\ell2\left[(\Xi+1)^2+\frac1\alpha(\Xi-1)^2\right]\lp\frac{p}{\sqrt2}\rp^3,
\label{wHairyI}
\eeq
where $\Xi=2\sqrt{2}\;\pi \ell T/p$, in complete agreement with the free energy obtained in eqn.~(5.6) of~\cite{Bardoux:2012tr} using a (real time) Hamiltonian approach to the thermodynamics. This means that we can use the thermodynamic analysis of \cite{Bardoux:2012tr} and, in particular, the results on the phase structure of the system obtained there.

If the variables $T,\mu,\mc B,\Pi$ were all independent, the expression \eqref{wHairyIb} for the free energy density could be used to determine the thermodynamic potentials conjugate to the magnetic and axion charge densities through the relations \eqref{conjugate}. However, these variables are not independent, due to the constraint \eqref{constraint2}. Nevertheless, since we know already the values for the entropy and electric charge densities, we can use a Lagrange multiplier for the constraint \eqref{constraint2} to obtain the potentials conjugate to the magnetic and axion charge densities. Considering the variation of $w+\l \cc$, where $\l$ is a Lagrange multiplier and $\cc$ is the constraint \eqref{constraint2}, and identifying the coefficient of $\tx dT$ with $-s$ fixes the value of the Lagrange multiplier to 
\be
\l=\frac{4\ell^2(\sqrt{2}\;\ell^2\m\P-2(1-\a)q\sbtx{e})}{(1-\a)\m(4\pi T+\sqrt{2}\;\P)^2}.
\ee
This allows us to read off the potentials 
\eq
\mc M
=-\ell^3\frac{\qm}{r_0},\qquad
\varpi
=2p\ell\lp r_0+\frac{v^2}{r_0+v\sqrt\alpha}\rp.
\eeq
Using these results one can verify that the first law \eqref{first-law-density} and the equation of state \eqref{eqState} hold.

\paragraph{BPS-like structure} Intriguingly, all thermodynamic variables of these solutions, including the temperature, are completely fixed by the charges. In particular, the energy density is given by 
\eq
\varepsilon(\rho,\mc B,\Pi)=\lp\frac{2\alpha}{1-\alpha}\rp^{\frac12}\,\ell\Pi
\sqrt{\rho^2+\mc B^2\ell^6}\;.
\eeq
Similarly the entropy density is also completely determined in terms of $\rho,\mc B$ and $\Pi$, but the corresponding expression is too complicated to usefully reproduce it here. 

This evokes the analogous property of extremal black holes, and despite having a non-vanishing temperature, the hairy black holes of Theory~I behave along the constraint \eqref{constraint} as extremal black holes. The energy density itself is linear in the charges, and we can think of the black hole as composed of elementary blocks carrying unit axionic and electric charges, $(\Pi,\rho,\mc B)=(1,1,0)$, and magnetic elementary blocks with charges $(1,0,1)$ (in suitable units). We can thus investigate the stability of such black holes towards fragmentation of the charges by comparing the entropies of the system before and after fragmentation. We find
\begin{align}
s(\Pi,\rho_1,0)+s(\Pi,\rho_2,0)&\geq s(\Pi,\rho_1+\rho_2,0),\NO\\
s(\Pi,0,\mc B_1)+s(\Pi,0,\mc B_2)&\geq s(\Pi,0,\mc B_1+\mc B_2),\NO\\
s(\Pi_1,\rho,0)+s(\Pi_2,\rho,0)&\leq s(\Pi_1+\Pi_2,\rho,0),\NO\\
s(\Pi_1,0,\mc B)+s(\Pi_2,0,\mc B)&\leq s(\Pi_1+\Pi_2,0,\mc B).
\end{align}
It is thus entropically favorable for these black holes to decay to a bound state of smaller black holes carrying a smaller electric charge/magnetic field. On the other hand, the axionic charge is stable against fragmentation.

\subsection{Thermodynamics of the hairy black branes of Theory~II}


\paragraph{Electrically charged solutions} Finally, we consider the thermodynamics of the hairy black holes of Theory~II. Starting with the magnetically neutral solutions in \eqref{bsolxi}, the  Fefferman-Graham radial coordinate $z$ is related to $r$ by
\begin{align}
r=&\frac{\ell^2}z-\frac{\xi v\sbtx{e}}2+\frac18\lp p^2+\frac{v\sbtx{e}^2}{2\ell^2}\xi(2-\xi)\rp z
+\frac1{12\ell^4}\lp2r_0^3h(r_0)^{2\xi}-\ell^2p^2(r_0+\xi v\sbtx{e})-\frac13\xi(1-\xi)(2-\xi)v\sbtx{e}^3\rp z^2\nonumber\\
&-\frac{\xi v\sbtx{e}}{16\ell^6}\lp r_0^3h(r_0)^{2\xi}-\frac12\ell^2p^2r_0h(r_0)-\frac14(1-\xi)^2(2-\xi)v\sbtx{e}^3\rp z^3
+\mc O(z^4),
\end{align}
and the corresponding Fefferman-Graham expansions are (see appendix \ref{asexp}) 
\begin{align}
g_{ij}&=
\lp
\begin{array}{ccc}
-1 & 0 & 0 \\
0 & \ell^2 & 0 \\
0 & 0 & \ell^2
\end{array}
\rp
+z^2
\lp
\begin{array}{ccc}
\frac{p^2}{4\ell^2}+\frac{\xi(2-\xi)v\sbtx{e}^2}{8\ell^4} & 0 & 0 \\
0 & \frac{p^2}{4}-\frac{\xi(2-\xi)v\sbtx{e}^2}{8\ell^2} & 0 \\
0 & 0 & \frac{p^2}{4}-\frac{\xi(2-\xi)v\sbtx{e}^2}{8\ell^2}
\end{array}
\rp\nonumber\\
&\quad
+z^3
\lp
\begin{array}{ccccc}
\frac{2r_0^3}{3\ell^6}h^{2\xi}(r_0)-\frac{p^2}{3\ell^4}(r_0+\xi v\sbtx{e})-\frac{\xi(1-\xi)(2-\xi)v\sbtx{e}^3}{9\ell^6}
\span\span & 0 & 0 \\
0 & \span
\frac{r_0^3}{3\ell^4}h^{2\xi}(r_0)-\frac{p^2}{6\ell^2}(r_0+\xi v\sbtx{e})+\frac{\xi(1-\xi)(2-\xi)v\sbtx{e}^3}{9\ell^4}\span
 & 0 \\
0 & 0 & \span\span
\frac{r_0^3}{3\ell^4}h^{2\xi}(r_0)-\frac{p^2}{6\ell^2}(r_0+\xi v\sbtx{e})+\frac{\xi(1-\xi)(2-\xi)v\sbtx{e}^3}{9\ell^4}
\end{array}
\rp+\mc O(z^4),
\nonumber\\
\phi&=-\sqrt{\xi(2-\xi)}\frac{v\sbtx{e}}{\ell^2}z+(1-\xi)\sqrt{\xi(2-\xi)}\frac{v\sbtx{e}^2}{2\ell^4}z^2+\mc O(z^3),
\qquad
\psi_I=px^I,\\
A_t&=\frac{q\sbtx{e}}{r_0 h(r_0)}
-\frac{q\sbtx{e}}{\ell^2}z
+\mc O(z^2).
\end{align}
As for the hairy black holes of Theory I, the asymptotic expansion for the scalar $\f$ determines the boundary conditions these black holes are compatible with. Comparing the relation $\dil1=\frac{1-\xi}{2\sqrt{\xi(2-\xi)}}\dilp02$ between the two scalar modes with the condition that the single trace source for the dual scalar operator vanishes, i.e. $J_\cf=-\ell^2\dil1-\cf'(\dil0)=0$, determines that the multi-trace deformation function $\cf(\vf_{(0)})$ is of the form \eqref{bc} with
\be\label{th-II-e}
\vartheta\sbtx{II}^{\tt e}=-\frac{(1-\xi)\ell^2}{2\sqrt{\xi(2-\xi)}}.
\ee

Introducing a radial cutoff at $z=\ep$, the bulk integration of the on-shell action gives 
\be
\hskip-0.3cmS_{\textsf{bulk}}=\int_{r_0}^{\bar r(\e)}\!\!\!dr\!\int\!d^{3}x\,\sqrt{-G}\left(V-\frac14F^2\right)
=-\frac{2\ell^4}{\ep^3}
-\frac{6\ell^2p^2-\xi(2-\xi)v\sbtx{e}^2}{8\ep}+r_0\lp\frac{p^2}2+\frac{r_0^2}{\ell^2}h^{2\xi}(r_0)\rp+\mc O(\ep).
\ee
We then find that the renormalized generating function  \eqref{Sren-mixed} is 
\eq\label{SrenII-e}
S_{\textsf{ren}}'^E=-S_{\textsf{ren}}'=-\b\cv\(\frac{r_0^3h^{2\xi}(r_0)}{\ell^2}+\frac{r_0p^2}2\).
\eeq
Moreover, the renormalized one-point functions \eqref{M-1pt-fns} are given by
\begin{align}\label{vevsIIe}
\langle \ct_{ij}\rangle&=
\lp
\begin{array}{ccccc}
\displaystyle
\frac1{\ell^4}{2r_0^3h^{2\xi}(r_0)}-\frac1{\ell^2}{p^2(r_0+\xi v\sbtx{e})}
\span\span & 0 & 0 \\[1ex]
0 & \span
\displaystyle
\frac1{\ell^2}{r_0^3h^{2\xi}(r_0)}-\frac12{p^2(r_0+\xi v\sbtx{e})}
\span & 0 \\[1ex]
0 & 0 & \span\span
\displaystyle
\frac1{\ell^2}{r_0^3h^{2\xi}(r_0)}-\frac12{p^2(r_0+\xi v\sbtx{e})}\end{array}
\rp, && \langle \cj^{i}\rangle=\lp
q\sbtx{e}/\ell^2
,\,0,\,0\rp, \NO\\
\langle\mc O_{\Delta_-}\rangle & =\dil0=-\sqrt{\xi(2-\xi)}\;v\sbtx{e}/\ell^2, && \langle\mc O_{\psi_I}\rangle=0.
\end{align}
Again, the stress tensor is traceless, in agreement with~\eqref{M-trace-WI} for the boundary condition~\eqref{bc}. 
Note that in the limit $v\sbtx{e}\rightarrow0$, $q\sbtx{e}\rightarrow0$, the $d=3$ solution~\eqref{bsolxi} reduces to the bald, uncharged solution~\eqref{bald} (with $\qe=0$ and $\qm=0$). One can easily check that in that limit both the renormalized on-shell action and the holographic stress tensor of these solutions nicely agree, with the identification $m=r_0^3/\ell^2-p^2r_0/2$ of their parameters. However, the solution~\eqref{bsolxi} also admits a neutral hairy limit, where $q\sbtx{e}\rightarrow0$, but $v\sbtx{e}\neq 0$.

%
The temperature and entropy density of these black holes were given respectively in  \eqref{ThairyII} and \eqref{sHairyII}. The energy density, chemical potential, and the electric and axionic charge densities are defined as for Theory I and take the values 
\eq\label{energyDensityIIe}
\varepsilon=\ell^2\langle \ct^{tt}\rangle=(1-\x)p^2v\sbtx{e}+\frac{q\sbtx{e}^2}{\x v\sbtx{e}},
\eeq
and
\eq
\mu=\lim_{r\rightarrow\infty}A_t=\frac{\rho}{r_0h(r_0)},\quad
\rho=\ell^2\langle \cj^t\rangle=q\sbtx{e},\quad \Pi=\frac{|p|}{\ell}.
\eeq

The Gibbs free energy is again obtained from the renormalized Euclidean action \eqref{SrenII-e} by invoking the definition \eqref{Wdef}, and it is straightforward to check that the resulting $w$ satisfies the thermodynamic relation \eqref{energyDensity}, as well as  
\eq
s=-\lp\frac{\p w}{\p T}\rp_{\mu,\Pi},
\qquad
\rho=-\lp\frac{\p w}{\p \mu}\rp_{T,\Pi}.
\eeq
Moreover, the free energy density allows us to obtain the thermodynamic potential conjugate to the axion charge density as 
\eq
\varpi=-\lp\frac{\p w}{\p \Pi}\rp_{T,\mu}\!=2p\ell\lp r_h+\frac{v\sbtx{e}\xi}2\rp,
\eeq
while the pressure \eqref{pressure} is given by 
\eq
\mc P=-\lp\frac{\p\mc E}{\p\mc V_2}\rp_{S,Q_{\textsf e},\Pi}\!
=\langle \ct_{xx}\rangle+p^2\lp r_0+\frac{v\sbtx{e}\xi}{2}\rp=\langle \ct_{xx}\rangle+\frac12\Pi\varpi,
\eeq
where again we have introduced the total energy $\mc E=\varepsilon\mc V$, electric charge $Q_{\textsf e}=\rho\mc V$, entropy $S=s\mc V$. This satisfies both \eqref{wP} and the Gibbs-Duhem relation \eqref{GD}. Finally, it is straightforward to verify that the first law \eqref{first-law-density} and the equation of state \eqref{eqState} (with $\cb=0$) hold. 

\paragraph{Magnetically charged solutions} Let us turn to the magnetically charged hairy black holes of Theory~II. Starting with the magnetically neutral solutions in \eqref{mag-sol-II}, the  Fefferman-Graham radial coordinate $z$ is related to $r$ by
\begin{align}
	r=&\frac{\ell^2}z-\frac{2-\xi}2v\sbtx{m}+\frac18\lp p^2+\frac{v\sbtx{m}^2}{2\ell^2}\xi(2-\xi)\rp z\NO\\
	&-\frac1{12\ell^2\xi v\sbtx{m}}\lp q_{\textsf{m}}^2+p^2v\sbtx{m}^2\xi(1-\xi)-\frac{v\sbtx{m}^4}{3L^2}\xi(1-\xi)(6-14\xi+7\xi^2)\rp z^2\nonumber\\
	&-\frac{1}{32\ell^4}\lp q_{\textsf{m}}^2-\frac{v\sbtx{m}^4}{2\ell^2}\xi(1-\xi)^2(2-\xi)\rp z^3
	+\mc O(z^4),
\end{align}
and the corresponding Fefferman-Graham expansions are (see appendix \ref{asexp}) 
\begin{align}
	g_{ij}&=
	\lp
	\begin{array}{ccc}
		-1 & 0 & 0 \\
		0 & \ell^2 & 0 \\
		0 & 0 & \ell^2
	\end{array}
	\rp
	+z^2
	\lp
	\begin{array}{ccc}
		\frac{p^2}{4\ell^2}+\frac{\xi(2-\xi)v\sbtx{m}^2}{8\ell^4} & 0 & 0 \\
		0 & \frac{p^2}{4}-\frac{\xi(2-\xi)v\sbtx{m}^2}{8\ell^2} & 0 \\
		0 & 0 & \frac{p^2}{4}-\frac{\xi(2-\xi)v\sbtx{m}^2}{8\ell^2}
	\end{array}
	\rp\nonumber\\
	&\quad
	+z^3
	\lp
	\begin{array}{ccccc}
		\frac{(1-\xi)(6-14\xi+7\xi^2)v\sbtx{m}^3}{9\ell^6}
		-\frac{q_{\textsf{m}}^2}{3\ell^4\xi v\sbtx{m}}
		-\frac{(1-\xi)p^2v\sbtx{m}}{3\ell^4}
		\span\span & 0 & 0 \\
		0 & \span
		\frac{(1-\xi)(3-10\xi+5\xi^2)v\sbtx{m}^3}{9\ell^4}
		-\frac{q_{\textsf{m}}^2}{6\ell^2\xi v\sbtx{m}}
		-\frac{(1-\xi)p^2v\sbtx{m}}{6\ell^2}
		\span
		& 0 \\
		0 & 0 & \span\span
		\frac{(1-\xi)(3-10\xi+5\xi^2)v\sbtx{m}^3}{9\ell^4}
		-\frac{q_{\textsf{m}}^2}{6\ell^2\xi v\sbtx{m}}
		-\frac{(1-\xi)p^2v\sbtx{m}}{6\ell^2}
	\end{array}
	\rp+\mc O(z^4),
	\nonumber\\
	\phi&=-\sqrt{\xi(2-\xi)}\frac{v\sbtx{m}}{\ell^2}z-(1-\xi)\sqrt{\xi(2-\xi)}\frac{v\sbtx{m}^2}{2\ell^4}z^2+\mc O(z^3),\\
	\psi_I&=px^I,\qquad
	A_i=(0,0,q\sbtx{m}x).
\end{align}
Once more, the asymptotic expansion for the scalar $\f$ determines the boundary conditions these black holes are compatible with. Comparing the relation $\dil1=\frac{1-\xi}{2\sqrt{\xi(2-\xi)}}\dilp02$ between the two scalar modes with the condition that the single trace source for the dual scalar operator vanishes, i.e. $J_\cf=-\ell^2\dil1-\cf'(\dil0)=0$, determines that the multi-trace deformation function $\cf(\vf_{(0)})$ is of the form \eqref{bc} with
\be\label{th-II-m}
\vartheta\sbtx{II}^{\tt m}=\frac{(1-\xi)\ell^2}{2\sqrt{\xi(2-\xi)}}.
\ee
Notice that this is the same as the boundary condition \eqref{th-II-e} for the electrically charged solutions, except for the sign.

Introducing a radial cutoff at $z=\ep$, the bulk integration of the on-shell action gives 
\begin{align}
	\hskip-0.3cmS_{\textsf{bulk}}=&\int_{r_0}^{\bar r(\e)}\!\!\!dr\!\int\!d^{3}x\,\sqrt{-G}\left(V-\frac14F^2\right)
	=-\frac{2\ell^4}{\ep^3}
	-\frac{6\ell^2p^2-\xi(2-\xi)v\sbtx{m}^2}{8\ep}
	\nonumber\\
	&+
	\frac1{6\ell^2\xi v\sbtx{m}r_0}
	\lp
	3\ell^2(p^2v\sbtx{m}^2 r_0\xi(1-\xi)+q_{\textsf{m}}^2(r_0-v\sbtx{m}\xi)) + 
	2v\sbtx{m}r_0\xi(6r_0^3+9v\sbtx{m}r_0^2(2-\xi)
	\right.
	\nonumber\\
	&\left.\qquad\qquad\qquad
	+ 3v\sbtx{m}^2r_0(6-7\xi+2\xi^2) + 
	v\sbtx{m}^3(3-2\xi-3\xi^2+2\xi^3))
	\rp
	+\mc O(\ep),
\end{align}
where $r_0$ is the largest zero of $f(r)$. The renormalized generating function  \eqref{Sren-mixed} then is given by 
\begin{align}\label{SrenII-m}
	S_{\textsf{ren}}'^E=-S_{\textsf{ren}}'=&-\b\cv\left[
	\frac1{\ell^2}\lp
	2r_0^3+3v\sbtx{m}r_0^2(2-\xi)+3v\sbtx{m}^2r_0(2-\xi)(1-\frac23\xi)
	+v\sbtx{m}^3(1-\xi^2)(1-\frac23\xi)\rp
	\vphantom{\frac{q_{\textsf{m}}^2}{2\xi v\sbtx{m}}}\right.
	\nonumber\\
	&\qquad\qquad\qquad\qquad\qquad\qquad\left.
	+\frac{q_{\textsf{m}}^2}{2\xi v\sbtx{m}}\lp1-\frac{\xi v\sbtx{m}}{r_0}\rp
	+\frac12(1-\xi)p^2v\sbtx{m}
	\right].
\end{align}
Moreover, the renormalized one-point functions \eqref{M-1pt-fns} are given by
\begin{align}\label{vevsIIm}
	\langle \ct_{ij}\rangle&=
	\lp
	\begin{array}{ccccc}
		\displaystyle
		\frac2{3\ell^4}(1-\xi)(1-2\xi)(3-2\xi)v\sbtx{m}^3
		-\frac{q_{\textsf{m}}^2}{\ell^2\xi v\sbtx{m}}-\frac1{\ell^2}(1-\xi)p^2v\sbtx{m}
		\span\span & 0 & 0 \\[1ex]
		0 & \span
		\displaystyle
		\frac2{6\ell^2}(1-\xi)(1-2\xi)(3-2\xi)v\sbtx{m}^3
		-\frac{q_{\textsf{m}}^2}{2\xi v\sbtx{m}}-\frac1{2}(1-\xi)p^2v\sbtx{m}
		\span & 0 \\[1ex]
		0 & 0 & \span\span
		\displaystyle
		\frac2{6\ell^2}(1-\xi)(1-2\xi)(3-2\xi)v\sbtx{m}^3
		-\frac{q_{\textsf{m}}^2}{2\xi v\sbtx{m}}-\frac1{2}(1-\xi)p^2v\sbtx{m}
	\end{array}
	\rp, && \NO\\
	\langle\mc O_{\Delta_-}\rangle & =\dil0=-\sqrt{\xi(2-\xi)}\;v\sbtx{m}/\ell^2,
	\qquad
	\langle \cj^{i}\rangle=\lp 0,\,0,\,0\rp,\qquad
	\langle\mc O_{\psi_I}\rangle=0.&&
\end{align}
Again, the stress tensor is traceless, in agreement with~\eqref{M-trace-WI} for the boundary condition~\eqref{bc}. 
%

%
The temperature and entropy density of these black holes were given in  \eqref{TsTheoryIIm}. The energy density and the magnetic and axionic charge densities are defined as for Theory I and take the values 
\be\label{energyDensityIIm}
\ve=\frac23(1-\x)(2\x-3)(2\x-1)\frac{v\sbtx{m}^3}{\ell^2}-(1-\x)p^2v\sbtx{m}-\frac{q\sbtx{m}^2}{\x v\sbtx{m}},
\ee
\be
\mc B=\frac1{\ell^3}F_{(0)xy}=\frac\qm{\ell^3},
\qquad
\Pi=\frac{|p|}{\ell}.
\ee

The Gibbs free energy is again obtained from the renormalized Euclidean action \eqref{SrenII-e} by invoking the definition \eqref{Wdef}. It is straightforward to check that the resulting $w$ satisfies the thermodynamic relation \eqref{energyDensity},
as well as the density first law~\eqref{first-law-density}, with the $U(1)$ and axionic magnetizations $\mc M$ and $\varpi$ given by
\eq
\mc M=-\lp\frac{\p w}{\p \mc B}\rp_{T,\Pi}\!=-\frac{q_{\textsf{m}}\ell^3}{r_0},
\qquad
\varpi=-\lp\frac{\p w}{\p \Pi}\rp_{T,\mu}\!=2p\ell\lp r_0+\frac{(2-\xi){v\sbtx{m}}}2\rp.
\eeq
Finally, introducing the total energy $\mc E=\varepsilon\mc V$ and entropy $S=s\mc V$, we obtain that the pressure \eqref{pressure} is again related to the transverse components of the stress tensor by 
\eq
\mc P=-\lp\frac{\p\mc E}{\p\mc V_2}\rp_{S,\mc B,\Pi}\!
=\langle \ct_{xx}\rangle+\frac12\Pi\varpi+\mc M\mc B.
\eeq
with both magnetizations contributing. The resulting pressure satisfies both \eqref{wP} and the Gibbs-Duhem relation \eqref{GD} and it is straightforward to verify that the first law \eqref{first-law} and the equation of state \eqref{eqState} hold.

\section{Stability and phase transitions}
\label{sec:phases}

Having analyzed the thermodynamics of the black brane solutions of Theories I and II, we can now address the stability of these solutions. Besides thermodynamic stability and the corresponding phase structure, we will compute the holographic effective potential for the {\em vev} of the scalar operator dual to the dialton, which will tell us whether these solutions correspond to stable (thermal) vacua of the dual theory.  

\subsection{Dynamical stability and the energy density}

The quantum effective action for the scalar {\em vev} $\s=\<\co_{\Delta_-}\>=\vf\sub{0}$ is given by the Legendre transform of the generating function \eqref{Sren-mixed} with respect to the scalar source (we fix all other sources to their values in the solutions), namely
\be\label{effaction}
\G[\s]=S'_{\textsf{ren}}-\ell^2 \int d^3x  \s J_\cf=\ell^2 \int d^3x \;\Big(\cv\sbtx{QFT}(\s)+\tx{derivatives}\Big),
\ee
where we used the fact that the QFT is on Minkowski (with metric $g_{(0)}=\diag(-1, \ell^2, \ell^2)$) and 
$\cv\sbtx{QFT}(\s)$ is the quantum effective potential for $\s$ and we will not be interested in the derivative terms since we are focusing on homogeneous solutions. From \eqref{deltaS-f} and \eqref{effaction} follows that the source of $\co_{\Delta_-}$ is then given by 
\be
J_\cf=-\frac{\delta\G[\s]}{\delta\s},
\ee
and, hence, vacua of the theory are extrema of the effective action: 
\be
\left.\frac{\delta\G[\s]}{\delta\s}\right|_{\s=\s_*}=0.
\ee 
 
To compute the effective action we observe that from \eqref{Sf} and \eqref{Sren-mixed} follows that \cite{Papadimitriou:2007sj}
\be\label{effaction-e}
\G[\s]=S_{\textsf{ren}}+\ell^2 \int d^3x \cf(\s),
\ee
where $S_{\textsf{ren}}$ is the generating function of the Dirichlet theory in \eqref{Sren}, or equivalently the effective action of the Neumann theory. As for Poincar\'e domain walls \cite{Skenderis:1999mm, Bianchi:2001de, Freedman:2003ax,Papadimitriou:2004rz}, for the homogeneous solutions we are interested in here $S_{\textsf{ren}}$ can be expressed in terms of a fake superpotential that governs non-relativistic flows \cite{Lindgren:2015lia}. The details of this calculation are presented in appendix \ref{effpot}.

It turns out that the result of this calculation can be cast in a rather universal form, that applies to all hairy black holes we have been studying here. In particular, the effective potential for the scalar {\em vev} $\s$ in all cases takes the form
\bal\label{effpot-universal}
\cv\sbtx{QFT}(\s)&=\cv_0+\frac{\m q\sbtx{e}}{\ell^2}+\frac{\ve}{2\ell^2\s_*^3} \(\s^3-3\s_*^2\s\)\NO\\
&=\cv_0+\frac{\m q\sbtx{e}}{\ell^2}+\frac{\ve}{2\ell^2\s_*^3} \(-2\s^3_*+3\s_*(\s-\s_*)^2+(\s-\s_\ast)^3\),
\eal
where $\cv_0$ is a constant (see appendix \ref{effpot}) and
$\s_*$ is the value of the {\em vev} at the extremum, i.e. the value corresponding to the specific background solution, and $\ve$ is the corresponding energy density. 
Dynamical stability is now determined by the sign of the effective mass term, {\it i.e.} the coefficient of the quadratic term, and we see that it is equivalent to the positivity of the energy density, as one may have expected. This result also provides an alternative method for computing the energy density.  Using the specific expressions for the energy density in each of the solutions, therefore, determines the range of parameters for which they are dynamically stable.

\subsection{Thermodynamic stability and phase transitions}

We finally turn to the thermodynamic stability of the various solutions discussed above, and the study of the phase structure of the corresponding theories. To do so we need to compare solutions that have the same asymptotic charges {\em and} satisfy the same boundary conditions, including the boundary conditions of the dialton $\f$. Since bald solutions are compatible with any boundary condition for the scalar $\f$, they can potentially compete with any hairy solutions with the same asymptotic charges. In addition, there may exist small and large black hole solutions that have the same charges and temperature. 

As we have seen in section \ref{sec:thermo}, the Gibbs free energy density $w$ defined in \eqref{free-energy-density} is a function of the variables $T$, $\m$, $\cb$ and $\Pi$. In order to compare solutions with the same charge densities, therefore, we need to Legendre transform $w$ with respect to the chemical potential $\m$ to obtain the Helmholtz free energy density 
\be
\frak f=w+\m\r= \ve-T s.
\ee
The thermodynamic identity \eqref{dw} implies that
\be
d\frak f=-s\,\dd T+\m\,\dd\r-\mc M\,\dd\mc B-\varpi\,\dd\Pi,
\ee
and so $\frak f$ is indeed a function of the variables  $T,\r,\cb,\Pi$, as desired.

\paragraph{Phases of theory I}

The Helmholtz free energy density for the bald solutions of Theory I can be deduced from the on-shell action \eqref{SrenBaldI} and is given by  
\be
\frak f=  m-\frac{2r_0^3}{\ell^2}+\frac{\qep2+\qmp2}{2r_0}.
\ee
Notice that the magnetic and electric charges enter the same way in the Helmholtz free energy and so the thermodynamic stability properties of the dyonic solutions are qualitatively equivalent to those of the corresponding purely electric solutions. Moreover, as it was pointed out in ~\cite{Bardoux:2012tr}, planar black holes with axion charge are equivalent to black holes with horizons of constant negative curvature and no axion charge. As a result, the stability properties of the planar bald solutions of Theory I are analogous to those of bald black holes with hyperbolic horizons, which have been studied for example in \cite{Brill:1997mf,Caldarelli:1998hg,Vanzo:1997gw,Emparan:1999gf}.

For the hairy solutions of Theory I the Helmholtz free energy density can be read off \eqref{Sren-hairy-I} and takes the form
\be
\frak f = p^2v\sqrt{\alpha}
+\frac{2v^3}{\ell^2}\frac{r_0^3\sqrt\alpha}{(r_0+v\sqrt\alpha)^3}-\frac{2r_0^3}{\ell^2}+\frac{\qep2+\qmp2}{2r_0}.
\ee
As for the bald solutions, the dependence of the Helmholtz free energy on the magnetic and electric charges is identical and so the thermodynamic stability properties of the hairy dyonic solutions of Theory I are identical to those of the purely electric solutions studied in~\cite{Bardoux:2012tr}. However, the constraint \eqref{constraint} implies that the temperature is not an independent thermodynamic variable for the hairy solutions of Theory I, but rather a fixed function of the charge densities, namely
\be\label{ThairyI}
T(\rho,\mc B,\Pi)= \frac{ \P}{2\pi\sqrt{2}}\sqrt{1+4\sqrt{\frac{\a}{1-\a}}\frac{\sqrt{\r^2+\cb^2\ell^6}}{\P^2\ell^3}}.
\ee 
This means that, for given charge densities, one can only compare the free energy of the hairy solutions with that of the bald ones at a {\em fixed} temperature, which considerably restricts the useful information one can extract from such an analysis. Nevertheless, this analysis was performed in~\cite{Bardoux:2012tr} and reveals that at large temperatures the unbroken phase of bald black holes dominates, and as we lower the temperature ({\em together with the charge densities according to \eqref{ThairyI}}), the system undergoes a second order phase transition towards a phase of hairy black holes. As the temperature is lowered further, below the lower bound in~\eqref{Tbound}, the hairy solution becomes dynamically unstable, while at an even lower temperature it ceases to exist. 

As we mentioned above, we believe that there exist more general hairy solutions of Theory I whose temperature is not determined by the charge densities. Such solutions would allow one to explore the full phase diagram of Theory I. However, we have been unable to find this more general class analytically. It would be interesting to see if this more general class of hairy solutions can be found numerically.

\paragraph{Phases of electric solutions of Theory II} For the hairy solutions of Theory II, both electrically and magnetically charged, the temperature is an independent variable and so we can explore the full phase diagram. However, due to the coupling between the dialton and the gauge field, bald solutions of Theory II are necessarily electrically and magnetically neutral, and so they do not compete with the hairy solutions at non zero charge density. Nevertheless, for a given non zero charge density and temperature, there are up to three hairy solutions with different horizon radii and scalar {\em vevs} that compete thermodynamically. This leads to an intricate phase diagram that we now describe.   

We will only discuss the phase structure of the electric and magnetic solutions of Theory II for the case $\x=1$, since in that case the analysis can be done analytically. Other values of $\x$ can be addressed in an analogous way, but generically require solving transcendental equations numerically. It may be useful to point out that the case $\x=1$ leads to vanishing coupling for the multi-trace deformation in \eqref{th-II-e} and \eqref{th-II-m}, corresponding to Neumann boundary conditions on the scalar $\f$.\footnote{Note that the case $\x=1$ is also the only case for which the electric and magnetic solutions of Theory II satisfy the same boundary conditions, and hence may possess dyonic solutions.}  

One way to understand the structure of the electric solutions of Theory II analytically is to use the horizon equation $f(r_0)=0$ in \eqref{horizon-eq},
\be \label{horizon-eq_xi1}
\frac{r_0^2}{\ell^2} h^2(r_0) - \frac{p^2}{2} h(r_0) - \frac{q\sbtx{e}^2}{2 v\sbtx{e} r_0} =0,
\ee
in order to obtain two different expressions for the temperature. The two different expressions for $T$ correspond to using the equation $f(r_0)=0$ in two different ways.
The first expression is obtained by solving \eqref{horizon-eq_xi1} for $\frac{q_e^2}{2 v_e r_0}$ and substituting the result in 
\eqref{ThairyII} with $\xi=1$,
\be \label{ThairyIIxi1}
T = \frac{1}{4 \pi \ell^2} \left(2 r_0 + v\sbtx{e} + \frac{q\sbtx{e}^2 \ell^2}{2 v\sbtx{e} r_0^2 h^2(r_0)}\right),
\ee 
This yields
\be\label{Te1}
\t=3 r_0+v\sbtx{e}-\frac{p^2\ell^2}{2(r_0+v\sbtx{e})},
\ee
where we have defined $\t\equiv 4\pi\ell^2 T$. Alternatively, we can solve \eqref{horizon-eq_xi1} for $\frac{r_0^2}{\ell^2} h^2(r_0) $ and substitute in \eqref{ThairyIIxi1}, finding
\be\label{Te2}
\t=2 r_0+v\sbtx{e} +\frac{q\sbtx{e}^2r_0}{q\sbtx{e}^2+p^2v\sbtx{e}(r_0+v\sbtx{e})}.
\ee
Eliminating the quadratic terms in $r_0$ we obtain an explicit formula for the radius of the horizon as a function of the charge densities, the temperature and the scalar {\em vev} parameter $v\sbtx{e}$, namely
\be\label{r0-e}
r_0=-v\sbtx{e}+\frac{(6q\sbtx{e}^2-\ell^2 p^4)v\sbtx{e}+3q\sbtx{e}^2\t}{9q\sbtx{e}^2+p^2v\sbtx{e}(v\sbtx{e}-\t)}.
\ee
Finally, inserting this result in either \eqref{Te1} or \eqref{Te2} we obtain the {\em characteristic curve}
\be\label{charIIe}
(\ell^2 p^4-8q\sbtx{e}^2)v\sbtx{e}^4+(2\ell^4p^6-\ell^2p^4\t^2-18\ell^2p^2q\sbtx{e}^2+6q\sbtx{e}^2\t^2)v\sbtx{e}^2+2q\sbtx{e}^2\t^3v\sbtx{e}-27\ell^2 q\sbtx{e}^4=0,
\ee
which is an equation relating physical observables only. 

\begin{figure}
	\begin{tabular}{cc}
		\scalebox{0.7}{\includegraphics{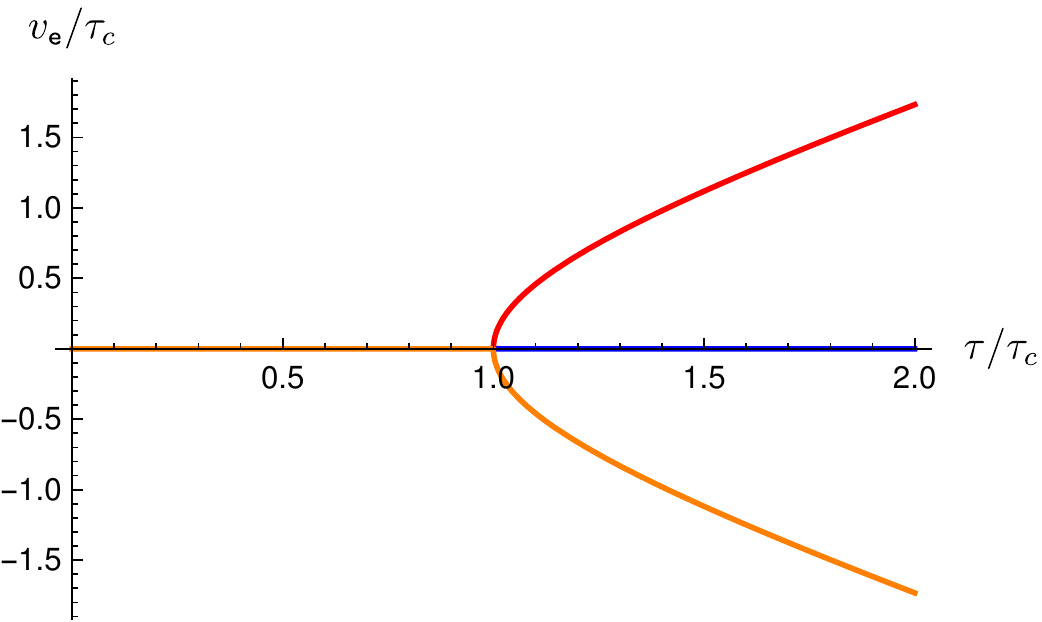}} &
		\scalebox{0.7}{\includegraphics{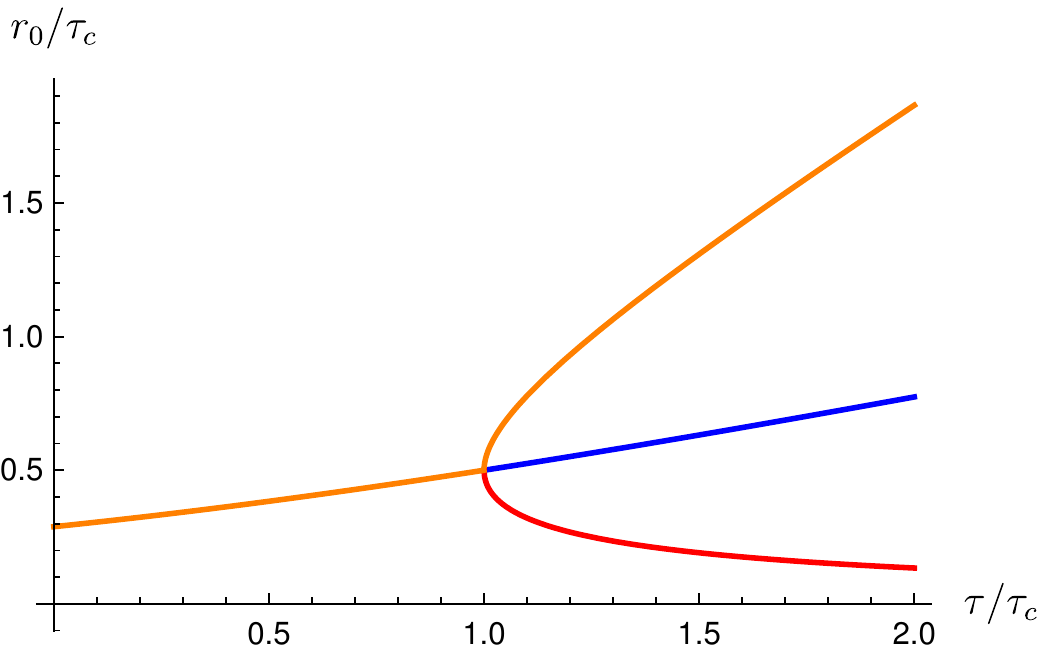}}\\
		\scalebox{0.7}{\includegraphics{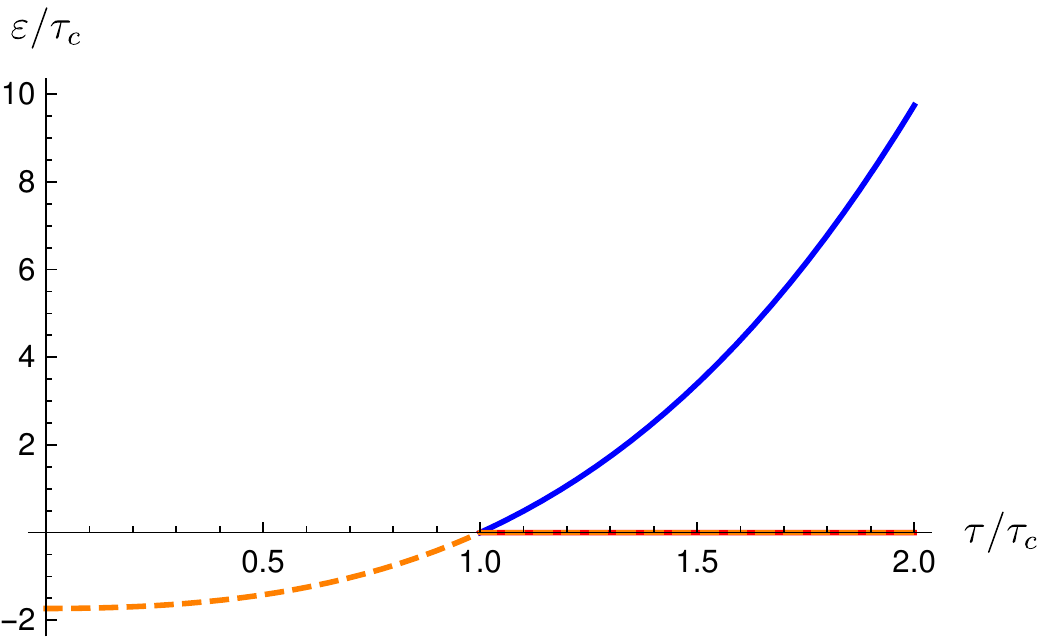}} &
		\scalebox{0.7}{\includegraphics{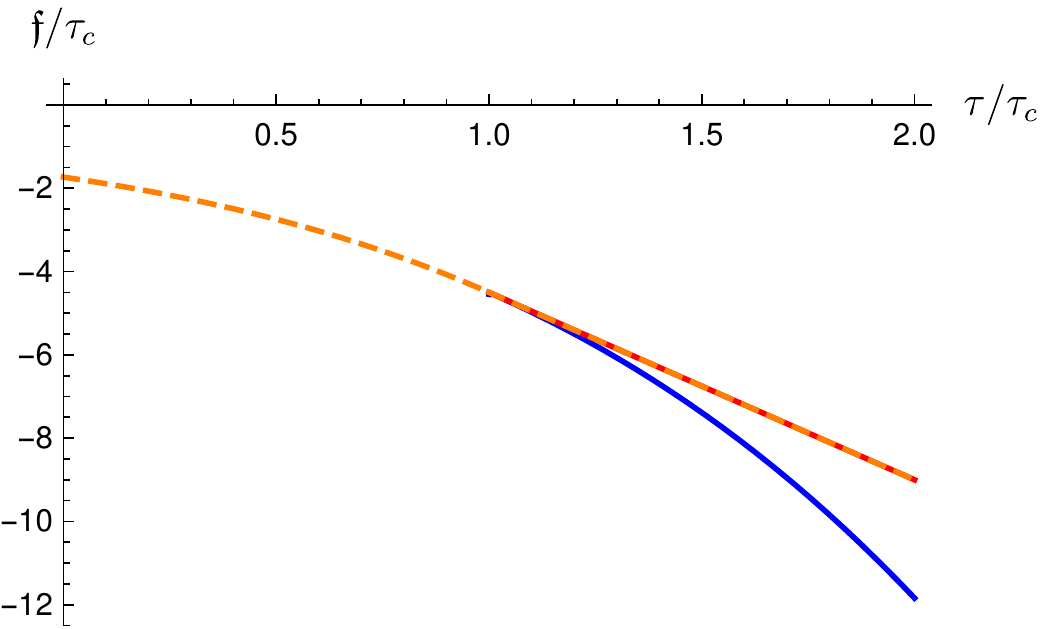}}\\
	\end{tabular}
	\centering
	\caption{Plot of the perturbative solutions \eqref{small-charge-e} for $\r=0.003 \r_c$, together with the corresponding energy $\ve$ and Helmholtz free energy $\frak f$ densities.} 
	\label{fig::xi=1::qe=0.00314qc}
\end{figure}

The expression \eqref{r0-e} for the horizon radius and the characteristic curve are valid everywhere except in the regime where $q\sbtx{e}\to 0$ and $v\sbtx{e}\to0$ simultaneously, in which case the manipulations that lead to \eqref{r0-e} and \eqref{charIIe} starting from \eqref{Te1} and \eqref{Te2} break down. This limit must be taken so that $v\sbtx{e}/q\sbtx{e}^2$ is kept fixed and corresponds to the bald solutions that only exist for Theory II at zero charge density.
Being a quartic equation in $v\sbtx{e}$, the characteristic equation admits four roots at fixed temperature and fixed electric and axionic charge densities. At most three of these roots are real and have $r_0\geq0$: we checked numerically that there is always either a root corresponding to a singular metric with $r_0<0$, or at least two complex conjugate roots.
From the two expressions \eqref{Te1} and \eqref{Te2} we find that for small $q\sbtx{e}$ there are three physical black hole solutions, which we will refer to as `red', `blue' and `orange', and whose scalar {\em vev} parameter and horizon radius are given by 
\bsub
\label{small-charge-e}
\bal
&\text{Red:} && v\sbtx{e}=\sqrt{\t^2-\t_c^2}+\co(q\sbtx{e}^2),&& r_0=\frac12\Big(\t-\sqrt{\t^2-\t_c^2}\Big)+\co(q\sbtx{e}^2),&& \t>\t_c,\\
&\text{Blue:} &&v\sbtx{e}=\frac{q\sbtx{e}^2\ell^2}{r_0(2r_0^2-p^2\ell^2)}+\co(q\sbtx{e}^4), && r_0=\frac16\Big(\t+\sqrt{\t^2+3\t_c^2}\Big)+\co(q\sbtx{e}^2),&& \t>\t_c,\\
&{\text{Orange:}} && v\sbtx{e}=\left\{\begin{matrix}
	\frac{q\sbtx{e}^2\ell^2}{r_0(2r_0^2-p^2\ell^2)}+\co(q\sbtx{e}^4), \\
	-\sqrt{\t^2-\t_c^2}+\co(q\sbtx{e}^2), 
\end{matrix}\right.  && r_0=\left\{\begin{matrix}
\frac16\Big(\t+\sqrt{\t^2+3\t_c^2}\Big)+\co(q\sbtx{e}^2),\\
\frac12\Big(\t+\sqrt{\t^2-\t_c^2}\Big)+\co(q\sbtx{e}^2),
\end{matrix}\right.&& \begin{matrix} \t<\t_c,\rule{0cm}{.4cm}\\ \t>\t_c.\rule{0cm}{.6cm}\end{matrix}
\eal
\esub
Here
\eq
\t_c=4\pi\ell^2 T_c=\sqrt{2}p \ell,
\label{Tc}\eeq
denotes the critical temperature at {\em zero} charge density. The true critical temperature increases slightly with increasing charge density, but we find it convenient to use $\t_c$ as a reference temperature at arbitrary charge density. The solutions \eqref{small-charge-e} are plotted in  
figure \ref{fig::xi=1::qe=0.00314qc}, together with the corresponding energy density \eqref{energyDensityIIe} and Helmholtz free energy (see \eqref{SrenII-e})
\be
\frak f=\frac{q\sbtx{e}^2}{r_0 h(r_0)}-\frac{r_0^3h^{2}(r_0)}{\ell^2}-\frac{r_0p^2}2.
\ee  

\begin{figure}
	\begin{tabular}{cc}
		\scalebox{0.7}{\includegraphics{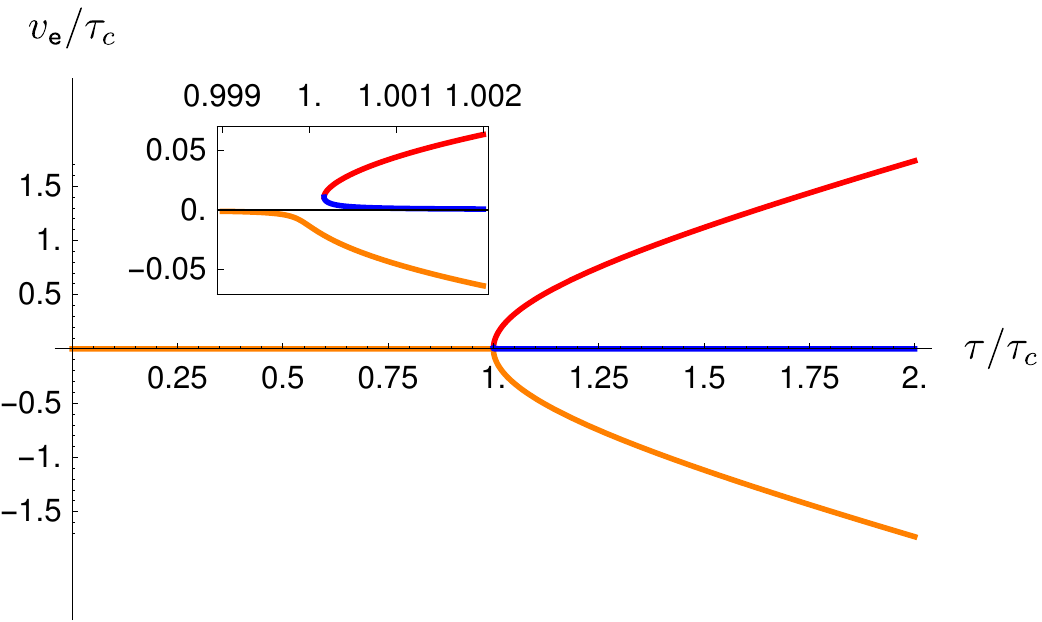}} &
		\scalebox{0.7}{\includegraphics{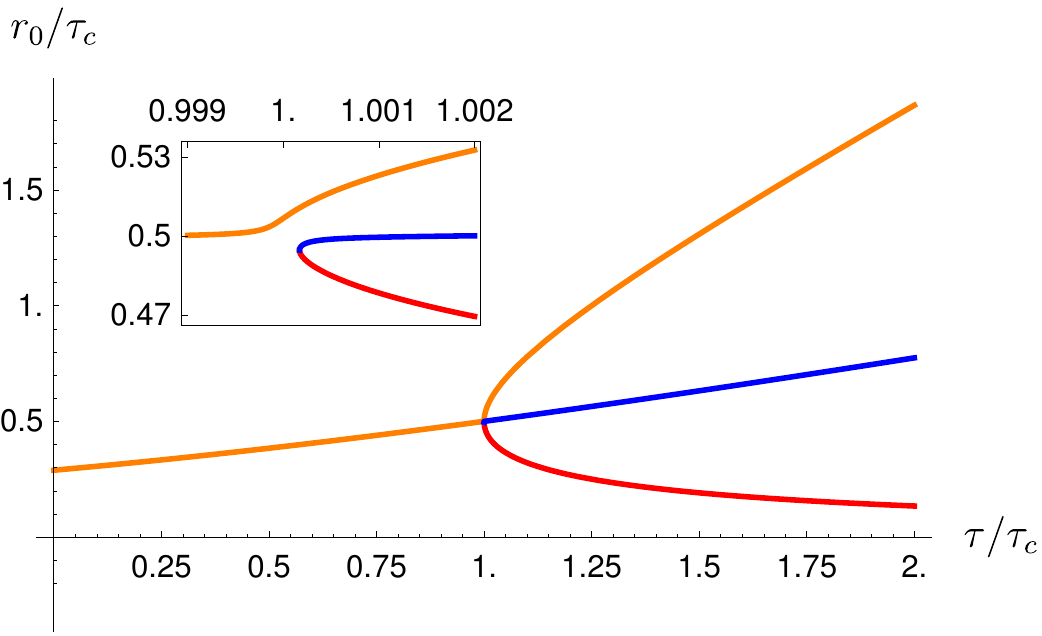}}\\
		\scalebox{0.7}{\includegraphics{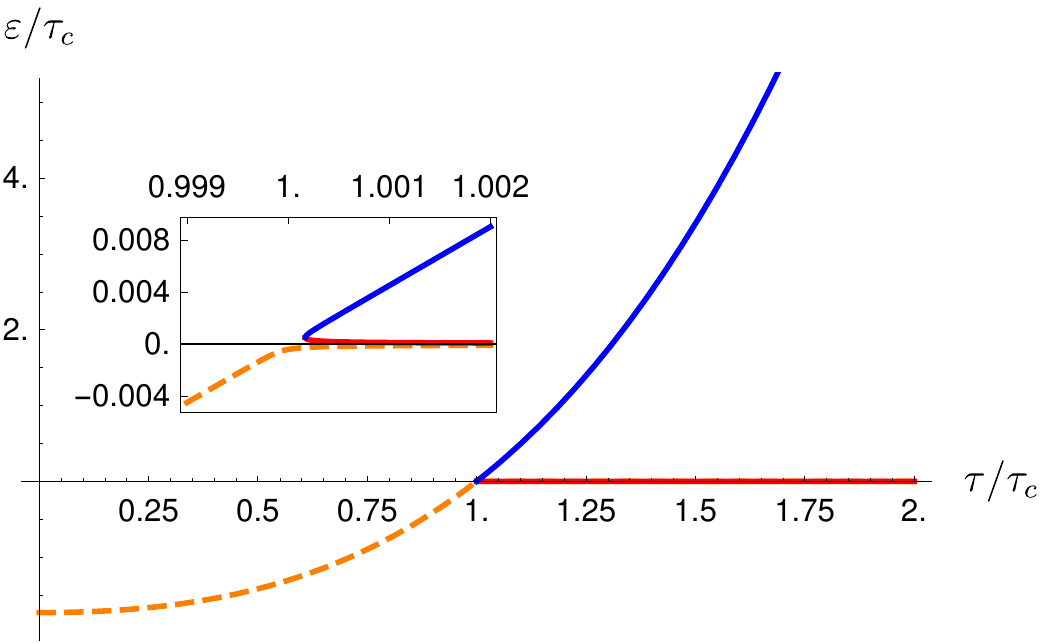}} &
		\scalebox{0.7}{\includegraphics{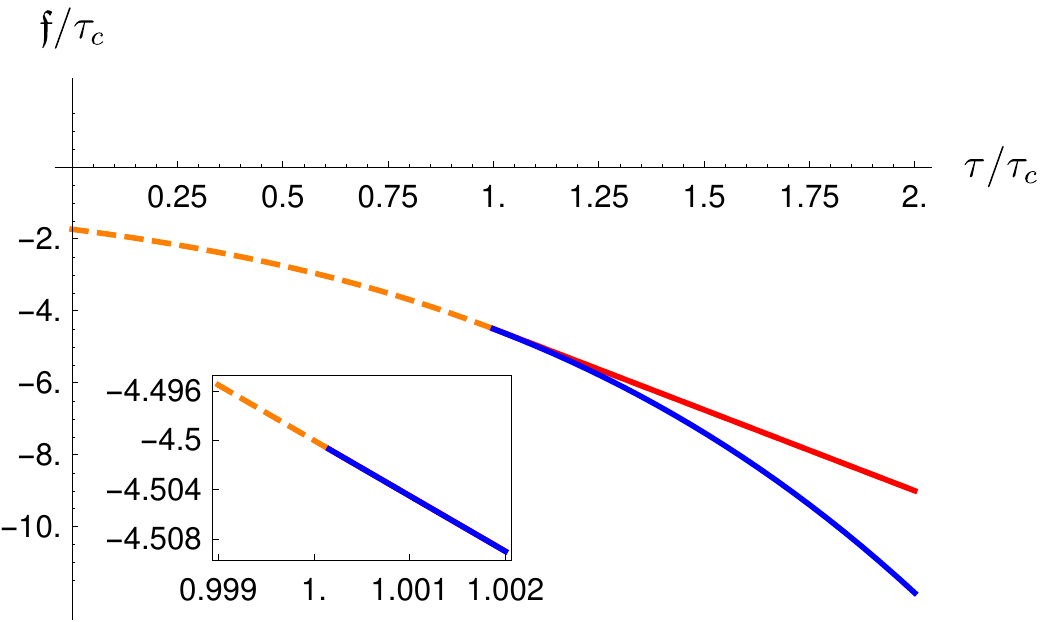}}\\
	\end{tabular}
	\centering
	\caption{Plot of the solutions of \eqref{charIIe} for $\r=0.003 \r_c$, together with the corresponding energy $\ve$ and Helmholtz free energy $\frak f$ densities. Notice that the solutions look identical to the perturbative ones in figure \ref{fig::xi=1::qe=0.00314qc-inset} for the same charge density, but zooming in near the critical temperature shows that the exact solutions are in fact smooth.} 
	\label{fig::xi=1::qe=0.00314qc-inset}
\end{figure}

\begin{figure}
	\begin{tabular}{cc}
		\scalebox{0.7}{\includegraphics{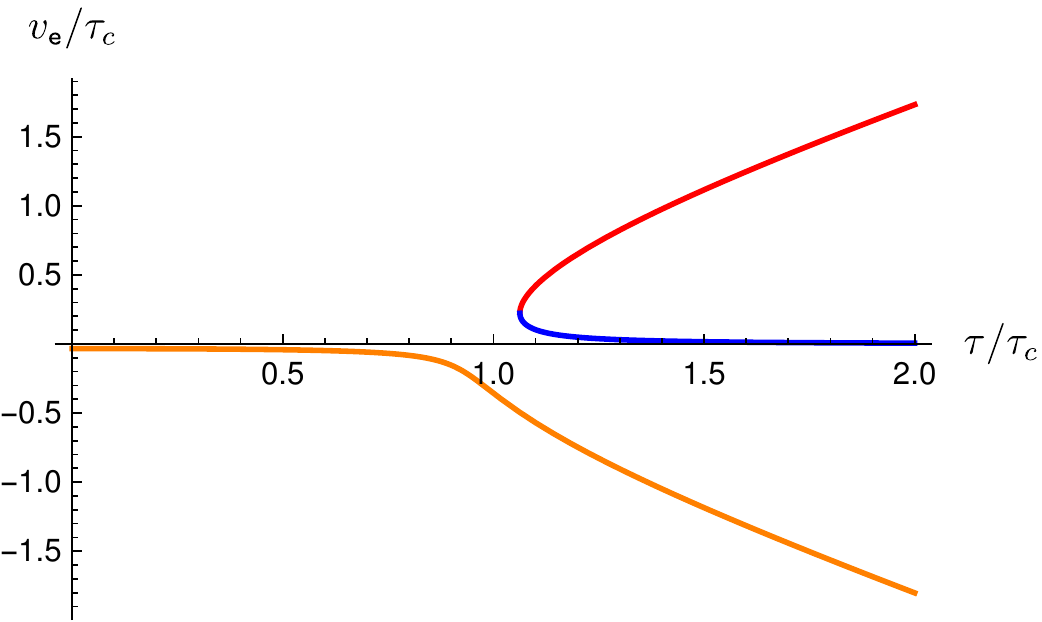}} &
		\scalebox{0.7}{\includegraphics{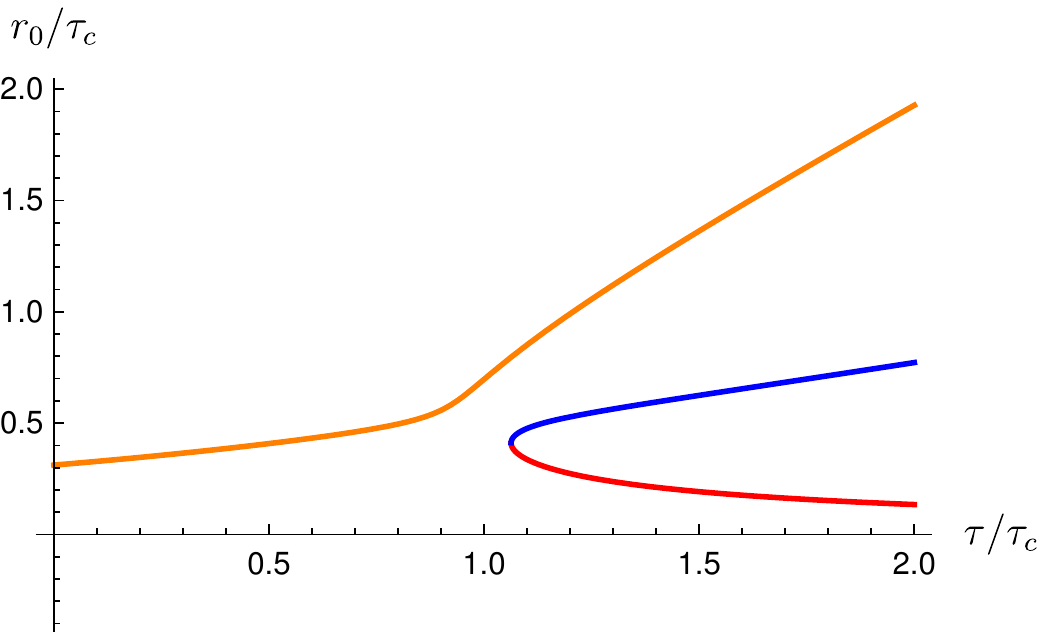}}\\
		\scalebox{0.7}{\includegraphics{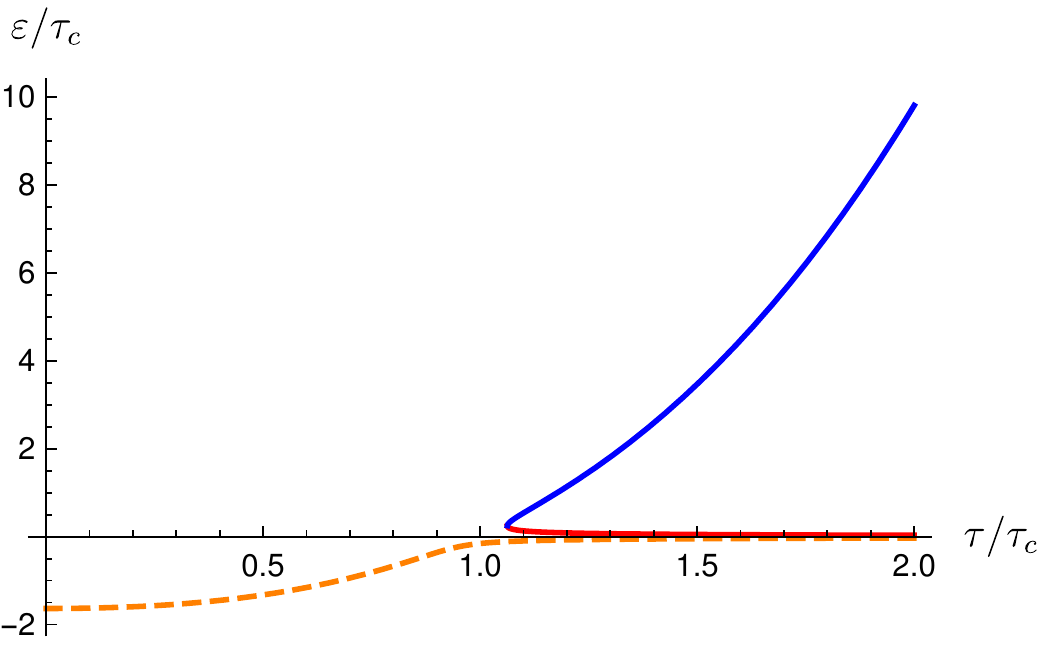}} &
		\scalebox{0.7}{\includegraphics{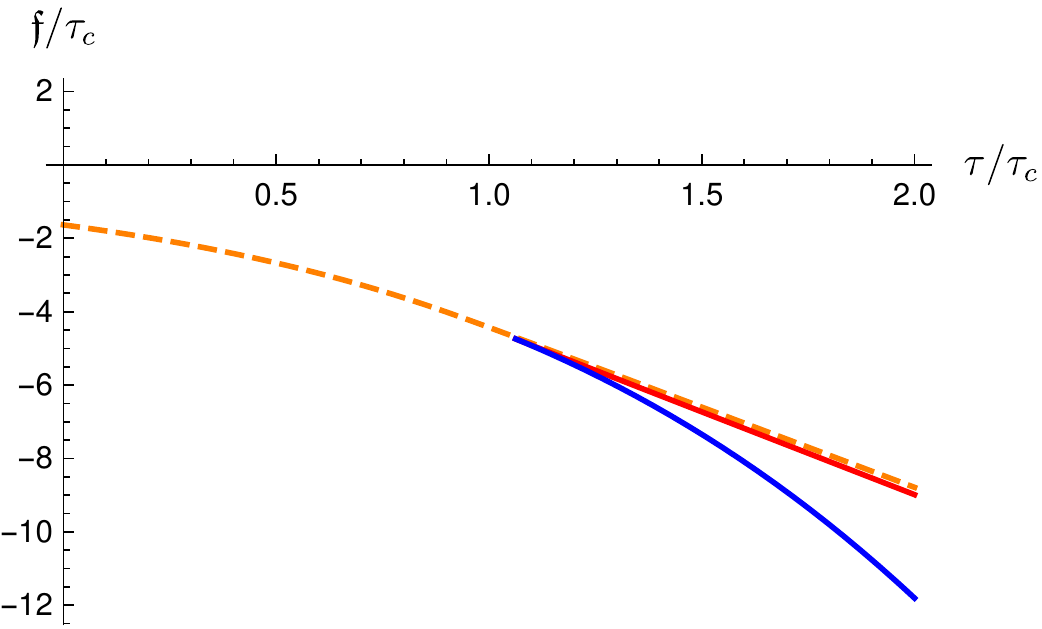}}\\
	\end{tabular}
	\centering
	\caption{Plot of the solutions of \eqref{charIIm} for $\r=0.314\r_c$, together with the corresponding energy $\ve$ and Helmholtz free energy $\frak f$ densities.}
	\label{fig::xi=1::qe=0.314qc}
\end{figure}

\begin{figure}
	\begin{tabular}{cc}
		\scalebox{0.7}{\includegraphics{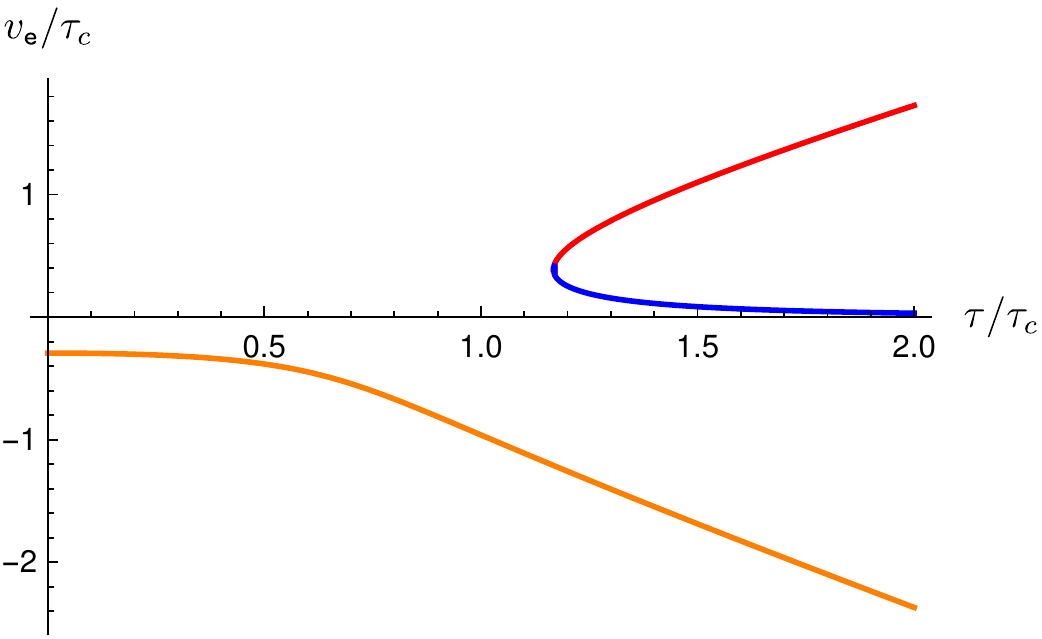}} &
		\scalebox{0.7}{\includegraphics{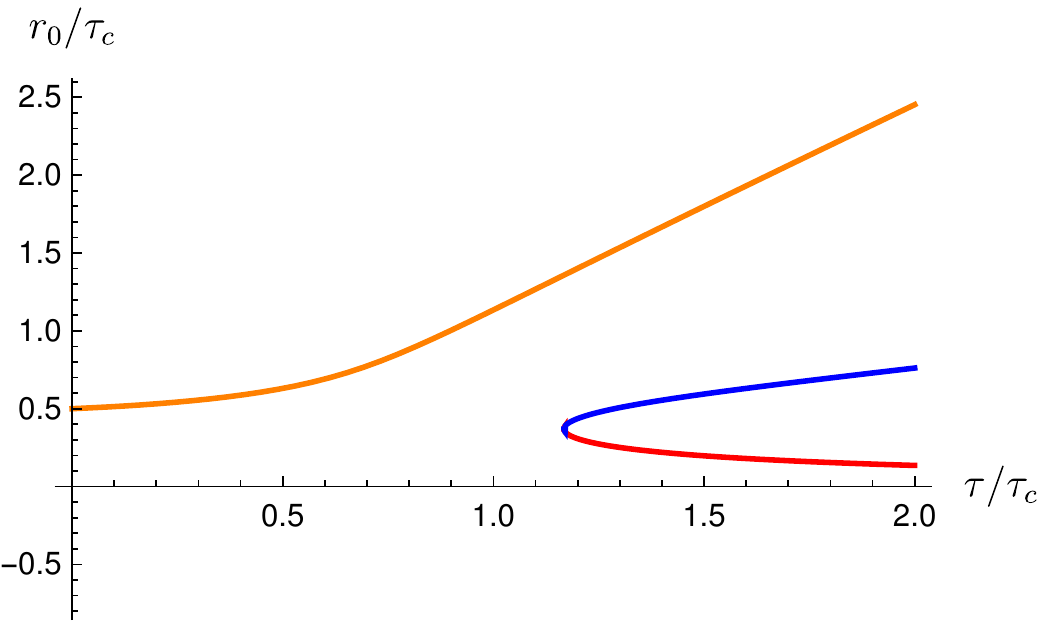}}\\
		\scalebox{0.7}{\includegraphics{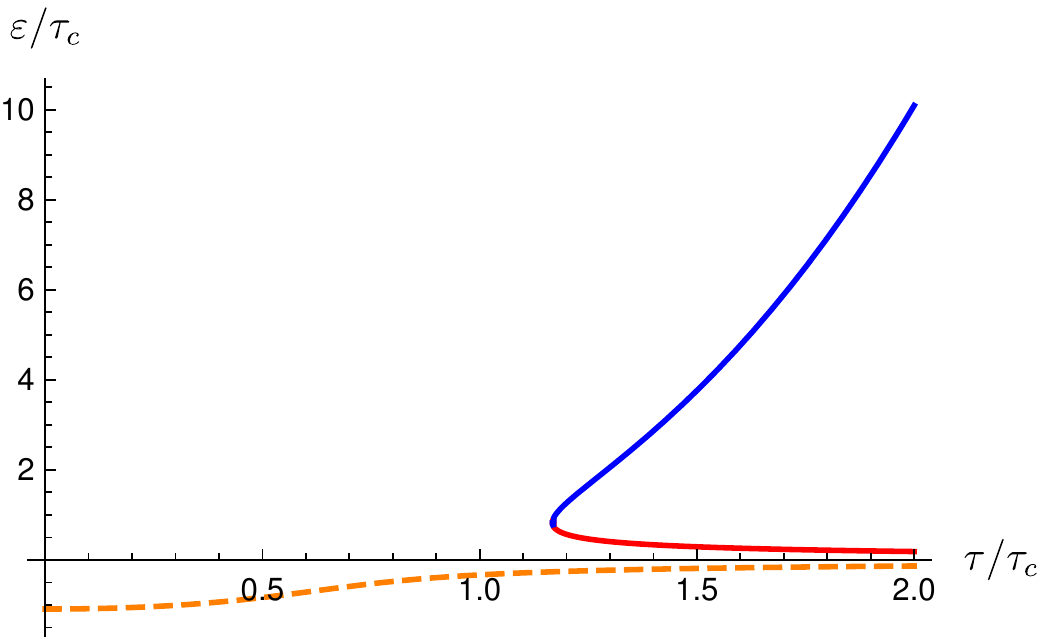}} &
		\scalebox{0.7}{\includegraphics{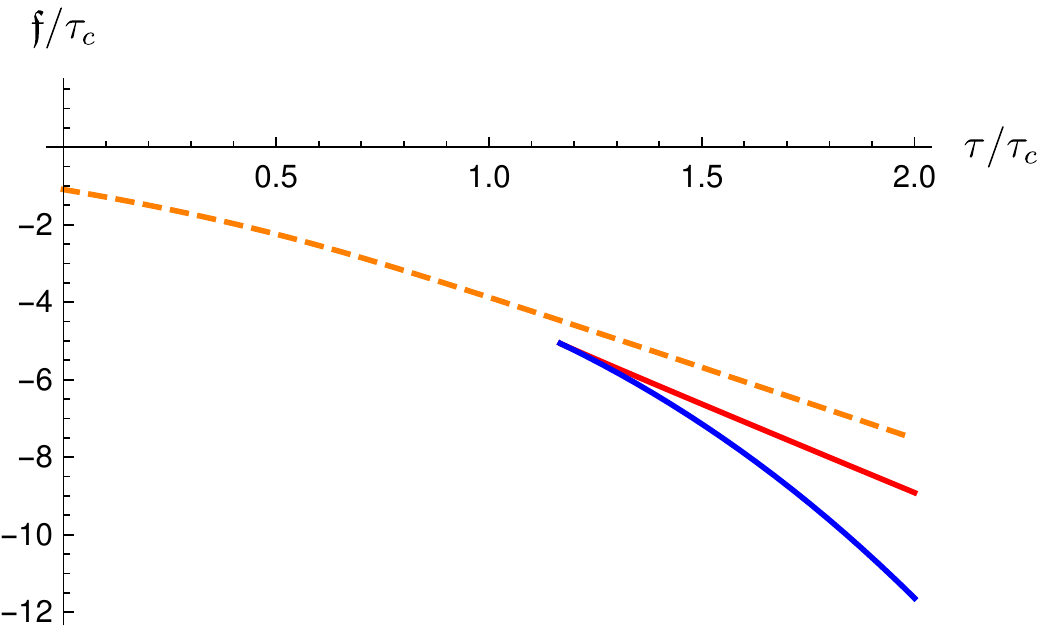}}\\
	\end{tabular}
	\centering
	\caption{Plot of the solutions of \eqref{charIIm} for $\r=0.786\r_c$, together with the corresponding energy $\ve$ and Helmholtz free energy $\frak f$ densities.}
	\label{fig::xi=1::qe=0.786qc}
\end{figure}

\begin{figure}
	\begin{tabular}{cc}
		\scalebox{0.7}{\includegraphics{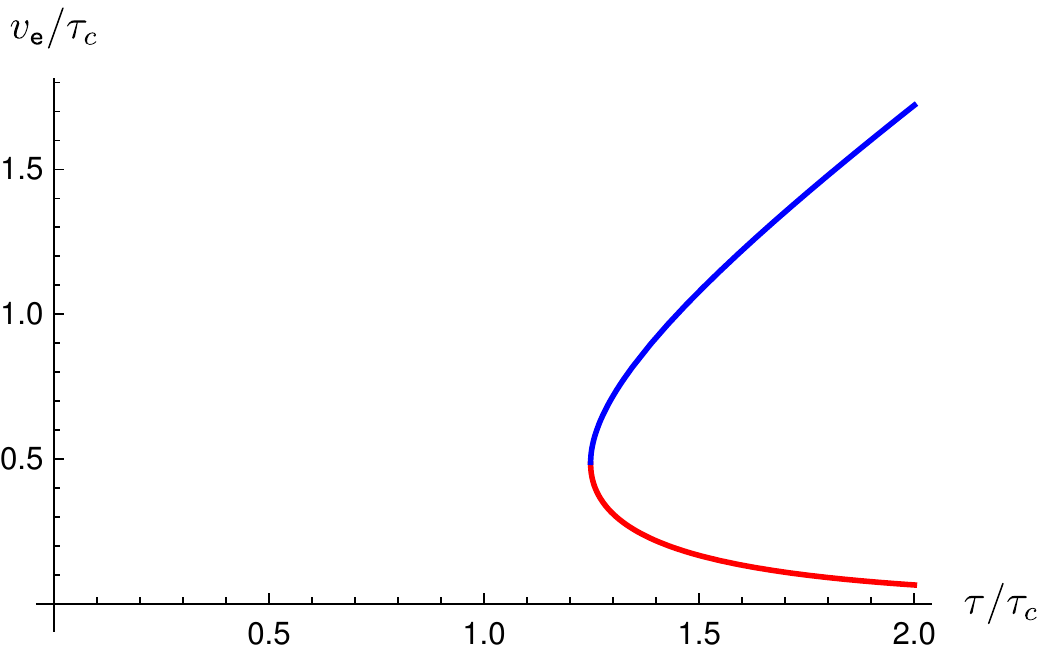}} &
		\scalebox{0.7}{\includegraphics{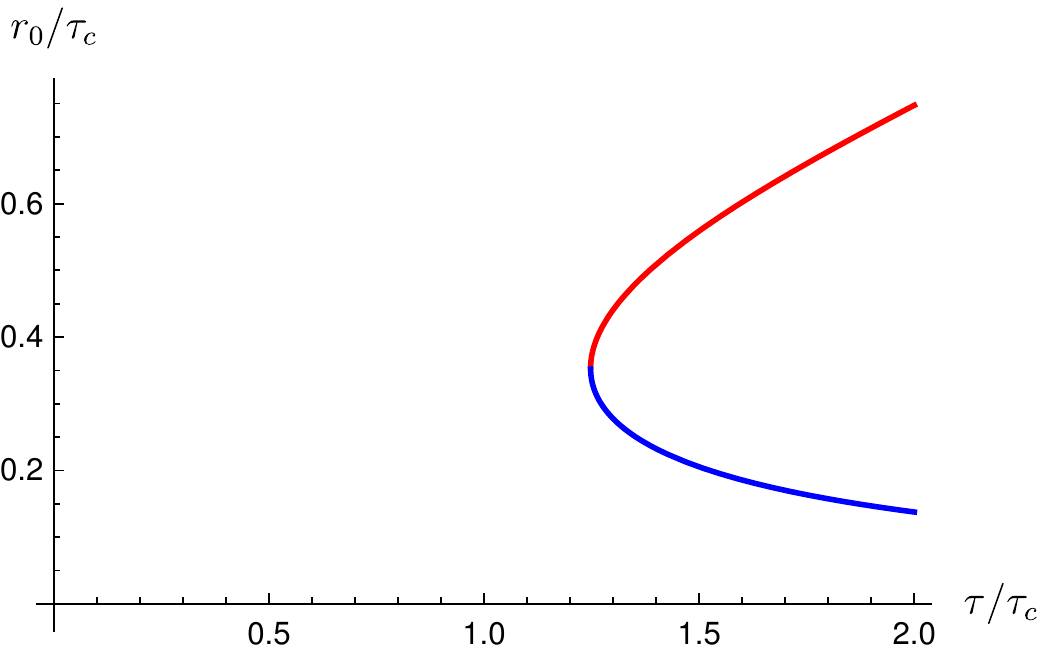}}\\
		\scalebox{0.7}{\includegraphics{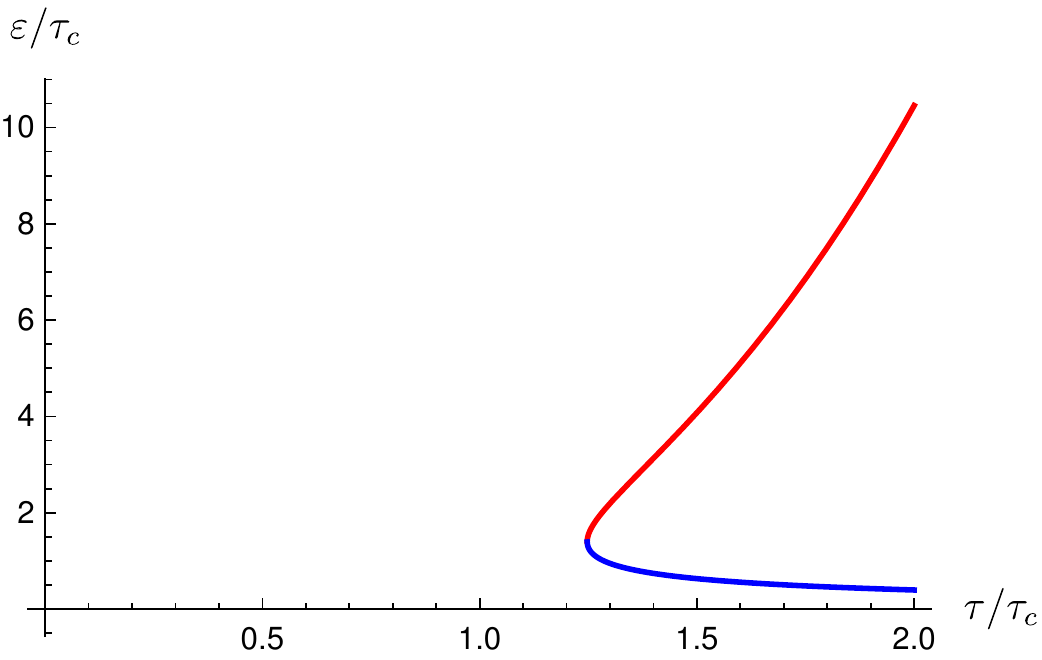}} &
		\scalebox{0.7}{\includegraphics{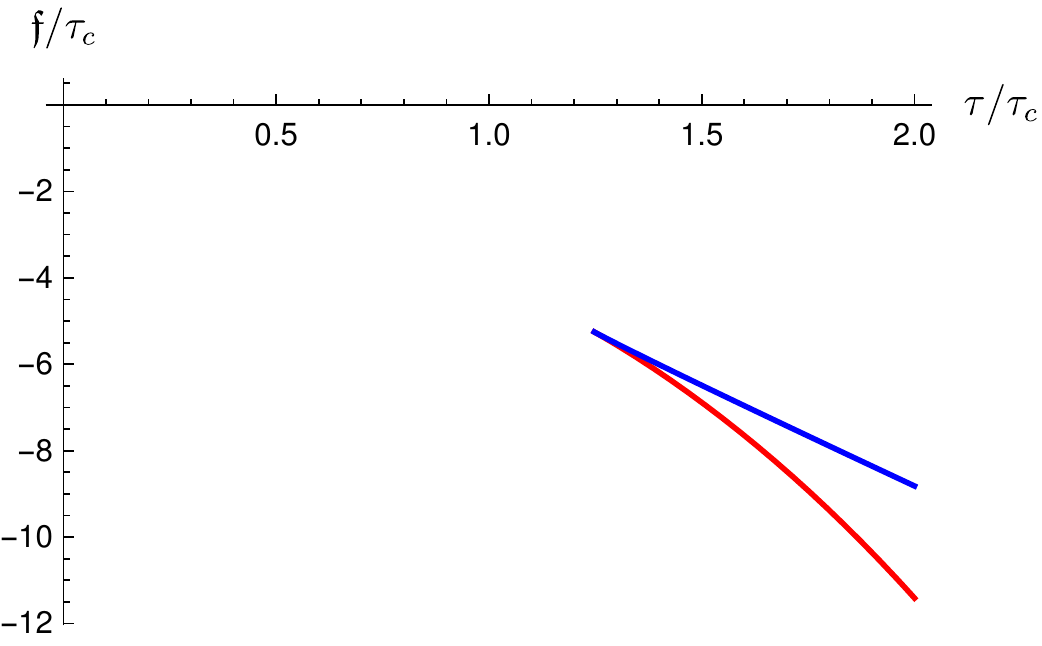}}\\
	\end{tabular}
	\centering
	\caption{Plot of the solutions of \eqref{charIIm} for $\r=1.1\r_c$, together with the corresponding energy $\ve$ and Helmholtz free energy $\frak f$ densities.}
	\label{fig::xi=1::qe=1.1qc}
\end{figure}

Several features of these solutions persist at higher charge densities, but there are also qualitative changes as the charge density is increased, as can be seen in figures \ref{fig::xi=1::qe=0.314qc}, \ref{fig::xi=1::qe=0.786qc} and \ref{fig::xi=1::qe=1.1qc}. An important observation regarding the perturbative solutions \eqref{small-charge-e} that is clear from the plots in figure~\ref{fig::xi=1::qe=0.00314qc} is that perturbation theory for small $q\sbtx{e}$ breaks down near the critical temperature $\t_c$. The corner in the orange curve as well as the apparent pole in $v\sbtx{e}$ at $\t_c$ for the orange and blue solutions in \eqref{small-charge-e} are indications that perturbation theory breaks down at $\t_c$. However, at small but non zero $q\sbtx{e}$ one can zoom in near the critical temperature using the analytical solutions of the characteristic curve \eqref{charIIe} as is done in figure \ref{fig::xi=1::qe=0.00314qc-inset}, which shows that all curves are in fact smooth. The corners exist only in the strict $q\sbtx{e}=0$ limit, or as an artifact of perturbation theory at small $q\sbtx{e}$. 

The general structure of the physical solutions that emerges from the plots in figures \ref{fig::xi=1::qe=0.00314qc}, \ref{fig::xi=1::qe=0.314qc}, \ref{fig::xi=1::qe=0.786qc} and \ref{fig::xi=1::qe=1.1qc} is as follows. Below the critical charge density  
\eq
|\rho|<\rho_c = \frac{\ell^3}{2\sqrt2}\,\Pi^2=\frac{p^2\ell}{2\sqrt2},
\eeq
there always exists one solution for all temperatures (orange), while two additional solutions appear above a (charge density dependent) critical temperature that equals $T_c$ at zero charge density. For $|\r|>\r_c$ the orange solution disappears, leaving only the other two branches above the critical temperature. There are no solutions above the critical charge density and below the critical temperature. This is depicted in the left plot in figure~\ref{fig::xi=1}. The right plot in figure~\ref{fig::xi=1} shows the number of solutions as a function of temperature and electric chemical potential, instead of charge density. Notice that there is no critical value for the chemical potential, which reflects the fact that $|q\sbtx{e}(\m,p,T)|$ for the orange solution is bounded by $\rho_c$. 

The dynamic and thermodynamic stability properties of the solutions can be read off respectively the energy density and Helmholtz free energy density plots in figures \ref{fig::xi=1::qe=0.00314qc}, \ref{fig::xi=1::qe=0.314qc}, \ref{fig::xi=1::qe=0.786qc} and \ref{fig::xi=1::qe=1.1qc}. From the energy density plots we deduce that the orange solution is always dynamically unstable, while the blue and red solutions are always dynamically stable. Above $T_c$, however, in the limit of vanishing charge density the red solution becomes marginally stable while the orange one becomes marginally unstable. Moreover, from the free energy plots follows that when they coexist, the blue solution is thermodynamically stable, while both the red and the orange solutions are thermodynamically unstable. Nevertheless, the orange solution has the largest radius, while the red solution has the smallest radius. We will see below that this last property is reversed in the magnetically charged solutions. 

\begin{figure}[tb]
\hspace{5pt}
		\begin{center}
			\begin{tikzpicture}[scale=0.7, every node/.style={scale=0.7}]

			\fill[color=orange!70!white, opacity=1.] (0,0)--(2.5,0) to [out=60, in= 250] (3.96 ,3) -- (2.87,3) -- (2.87,3) -- (0,3) -- (0,0) -- (2.5,0);
			\fill[color=blue!20!white, opacity=0.5] (3.96,3) to [out = 70, in=260] (5,7) -- (11.5,7) -- (11.5,0) -- (2.5,0);
			
			\node [anchor=south] at (6.5,7.5) {\large $\xi = 1$};
			
			\draw[->] (0,0) -- (12,0) node[anchor=north] {$T$};
			\draw[->] (0,0) -- (0,7.5) node[anchor=east] {$|\r|$};
			
			\draw (2pt,3) -- (0,3) node [left]  {$\r_c$};
			\draw[dashed] (0,3) -- (3.96,3);
			\draw (2.5, 2pt) --(2.5,0) node [below] {$T_c$};
			\draw[thick] (3.96,3) to [out=70, in= 260] (5,7);
			\draw[thick] (2.5,0) to [out=61, in =247] (3.96,3);			
			
			\node at (1.8,1.5) {Phase I};
			\node at (6.5,1.5) {Phase II};

			\end{tikzpicture}
		\end{center}
	
	\caption{Phase diagram for the electric solutions of Theory II with $\xi=1$ as a function of charge density $\rho$ and temperature $T$. Below a critical charge density $\r_c$ and a critical temperature $T_c$ there is only one, dynamically unstable, black hole solution. This solution exists for arbitrary temperature, but it disappears above $\r_c$. Above the critical temperature $T_c$ there are either two or three solutions depending on whether $|\r|>\r_c$ or $|\r|<\r_c$ respectively. The two solutions that exist only above $T_c$ are both dynamically stable and the one with larger radius is thermodynamically preferred. At the critical temperature $T_c$, therefore, there is a phase transition from a dynamically stable black hole with negative scalar {\em vev} above $T_c$, to a dynamically unstable solution with positive scalar {\em vev}. At non-zero charge density the Helmholtz free energy jumps at $T_c$ and so this is a {\em zeroth} order transition. At zero charge density the free energy is continuous across $T_c$, as is its first derivative, and so the phase transition becomes second order at $\r=0$. }
	\label{fig::xi=1-phase-diag}
\end{figure}
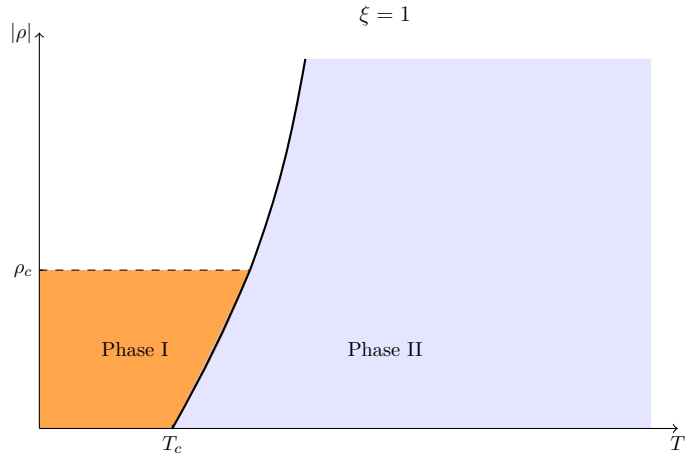

Putting everything together, we conclude that the electric solutions of theory II describe in general two distinct phases, as shown in figure~\ref{fig::xi=1-phase-diag}. Phase~I corresponds to the orange solutions and is the only possible phase below the critical temperature and critical charge density. Phase~II corresponds to the blue solutions and it dominates above the critical temperature, for any value of the charge density. There is no regime of parameters where the red solutions are thermodynamically dominant. At non zero charge density the Helmholtz free energy jumps at the critical temperature and so this is a {\em zeroth} order phase transition.  Zeroth order phase transitions have been predicted in the context of superfluidity and superconductivity and are related to the presence of metastable states \cite{MR2129341,Ibanez:2008xq}, as well as in higher dimensional black holes \cite{Altamirano:2013ane}. As the charge density approaches zero, however, the jump of the free energy across the critical temperature goes to zero, but at the same time its derivative is continuous across $T_c$ and, hence, the transition becomes second order. However, since the solutions of phase~I are dynamically unstable, this phase diagram is presumably not the complete picture. There are probably other solutions that are thermodynamically and dynamically stable below the critical temperature, that also continue to exist above the critical charge density. It would be interesting to identify these solutions.    
\begin{figure}[tb]
	\begin{minipage}{0.5\textwidth}
		\begin{center}
			\begin{tikzpicture}[scale=0.6, every node/.style={scale=0.7}]
			\node [anchor=south] at (6.5,7.5) {\large $\xi = 1$};
			\draw[->] (0,0) -- (12,0) node[anchor=north] {$T$};
			\draw[->] (0,0) -- (0,7.5) node[anchor=east] {$|\r|$};
			
			\node[anchor = east] at (0,0) {0};
			
			\draw (2pt,3) -- (0,3) node [left]  {$\r_c$};
			\draw[dashed] (0,3) -- (11.5,3);
			\draw (2.5, 2pt) --(2.5,0) node [below] {$T_c$};
			\draw[thick] (2.5,0) to [out=60, in= 260] (5,7);
			\fill[pattern=vertical lines, pattern color = red]
			(2.5,0) to [out=61, in =247] (3.96,3) -- (11.5,3) -- (11.5,0) -- (2.5,0);
			\fill[pattern=north east lines, pattern color = blue]
			(2.5,0) to [out=61, in =247] (3.96,3) -- (8.,3) -- (8.0,0) -- (2.5,0);

			\fill[pattern= north east lines, pattern color = blue] (8.,0) to  (8.,7) -- (11.5,7) -- (11.5,0) -- (11.5,0);

			\fill[pattern= north east lines, pattern color = blue] (3.96,3) to [out = 70, in=260] (5,7) -- (8.,7) -- (8.,3) -- (8.,3);

			\fill[pattern= vertical lines, pattern color = red] (3.96,3) to [out = 70, in=260] (5,7) -- (11.5,7) -- (11.5,3) -- (11.5,3);
			
			\fill[pattern= horizontal lines, pattern color = orange] (0,0)--(11.5,0) to 
			(11.5,3) -- (0,3) -- (0,0);

			\node[mybox,anchor =  west] at (0.1,1.5) {One Solution};
			\node[anchor=west] at (0.8,5.) {No Solutions};
			\node[mybox] at (8.2,1.5) {Three Solutions};
			\node[mybox] at (8.2,5.0) {Two Solutions};

			\end{tikzpicture}
		\end{center}
	\end{minipage}
	\hspace{5pt}
	\begin{minipage}{0.5\textwidth}
	\begin{center}
		\begin{tikzpicture}[scale=0.7, every node/.style={scale=0.7}]
		
		\fill[pattern= horizontal lines, pattern color = orange]
		(2,0) to [out = 80, in= -105] (3.5,6) -- (10,6)--(10,0) -- (2,0);
		\fill[pattern= vertical lines, pattern color = red]
		(2,0) to [out = 80, in= -105] (3.5,6) -- (10,6)--(10,0) -- (2,0);
		\fill[pattern= north east lines, pattern color = blue]
		(2,0) to [out = 80, in= -105] (3.5,6) -- (10,6)--(10,0) -- (2,0);

		\fill[pattern= horizontal lines, pattern color = orange]
		(2,0) to [out = 80, in= -105] (3.5,6) -- (0,6)--(0,0) -- (2,0);
		
		\draw[thick] (2,0) to [out = 80, in= -105] (3.5,6);
		
		\node [anchor=south] at (5,6.5) {\large $\xi = 1$};
		
		\draw[->,thick] (4,2) -- (4,4);
		\node[mybox, anchor=west] at (4.2,3){Increasing $\ve$, scalar {\em vev} and $s$};
		\draw[->,thick] (5,1.2) -- (7,1.2);
		\node[mybox, anchor=north] at (6,1){Decreasing scalar {\em vev}, increasing $\ve$ and $s$};

		\draw[->] (0,0) -- (10.5,0) node[anchor=north] {$T$};
		\draw[->] (0,0) -- (0,6.6) node[anchor=east] {$|\m|$};
		
		\draw (2., 2pt) --(2.,0) node [below] {$T_c$};
		
		\node[anchor = east] at (0,0) {0};
		
		\node [mybox, anchor=center] at (1.5,3.5) {One Solution};
		\node [mybox, anchor=center] at (8,5.5) {Three Solutions};
		
		\end{tikzpicture}
	\end{center}
	\end{minipage}

	\caption{Number of electric solutions of Theory II with $\xi=1$ as a function of charge density $\rho$ and temperature $T$ ({\bf left plot}), or chemical potential $\m$ and temperature $T$ ({\bf right plot}). The plots apply to any fixed $|p|>0$. There are three distinct solutions, indicated respectively by horizontal orange lines, red vertical lines, and blue diagonal lines. We refer to these solutions as `orange', `red' and `blue', respectively. The orange solution exists for all temperatures provided the (absolute value of the) charge density is below the critical value $\r_c=\ell p^2/2\sqrt{2}$. However, this solution is dynamically unstable since it always has negative energy density. Moreover, where it coexists with the red and blue solutions it has the largest radius, but it is thermodynamically unstable. The red and blue solutions appear simultaneously above a critical temperature                                                                                                                                                                                                                                                  $T_c\geq |p|/2\sqrt{2}\pi\ell$ and are always dynamically stable, since they have positive energy density. The blue solutions have larger radius than the red ones and are thermodynamically preferred. }
	\label{fig::xi=1}
\end{figure}
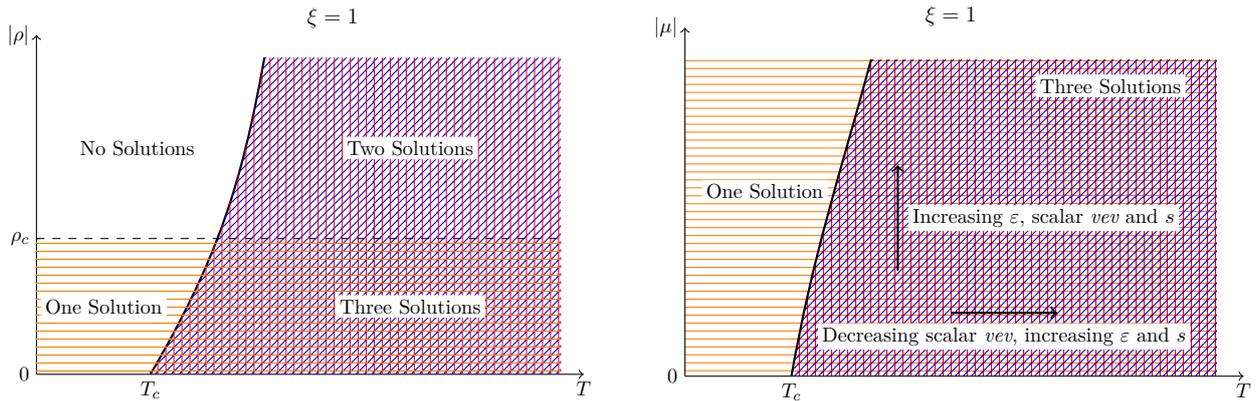

\paragraph{Phases of magnetic solutions of Theory II} The structure of the magnetic solutions of Theory II is very similar to that of the electric ones, except for a few minor features that we are going to highlight. As for the electric solutions, we can process the temperature in two different ways leading to 
\be\label{Tm1}
\t=3r_0+2v\sbtx{m}-\frac{\ell^2p^2}{2r_0},
\ee
and 
\be\label{Tm2}
\t=2v\sbtx{m}-\frac{3\ell^2q\sbtx{m}^2+\ell^2p^2v\sbtx{m}(v\sbtx{m}-2r_0)}{2r_0v\sbtx{m}(v\sbtx{m}+r_0)},
\ee
for the rescaled temperature $\t=4\pi\ell^2 T$. At generic values of the parameters these again correspond to two quadratic equations for $r_0$. Eliminating the quadratic term in $r_0$ by a suitable linear combination of these expressions, we obtain the general expression for the horizon radius
\be\label{r0-m}
r_0=\frac{9\ell^2q\sbtx{m}^2+\ell^2p^2v\sbtx{m}(v\sbtx{m}+\t)}{2v\sbtx{m}\(3\ell^2p^2+(2v\sbtx{m}-\t)(v\sbtx{m}+\t)\)}. 
\ee
Inserting this back in either \eqref{Tm1} or \eqref{Tm2} gives the characteristic curve
\be\label{charIIm}
(\ell^2 p^4-8q\sbtx{m}^2)v\sbtx{m}^4+(2\ell^4p^6-\ell^2p^4\t^2-18\ell^2p^2q\sbtx{m}^2+6q\sbtx{m}^2\t^2)v\sbtx{m}^2-2q\sbtx{m}^2\t^3v\sbtx{m}-27\ell^2 q\sbtx{m}^4=0.
\ee
Notice that although the expression for the horizon \eqref{r0-m} is different from the corresponding expression for the electric solutions in \eqref{r0-e}, the characteristic curves are identical under the map $q\sbtx{e}\to q\sbtx{m}$, $v\sbtx{e}\to -v\sbtx{m}$. It follows that the solutions for the scalar {\em vev} are identical to those for the electric solutions, except for an overall sign change. 

\begin{figure}
	\begin{tabular}{cc}
		\scalebox{0.7}{\includegraphics{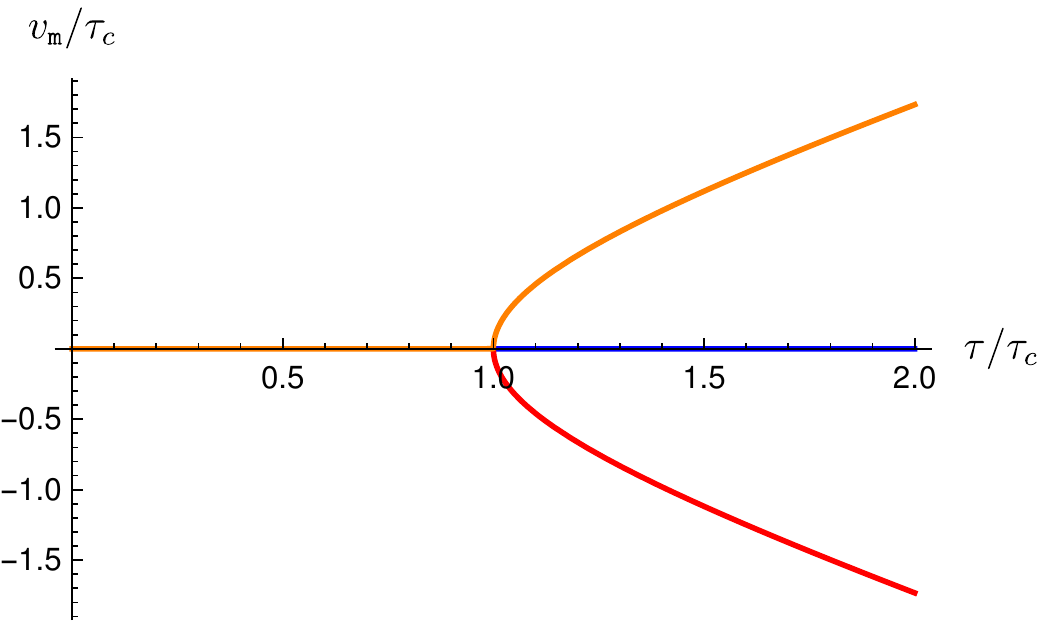}} &
		\scalebox{0.7}{\includegraphics{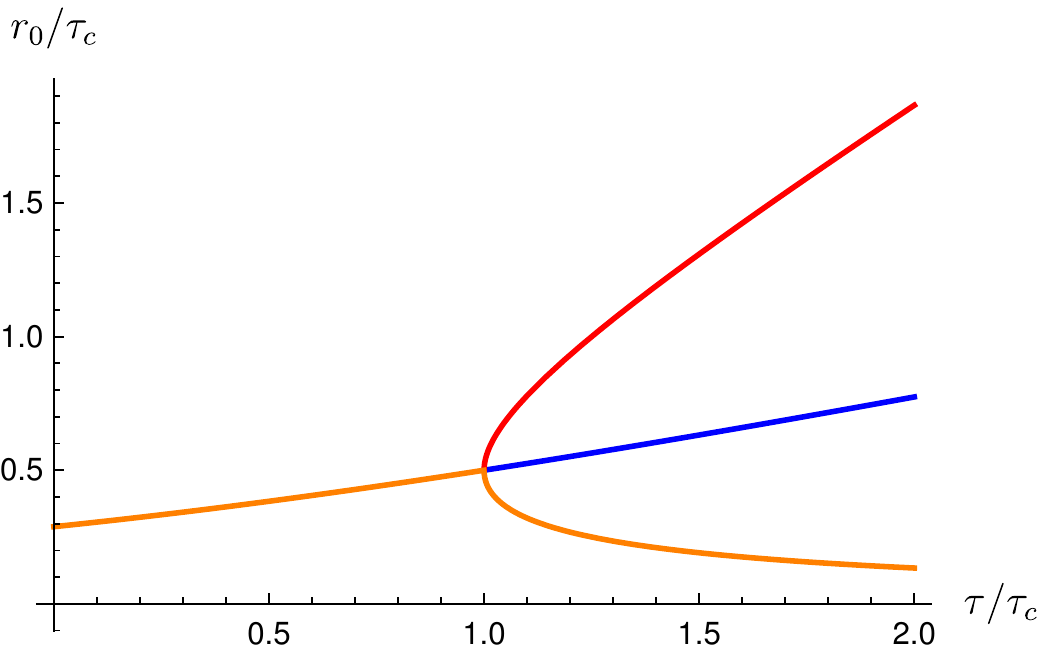}}\\
		\scalebox{0.7}{\includegraphics{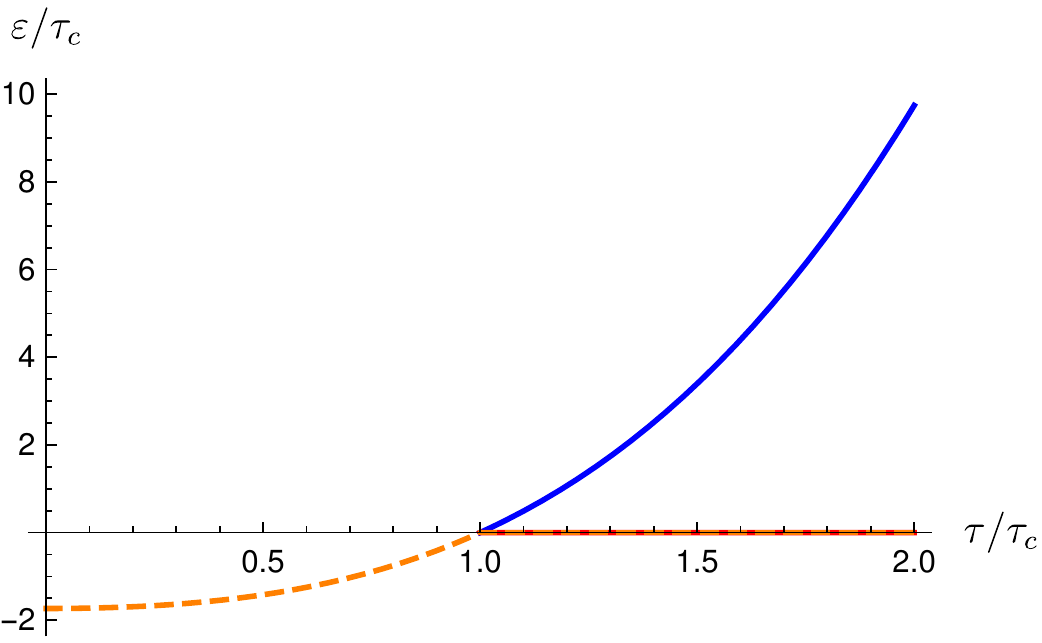}} &
		\scalebox{0.7}{\includegraphics{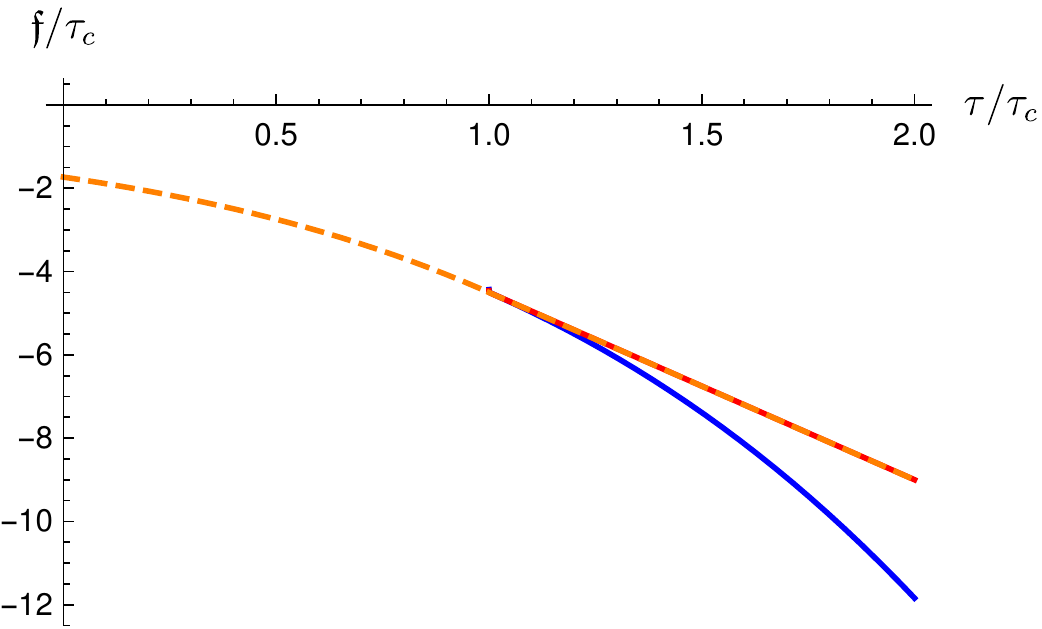}}\\
	\end{tabular}
	\centering
	\caption{Plot of the perturbative solutions \eqref{small-charge-m} for $\cb=0.003\cb_c$, together with the corresponding energy $\ve$ and Helmholtz free energy $\frak f$ densities.}
	\label{fig::xi=1::qm=0.00314qc}
\end{figure}

\begin{figure}
	\begin{tabular}{cc}
		\scalebox{0.7}{\includegraphics{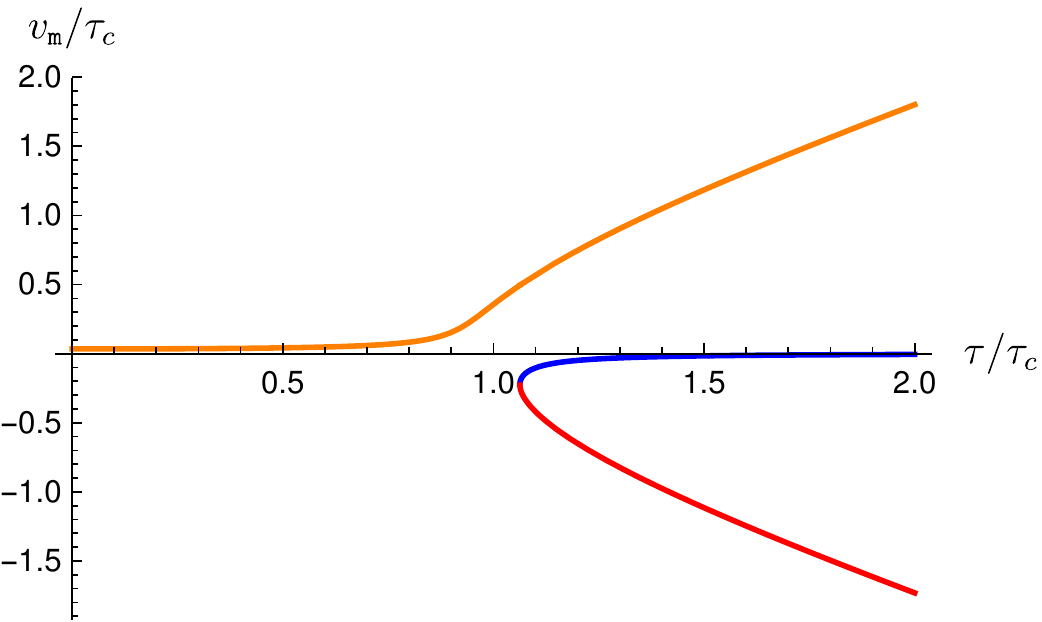}} &
		\scalebox{0.7}{\includegraphics{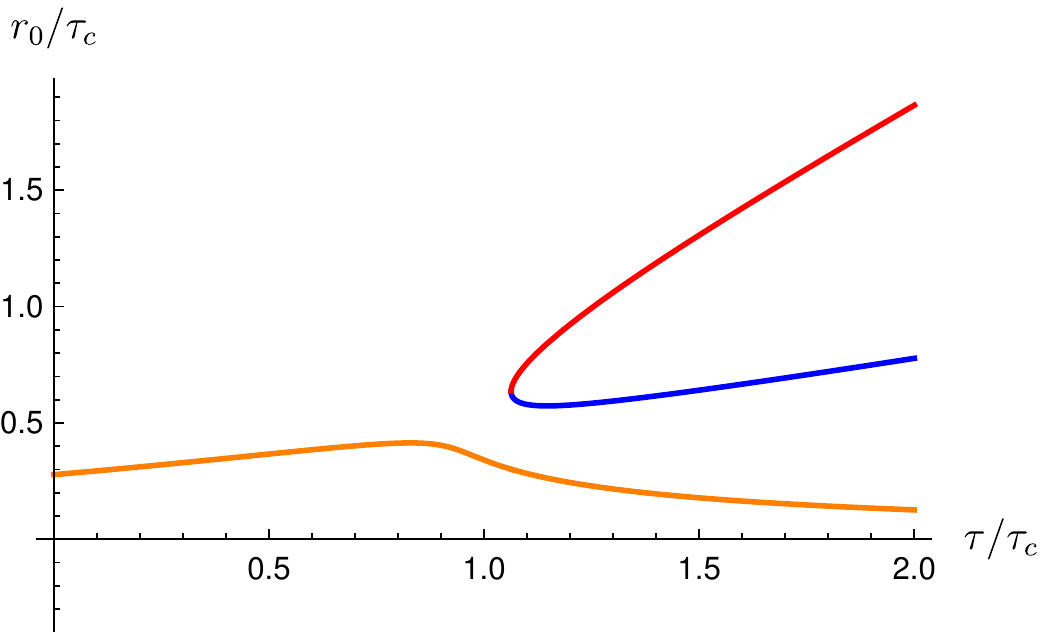}}\\
		\scalebox{0.7}{\includegraphics{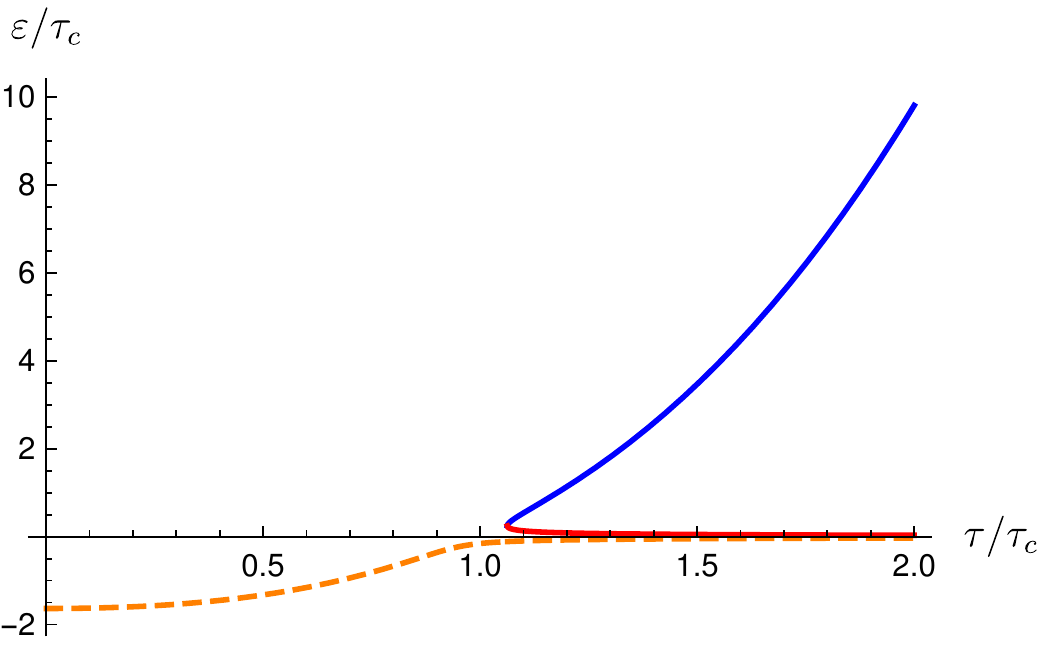}} &
		\scalebox{0.7}{\includegraphics{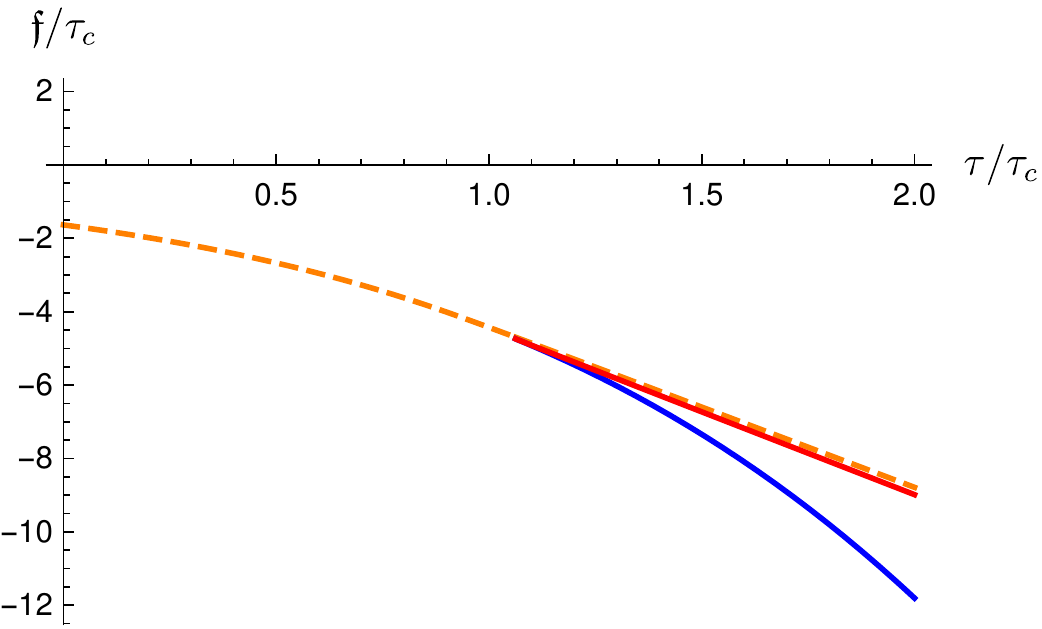}}\\
	\end{tabular}
	\centering
	\caption{Plot of the solutions of \eqref{charIIm} for $\cb=0.314\cb_c$, together with the corresponding energy $\ve$ and Helmholtz free energy $\frak f$ densities.}
	\label{fig::xi=1::qm=0.314qc}
\end{figure}

\begin{figure}
	\begin{tabular}{cc}
		\scalebox{0.7}{\includegraphics{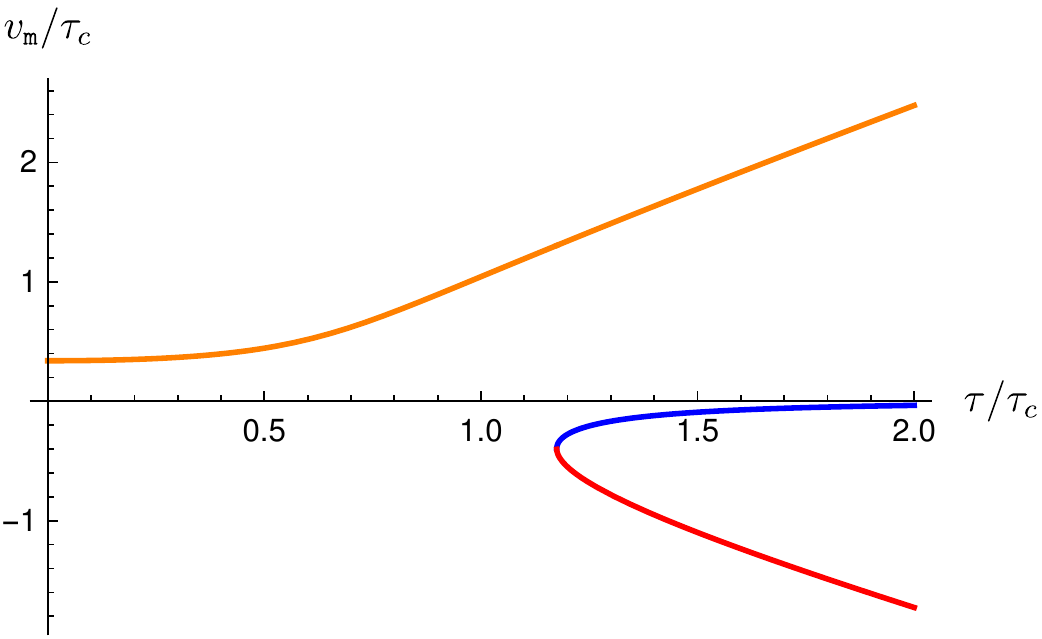}} &
		\scalebox{0.7}{\includegraphics{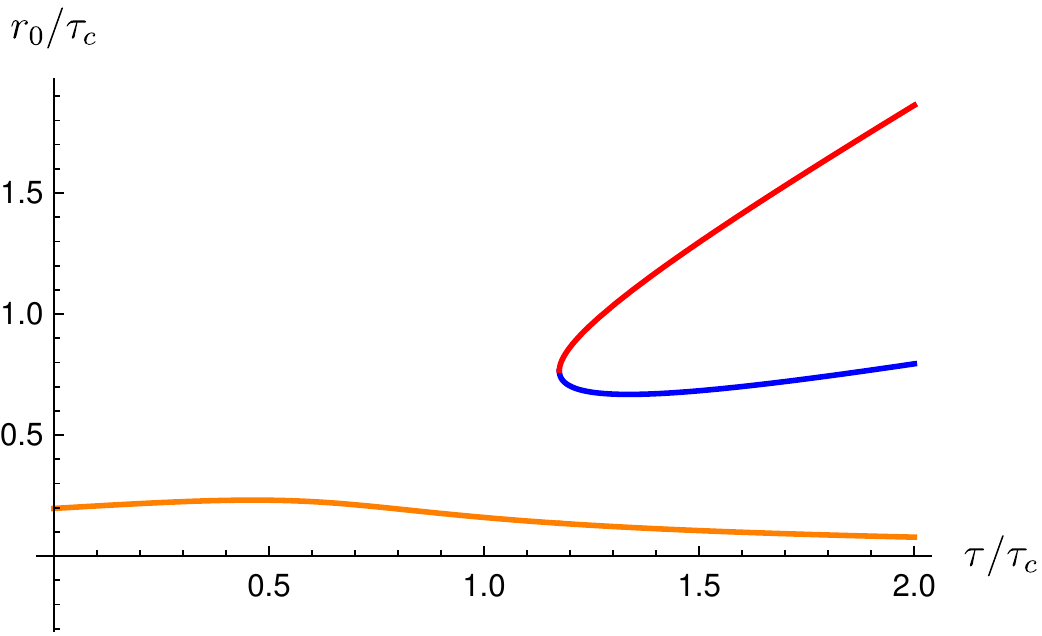}}\\
		\scalebox{0.7}{\includegraphics{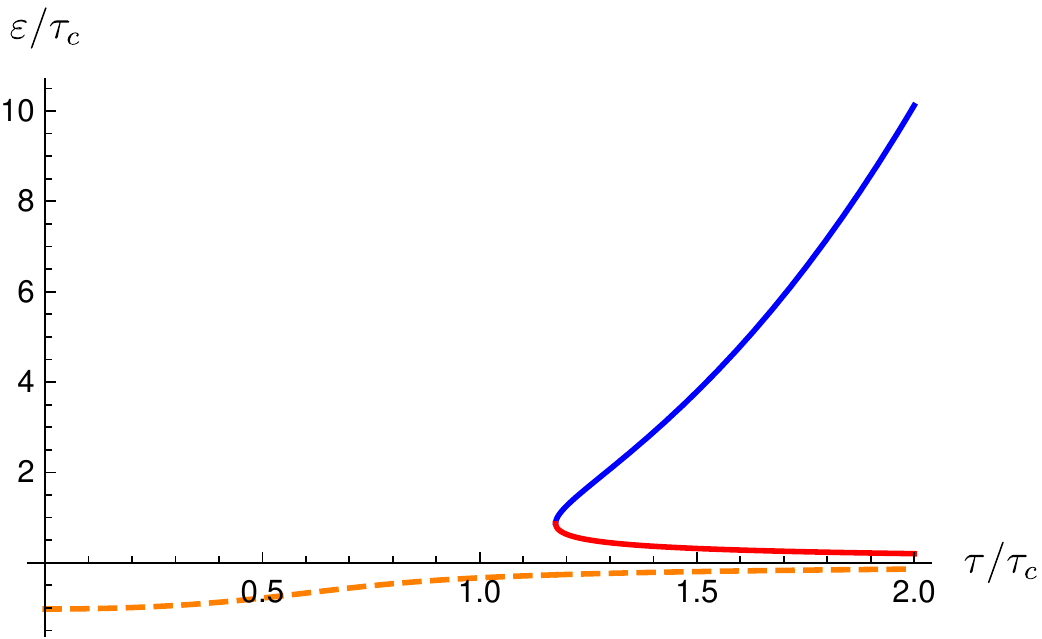}} &
		\scalebox{0.7}{\includegraphics{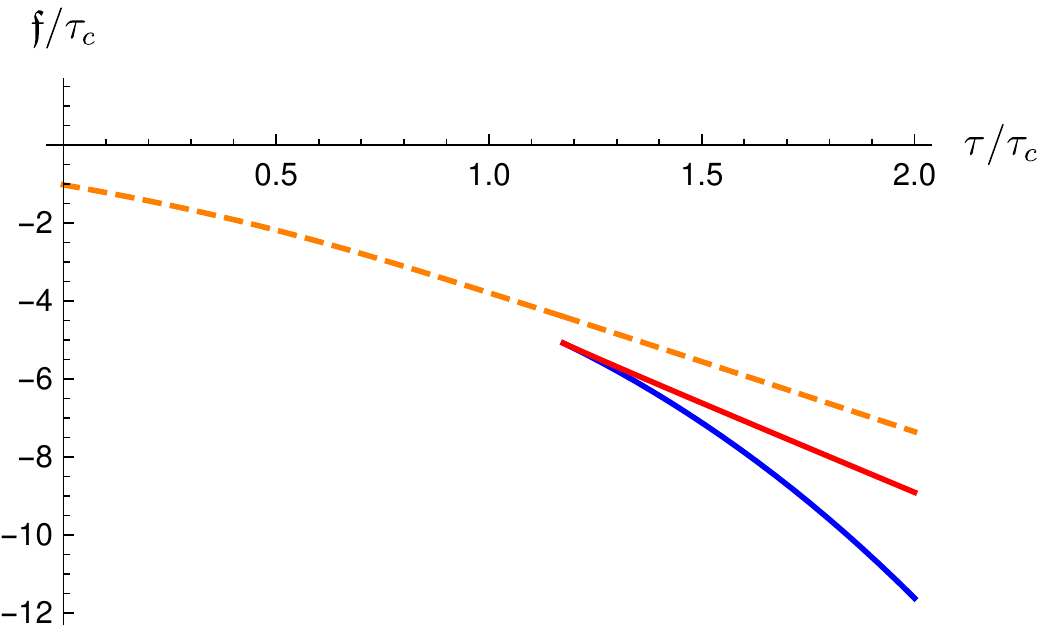}}\\
	\end{tabular}
	\centering
	\caption{Plot of the solutions of \eqref{charIIm} for $\cb=0.786\cb_c$, together with the corresponding energy $\ve$ and Helmholtz free energy $\frak f$ densities.}
	\label{fig::xi=1::qm=0.786qc}
\end{figure}

\begin{figure}
	\begin{tabular}{cc}
		\scalebox{0.7}{\includegraphics{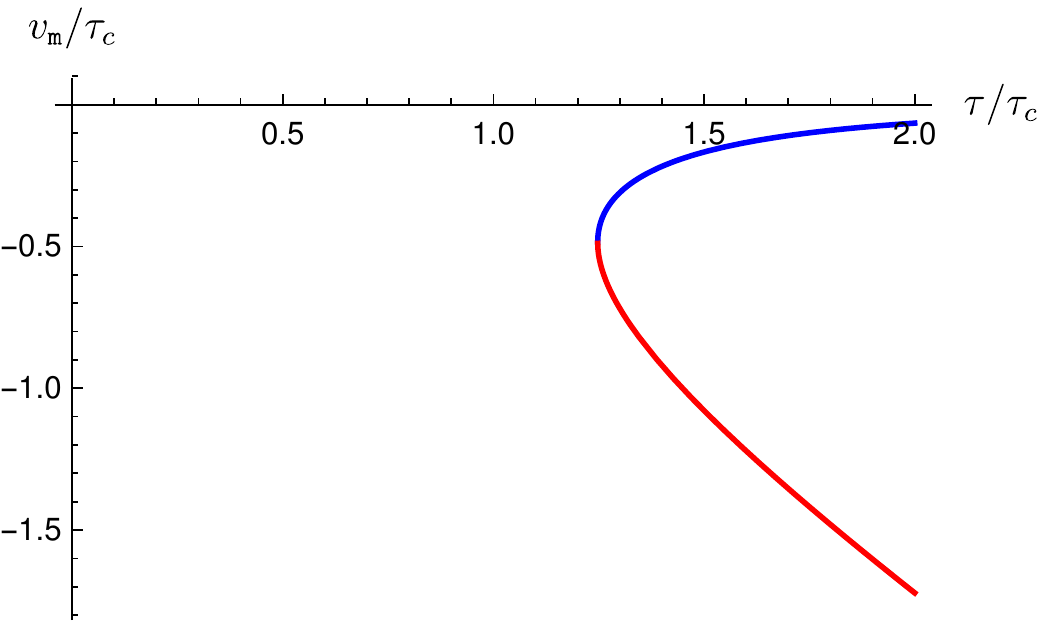}} &
		\scalebox{0.7}{\includegraphics{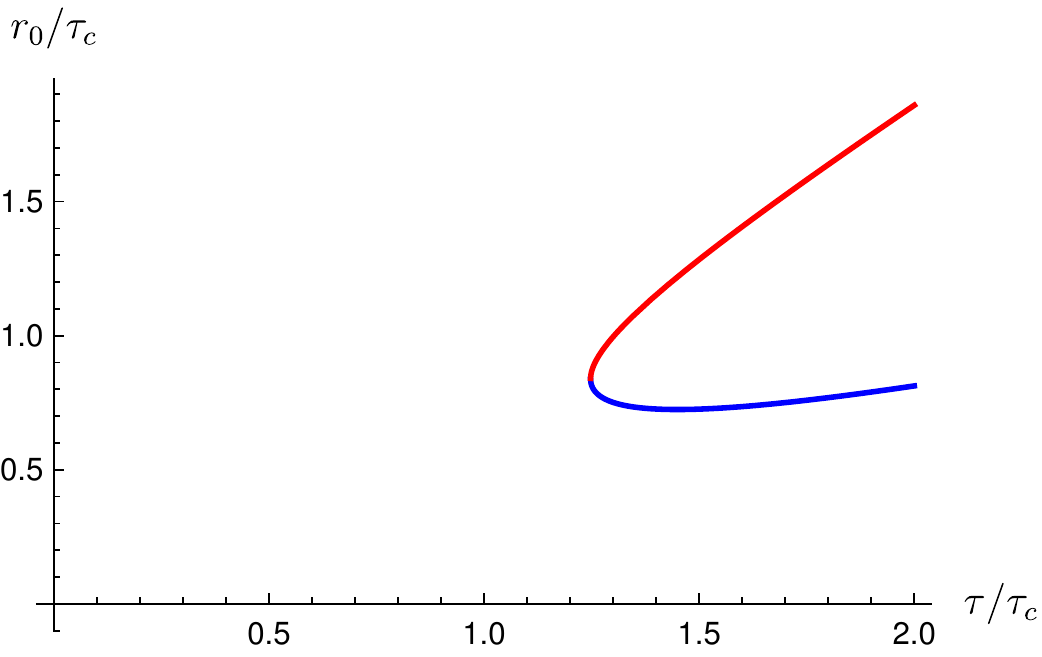}}\\
		\scalebox{0.7}{\includegraphics{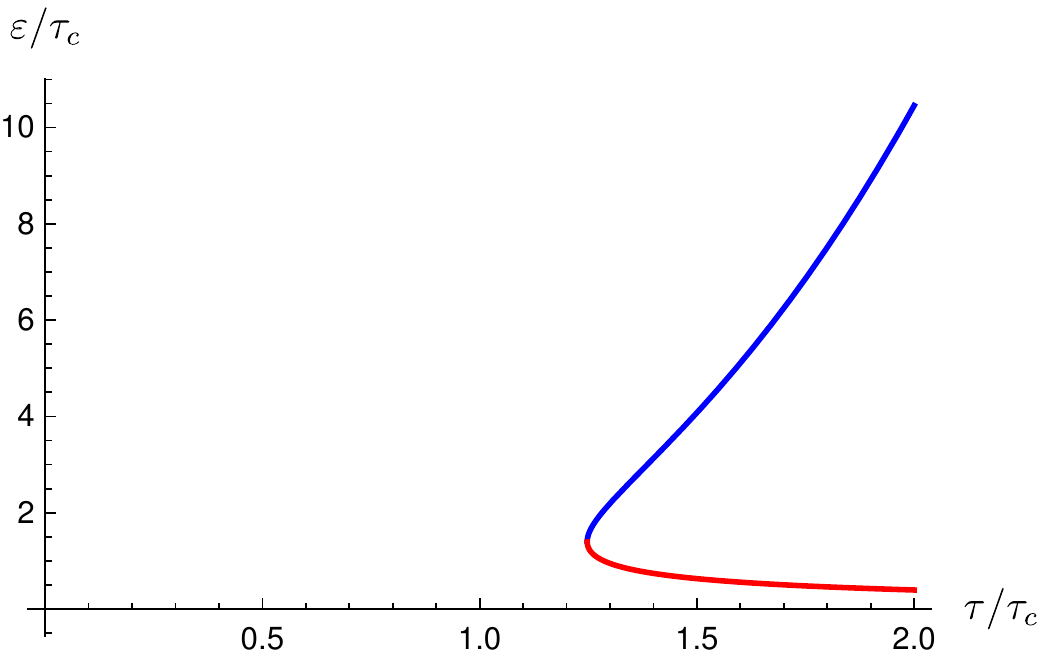}} &
		\scalebox{0.7}{\includegraphics{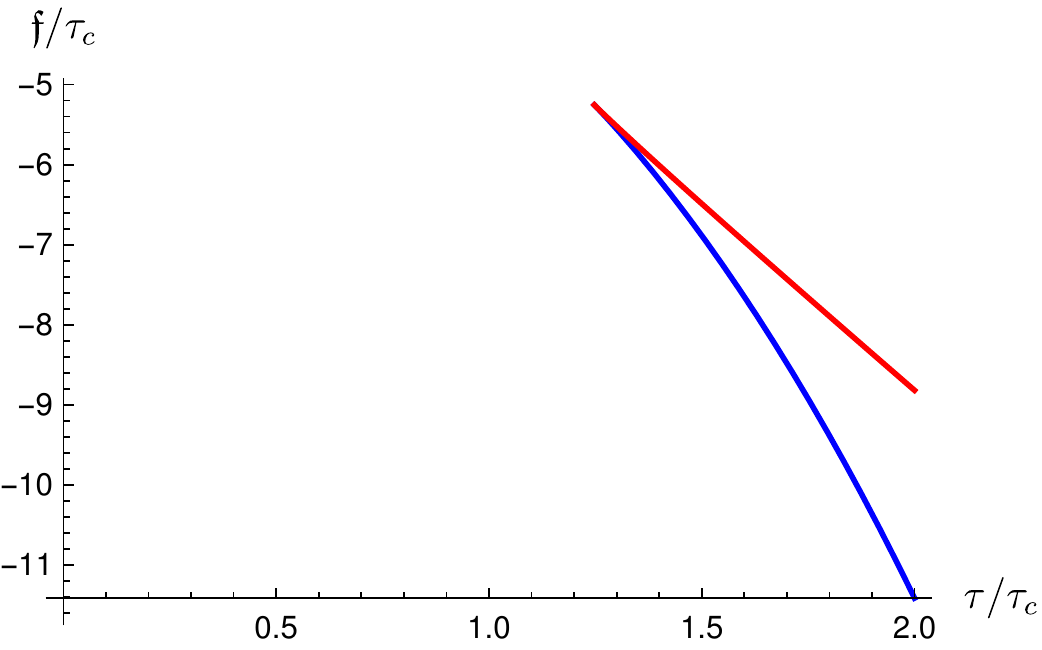}}\\
	\end{tabular}
	\centering
	\caption{Plot of the solutions of \eqref{charIIm} for $\cb=1.1\cb_c$, together with the corresponding energy $\ve$ and Helmholtz free energy $\frak f$ densities.}
	\label{fig::xi=1::qm=1.1qc}
\end{figure}

In particular, for small $q\sbtx{m}$ the perturbative solutions take the form  
\bsub
\label{small-charge-m}
\bal
&\text{Red:} && v\sbtx{m}=-\sqrt{\t^2-\t_c^2}+\co(q\sbtx{m}^2),&& r_0=\frac12\Big(\t+\sqrt{\t^2-\t_c^2}\Big)+\co(q\sbtx{m}^2),&& \t>\t_c,\\
&\text{Blue:} &&v\sbtx{m}=\frac{-q\sbtx{m}^2\ell^2}{r_0(2r_0^2-p^2\ell^2)}+\co(q\sbtx{m}^4), && r_0=\frac16\Big(\t+\sqrt{\t^2+3\t_c^2}\Big)+\co(q\sbtx{m}^2),&& \t>\t_c,\\
&{\text{Orange:}} && v\sbtx{m}=\left\{\begin{matrix}
	\frac{-q\sbtx{m}^2\ell^2}{r_0(2r_0^2-p^2\ell^2)}+\co(q\sbtx{m}^4), \\
	\sqrt{\t^2-\t_c^2}+\co(q\sbtx{m}^2), 
\end{matrix}\right.  && r_0=\left\{\begin{matrix}
\frac16\Big(\t+\sqrt{\t^2+3\t_c^2}\Big)+\co(q\sbtx{m}^2),
\\
\frac12\Big(\t-\sqrt{\t^2-\t_c^2}\Big)+\co(q\sbtx{m}^2),
\end{matrix}\right.&& \begin{matrix} \t<\t_c,\rule{0cm}{.4cm}\\ \t>\t_c,\rule{0cm}{.6cm}\end{matrix}
\eal
\esub
where again $\t_c=4\pi\ell^2 T_c=\sqrt{2}p \ell$. As for the electric solutions, perturbation theory breaks down near $\t_c$, as can be seen from the plots in figure \ref{fig::xi=1::qm=0.00314qc}. Comparing with the corresponding plots for the electric solutions in figure \ref{fig::xi=1::qe=0.00314qc} we see that at small charge densities the electric and magnetic solutions look identical, except that the orange and red branches of the solution are switched above the critical temperature. At higher charge densities the corners are again smoothened out as is shown in figures \ref{fig::xi=1::qm=0.314qc}, \ref{fig::xi=1::qm=0.786qc} and \ref{fig::xi=1::qm=1.1qc}.

As for the electric solutions, the orange branch exists for all temperatures but disappears above the critical magnetic field 
\eq
|\mc B|<\mc B_c = \frac{\Pi^2}{2\sqrt2}=\frac{p^2}{2\sqrt2\ell^2}.
\eeq
Above the critical temperature again there are two additional branches for any value of the magnetic field, while above the critical magnetic field and below the critical temperature there are no solutions. The number of solutions as a function of the temperature and the magnetic charge density matches the number of electric solutions as a function of of the temperature and the electric charge density depicted in figure \ref{fig::xi=1}. Moreover, the dynamic and thermodynamic stability properties of the three branches of solutions are identical to those of the electric solutions and hence the phase diagram in figure \ref{fig::xi=1-phase-diag} applies equally well to the magnetic solutions upon replacing the charge density $\r$ with the magnetic field $\cb$. As can be seen from the plots of the solutions, the only difference between the electric and magnetic ones is that the value of the scalar {\em vev} is opposite, while above the critical temperature the red branch of the electric solutions has the smallest radius and the orange has the largest, while the reverse holds for the magnetic solutions.

\section{Discussion}
\label{sec:conclusion}

In this paper we have explored the holographic dictionary and the thermodynamics of AdS black branes with axion charge and secondary scalar hair. We have focused on a number of concrete exact solutions, but our holographic and thermodynamics analysis is applicable more generally, including, in particular, AdS black holes with non-planar horizons. 

Three of the general conclusions our analysis allows us to draw are the following: 
\begin{itemize}
	
	\item Axion fields with a linear profile in the boundary coordinates, such as those discussed in the context of momentum relaxation \cite{Andrade:2013gsa}, anisotropic plasmas \cite{Mateos:2011ix} or $\t$-lattices \cite{Donos:2016zpf}, should be understood as 0-forms carrying magnetic charge. In particular, such axions correspond to {\em primary} hair and their holographic description includes {\em global} Ward identities, which are not present for standard scalar operators.
	
	\item Mixed and Neumann boundary conditions on running scalar hair modify the expressions for the asymptotic conserved charges and free energy, but {\em do not} lead to independent charges associated with the scalars themselves. Hence, such scalars correspond to {\em secondary} hair. Correctly taking into account the scalar boundary conditions one recovers the standard first law and other thermodynamic relations, despite claims to the contrary in part of the recent literature.   
	
	\item The local dynamic stability of hairy black branes with respect to scalar perturbations is equivalent to the positivity of the energy density. This can be demonstrated by computing the holographic off-shell quantum effective potential for the dual scalar operator, as is described in appendix \ref{effpot}. However, this result provides a general stability criterion based only on the energy density, without the need to compute the full effective potential.  
	
\end{itemize}

A number of open questions and future directions remain. Firstly, we have seen that the phase structure of Theory I remains partially understood due to the fact that the known hairy solutions must satisfy the constraint \eqref{constraint}, which fixes the temperature and the entropy as functions of the charge densities. This property resembles extremal solutions, yet the black branes we discuss have finite temperature. This leads us to believe that there is a more general family of hairy solutions that allows one to relax the constraint \eqref{constraint}. It would be interesting to find this wider class of solutions analytically or numerically.   

Another potentially interesting class of exact solutions to seek is dyonic solutions of Theory II. However, it seems likely that such solutions can only exist for the special case $\x=1$, since only in that case the scalar boundary condition \eqref{th-II-e} and \eqref{th-II-m} coincide. Moreover, since the relative size of the radii of the orange and red branches of the solutions are opposite for the electric and magnetic solutions, one expects that only the blue branch of black hole solutions will exist for a dyonic black brane of Theory II above the critical temperature. Below the critical temperature, however, the orange solutions should continue to exist even for a dyonic black brane. For $p=0$ dyonic solutions of Theory II have been found in \cite{Anabalon:2013sra}. It would be interesting to generalize these solutions to non-zero axion charge.      

Finally, in the present paper we have focused only on properties of the background solutions, without discussing fluctuations around them. It would be very interesting, for example, to compute the thermoelectric and Hall conductivities for these black branes. We hope to address some of these questions in the near future.

\FloatBarrier
\acknowledgments

We would like to thank Tom\'as Andrade, Keun-Young Kim and Ben Withers for useful discussions. We also thank the Galileo Galilei Institute for Theoretical Physics for the hospitality and the INFN for partial support during the completion of this work.
The work of M.M.C.~is supported by the Marie Curie Intra-European Fellowship nr~628104 within the 7th European Community Framework Programme FP7/2007-2013.
AC is supported by EPSRC Award 1376105.
IP acknowledges partial support through COST Action MP 1210 Short Term Scientific Mission COST-STSM-ECOSTSTSM-MP1210-301114-051520 and thanks the University of Southampton for hospitality during the early stages of this work. KS is supported in part by the Science and Technology Facilities Council (Consolidated Grant “Exploring the Limits of the Standard Model and Beyond”). This project has received funding from the European Union’s Horizon 2020 research and innovation programme under the Marie Skłodowska-Curie grant agreement No 690575.

\appendix

\section*{Appendices}

\section{Asymptotic expansions and boundary counterterms}
\label{asexp}

In this appendix we derive the Fefferman-Graham expansions for asymptotically AdS solutions of theory \eqref{genAction}, and derive the relevant boundary counterterms in the case $d=3$ and $\Delta_-=1$,  following \cite{Henningson:1998gx, Henningson:1998ey, deHaro:2000vlm, Bianchi:2001kw} (see also the review \cite{Skenderis:2002wp}). 
These counterterms are a special case of those obtained using a radial Hamiltonian approach in \cite{Papadimitriou:2011qb,Lindgren:2015lia}. For a recent review see \cite{Papadimitriou:2016yit}.

\paragraph{Fefferman-Graham gauge} To construct the general asymptotic expansions in AdS$_{d+1}$, we gauge-fix the bulk metric to be of the form 
\be
ds^2=\frac{\ell^2}{z^2}\(dz^2+g_{ij}(x,z)dx^idx^j\),\label{FGmetric}
\ee
where $g_{ij}(x,z)$ is a $d$-dimensional metric.  We denote derivatives with respect to the Fefferman-Graham radial coordinate $z$ with a prime. Latin indices $i,j,\ldots,$ are reserved for the $d$-dimensional spacetime coordinates, which are raised/lowered with the metric $g_{ij}$.\footnote{This preserves diffeomorphism covariance on the boundary but breaks bulk covariance. To preserve the latter one should work instead with the induced metric $\gamma_{ij}=\ell^2 g_{ij}/z^2$. }
Moreover, $D_i$ stands for the covariant derivative with respect to the metric $g_{ij}$ (or equivalently the induced metric $\gamma_{ij}$), and we denote the corresponding Laplace operator by $\Box_{g} \equiv g^{ij}D_iD_j$ (and $\Box_\gamma=\gamma^{ij}D_iD_j$ for the covariant Laplacian).
Similarly, for the Maxwell field we choose the radial gauge $A_z=0$, so that 
\eq
F_{zi}=A'_i,\qquad
F_{ij}=2\p_{[i}A_{j]},
\eeq
and for later convenience we introduce the shorthand notation
\eq
A'^2\equiv g^{ij}A'_iA'_j,\qquad
\hat F^2\equiv g^{ij}g^{kl}F_{ik}F_{jl}.
\eeq

\paragraph{Asymptotic expansions} The Fefferman-Graham expansions take the form,
\begin{subequations}
	\begin{align}
		g_{ij}(x,z)&=g_{(0)ij}+zg_{(1)ij}+z^2g_{(2)ij}+\cdots,\label{FGg}\\
		A_i(x,z)&=\A0i+z\A1i+z^2\A2i+\cdots,\\
		\psi_I(x,z)&=\psi^{(0)}_I+z\psi^{(1)}_I+z^2\psi^{(2)}_I+\cdots,\\
		\phi(x,z)&=z^{\Delta_-}\varphi(x,z) =z^{\Delta_-}(\dil0+z\dil1+z^2\dil2+\cdots),\label{FGphi}
	\end{align}
	\label{FG}
\end{subequations}
where all coefficients are yet undetermined functions of the transverse coordinates $x^i$. It should be emphasized that these are not the most general asymptotic expansions for the fields in the action \eqref{genAction}, but they suffice for the purposes of this article. In particular, we are not covering cases where logarithmic terms appear in the Fefferman-Graham expansions (e.g. when $d$ is even or when $\Delta_\pm=d/2$). Moreover, the fact that only integer powers of $z$ appear in the expansions \eqref{FG} is a consequence of our assumption that $\Delta_\pm$ are non negative integers. All solutions we study in this article have $d=3$ and $\Delta_-=1$ and so both assumptions are justified. However, our analysis in this appendix is slightly more general and applies to any boundary dimension of the form $d=2n+1$, with $n=1,2,\ldots$, and for $\Delta_- < d/2$, $d/2 <\Delta_+ \leq d$. Note that imposing in addition the unitarity bound for $\Delta_-$, which is equivalent to the condition that the scalar mass lies in the window \eqref{window}, would uniquely determine $\Delta_-=n$ and $\Delta_+=n+1$. However, we need not impose the unitarity bound on $\Delta_-$ in general.

\paragraph{Equations of motion} Given the gauge-fixed form \eqref{FGmetric} of the bulk metric, the field equations following from the action \eqref{genAction} can be decomposed into radial and transverse directions. In particular, we obtain the following set of equations. \\\vskip0.1cm
\begin{flushleft}
{\em Einstein equations:}  
\begin{subequations}
\begin{align}
&-\frac12\tr(\ginv g'')+\frac14\tr(\ginv g'\ginv g')+\frac1{2z}\tr(\ginv g')
=
\label{geqScalar}\NO\\
&
\qquad\qquad
\frac12\psi'_I\psi'_I+\frac12\lp\phi'^2+\frac{2\ell^2}{(d-1)z^2}(V(\phi)-2\Lambda)\rp
-\frac{z^2}{4(d-1)\ell^2}\lp \hat F^2-2(d-2)A'^2\rp,\\
&\frac12g^{jk}\lp D_jg'_{ki}-D_ig'_{jk}\rp
=\frac12\psi'_I\p_i\psi_I+\frac12\phi'\p_i\phi+\frac{z^2}{2\ell^2}Z(\phi)g^{jk}F_{ij}A'_{k},
\label{geqVector}\\
& R_{ij}[g]-\frac1{2}g''_{ij}+\frac1{2z}\lp(d-1)g'_{ij}+\tr(\ginv g')\,g_{ij}\rp
-\frac1{4}\tr(\ginv g')\,g'_{ij}+\frac1{2}(g'\ginv g')_{ij}\nonumber\\
& \qquad\qquad\quad
=\frac12\p_i\psi_I\p_j\psi_I+\frac12\lp\p_i\phi\p_j\phi+\frac{2\ell^2}{(d-1)z^2}\lp V(\phi)-2\Lambda\rp g_{ij}\rp
\label{geqTensor}\NO\\
&\qquad\qquad\qquad\qquad\quad
+\frac{z^2}{2\ell^2}Z(\phi)\lp
g^{kl}F_{ik}F_{jl}+A'_iA'_j-\frac1{2(d-1)}(\hat F^2+2A'^2)g_{ij}
\rp.
\end{align}
\end{subequations}
\flushleft{\em Scalar equations:}
\bsub
\begin{align}
&\psi''_I-\frac{d-1}z\psi'_I+\p_z(\log\sqrt{-g})\psi'_I+\Box_g\psi_I=0,\label{axioneq}\\
&\phi''-\frac{d-1}z\phi'+\p_z(\log\sqrt{-g})\phi'+\Box_g\phi-\frac{\ell^2}{z^2}V'(\phi)=
\frac{z^2}{4\ell^2}Z'(\phi)\lp\hat F^2+2A'^2\rp.\label{dilatoneq}
\end{align}
\esub
\flushleft{\em Maxwell equations:}
\bsub
\begin{align}
&D^i\lp Z(\phi)A'_i\rp=0,\label{Aeom1}\\
&D^j\lp Z(\phi)F_{ji}\rp+\p_z\lp Z(\phi)A'_i\rp-Z(\phi)\lp\frac{d-3}zA'_i-\frac12\tr(\ginv g')A'_i+g'_{ij}g^{jk}A'_k\rp=0.\label{Aeom2}
\end{align}
\esub
\end{flushleft}

\paragraph{Recursive solution of the field equations} These equations can be solved iteratively up to the desired order by inserting the formal expansions \eqref{FG}. Requiring that the equations of motion admit an AdS solution with $\f=0$ implies that the linear term $V_1$ in the Taylor expansion of the potential $V(\f)$ vanishes. We therefore need to set $V_1=0$ from the outset in order to obtain the correct values for the coefficients of the expansions \eqref{FG}.

\begin{flushleft}
{\em Einstein equations:}
\end{flushleft}
Multiplying equation \eqref{geqScalar} by $z$ and taking the $z\rightarrow0$ limit we obtain
\eq
\tr(\gi0\g1)=0.
\label{trg1}\eeq
Moreover, to keep the limit $z\rightarrow0$ of the tensor equation~\eqref{geqTensor} multiplied by $2z$ finite, we have to impose $\lim_{z\rightarrow0}V(\phi)=V_0=2\Lambda$. This leads to
\eq
(d-1)\gij1+\tr(\gi0\g1)\gij0=0,
\eeq
which, in combination with \eqref{trg1}, implies 
\eq
\gij1=0.
\eeq

Next, taking the derivative of equation~\eqref{geqScalar} with respect to $z$ and evaluating the limit $z\rightarrow0$, we find that the metric dependence cancels out and the equation requires that\footnote{The vanishing of $\ax1$ can alternatively be obtained from the equation of motion $\Box\psi_I=0$.}
\eq
\ax1=0,
\label{ax1}\eeq
as well as
\bsub
\begin{align}
V_2\varphi_{(1)}^2&=0 \qquad\text{for $\Delta_-=0$,}\\
\lp1+\frac{\ell^2V_2}{d-1}\rp\varphi_{(0)}^2&=0 \qquad\text{for $\Delta_-=1$.}
\end{align}
\esub
Hence, for $\Delta_-=1$, to keep the dialton source unconstrained we need
\eq
V_2=-\frac{d-1}{\ell^2},
\label{v2}\eeq
in agreement with \eqref{AdSmass} for $d=3$. Moreover, when $\Delta_-=0$, or equivalently $\Delta_+=d$, the scalar is massless (i.e. $V_2=0$ as we confirm below) and so the first condition is automatically satisfied. 

Taking the derivative of equation~\eqref{geqTensor} with respect to $z$ gives in the limit $z\rightarrow0$
\begin{align}
(d-2)\gij2&=-\lp R^{(0)}_{ij}-\frac1{2(d-1)}R^{(0)}\gij0\rp
-\frac{d-2}{4(d-1)}\delta_{\Delta_-,1}\varphi^2_{(0)}\gij0
\nonumber\\
&\qquad
+\frac12\lp\p_i\ax0\p_j\ax0-\frac1{2(d-1)}g^{kl}_{(0)}\p_k\ax0\p_l\ax0\gij0\rp,\label{g2}
\end{align}
where $\delta_{k,l}$ is the Kronecker delta.\footnote{\label{footnote} As we mentioned above, the form \eqref{FG} of the asymptotic expansions is not the most general possible. A concrete demonstration of this is provided by the expression \eqref{g2} when applied to the case $d=2$. The first line on the right hand side automatically vanishes for $d=2$, but the second line seems to impose a constraint on the sources for the axions. However, this conclusion is incorrect. It simply indicates that in the case $d=2$ there are additional logarithmic terms in the asymptotic expansions, which we have not included. We will encounter additional instances of this phenomenon below. For $d>2$, however, \eqref{g2} is correct.} Similarly, taking the derivative of equation~\eqref{geqVector} with respect to $z$ and setting $z=0$ we find the divergence condition
\be
	\ggi0{jk}\lp D_{(0)j}g_{(2)ki}-D_{(0)i}g_{(2)jk}\rp=\ax2\p_i\ax0+\frac12\dil0\p_i\dil0\delta_{\Delta_-,1}+\lp\dil2\p_i\dil0+\frac12\dil1\p_i\dil1\rp\delta_{\Delta_-,0},
\ee
where $D_{(0)i}$ denotes the covariant derivative with respect to the metric $\g0_{ij}$.

For $d=3$, which is the case we are mostly interested in here, the recursive procedure for equation \eqref{geqTensor} breaks down at the next order and it does not determine $g_{(3)ij}$. When $d=3$, $g_{(3)ij}$ is related to the stress tensor of the dual theory (see eq. \eqref{D-1pt-fns} or \eqref{M-1pt-fns}), and so it can only be determined by infrared data. However, the trace of $g_{(3)}$ can be determined by multiplying the scalar equation by $z$, taking two derivatives with respect to $z$, and the setting $z=0$. The result is
\eq
-3\tr(\gi0\g3)=\lp2\dil0\dil1+\frac{\ell^2V_3}{3(d-1)}\dilp03\rp\delta_{\Delta_-,1}+4\dil1\dil2\delta_{\Delta_-,0}.
\label{trg3}\eeq
Moreover, evaluating the second derivative of equation~\eqref{geqVector} with respect to $z$ at $z=0$ we obtain the divergence condition
\begin{align}
	\ggi0{jk}\lp D_{(0)j}g_{(3)ki}-D_{(0)i} g_{(3)jk}\rp&=\ax3\p_i\ax0+\frac{Z_0}{3\ell^2}\ggi0{jk}F_{ij}^{(0)}\A1k
	+\frac13\lp2\dil1\p_i\dil0+\dil0\p_i\dil1\rp\delta_{\Delta_-,1}\nonumber\\
	&\qquad+\lp\dil3\p_i\dil0+\frac23\dil2\p_i\dil1+\frac13\dil1\p_i\dil2\rp\delta_{\Delta_-,0}.
\end{align}
For $\Delta_-=1$ and using \eqref{trg3} for the trace of $\g3$ this simplifies to
\eq
D_{(0)}^kg_{(3)ki}=\ax3\p_i\ax0+\frac{Z_0}{3\ell^2}\ggi0{jk}F_{ij}^{(0)}\A1k-\frac13\dil0\p_i\dil1
-\frac{\ell^2V_3}{3(d-1)}\dilp02\p_i\dil0.
\label{divg3}\eeq
For $d=3$ this condition corresponds to the diffeomorphism Ward identity in \eqref{D-trace-WI} or \eqref{M-trace-WI}.

\begin{flushleft}
	{\em Axion equations:}
\end{flushleft}
Multiplying equation~\eqref{axioneq} by $z$ and setting $z=0$ requires $\ax1=0$, in agreement with~\eqref{ax1}. Taking a derivative with respect to $z$ leads to the condition
\eq
\ax2=\frac1{2(d-2)}\Box_{(0)}\ax0,
\eeq
while at the next order we find
\eq
(d-3)\ax3=0.
\eeq
Hence, for $d\neq3$ we have $\ax3=0$, but when $d=3$ $\ax3$ is left unconstrained and it corresponds to the one-point of the scalar operator dual to the axions (see eq. \eqref{D-1pt-fns} or \eqref{M-1pt-fns}).
Going one order higher gives
\eq
4(d-4)\ax4=\frac1{2(d-2)}\Box_{(0)}^2\ax0+\frac1{d-2}\tr(\gi0\g2)\Box_{(0)}\ax0+\Box_{(2)}\ax0,
\eeq
where
\be
\Box_{(2)}\ax0=\frac12g^{ij}_{(0)}\p_i\tr(\gi0\g2)\p_j\ax0-\frac1{\sqrt{-\g0}}\p_i\lp\sqrt{-\g0}(\gi0\g2\gi0)^{ij}\p_j\ax0\rp.
\ee
This expression for $\ax4$ provides another instance of the phenomenon pointed out in footnote \ref{footnote}. Namely, this equation is not correct for the case $d=4$, since additional logarithmic terms in the asymptotic expansions must be included in that case.

\begin{flushleft}
	{\em Maxwell equations:}
\end{flushleft}
Multiplying equation~\eqref{Aeom2} by $z$ and taking the limit $z\rightarrow0$ gives
\eq
(d-3)Z_0\A1i=0.
\eeq
Hence, assuming $Z_0>0$, we have $\A1i=0$ unless $d=3$. When $d=3$ $\A1i$ corresponds to the one-point function of the global $U(1)$ current in the dual theory (see eq. \eqref{D-1pt-fns} or \eqref{M-1pt-fns}) and so it is left undetermined by the asymptotic analysis. The $z\rightarrow0$ limit of equation~\eqref{Aeom1} gives the divergence constraint
\eq
Z_0 D_{(0)}^i\A1i=-Z_1 g_{(0)}^{ij}\p_i\dil0\A1j\delta_{\Delta_-,0},\label{divA}
\eeq
which in turn leads to the $U(1)$ Ward identity in \eqref{D-trace-WI} or \eqref{M-trace-WI}. 

As long as $d\leq 3$ we need not go to higher order in the expansion for the Maxwell field. However, for $d>3$ at the next order \eqref{Aeom2} yields 
\be
2(d-4)Z_0\A2i=Z_0D_{(0)}^jF_{ji}^{(0)}+Z_1\(D_{(0)}^j\dil0F^{(0)}_{ji}-(d-4)\vf_{(1)}\A1i\)\delta_{\Delta_-,0}-(d-4)Z_1\dil0\A1i\delta_{\Delta_-,1}.
\label{A2}
\ee
This determines $\A2i$ in terms of the lower order coefficients, except when $d=4$, in which case $\A2i$ corresponds to the undetermined current one-point function. Again, this equation cannot be applied to the case $d=4$, since it requires additional logarithmic  terms in the asymptotic expansions (see e.g. eqs. (6.69) and (6.70) in \cite{Bianchi:2001kw}).

\begin{flushleft}
	{\em Dialton equations:}
\end{flushleft}
Multiplying the dialton equation~\eqref{dilatoneq} by $z^{2-\Delta_-}$ and taking the limit $z\rightarrow0$ we obtain
\eq
\lp(d-\Delta_+)\Delta_++\ell^2V_2\rp\dil0=0.
\eeq
To keep the mode $\dil0$ unconstrained, therefore, we need to set
$V_2=-(d-\Delta_+)\Delta_+/\ell^2$, in agreement with the result~\eqref{v2} from the gravitational equations in the case $\Delta_-=1$, or equivalently $\Delta_+=d-1$.
We recall from \eqref{AdSmass} that $V_2=m_\f^2$ corresponds to the AdS mass of the scalar $\f$ and so this result for $V_2$ is the standard relation between the AdS mass of a scalar field and the scaling dimension of the dual operator.

Evaluating the derivative of the product of equation~\eqref{dilatoneq} and $z^{2-\Delta_-}$ at $z=0$ gives the next order equation 
\eq
\lp d-2\Delta_++1\rp\dil1=\frac12\ell^2V_3\dilp02\delta_{\Delta_-,1}.
\eeq
If $\Delta_-=1$, then $(d-3)\dil1=-\frac12\ell^2V_3\dilp02$, and hence
\eq
\dil1=-\frac1{2(d-3)}\ell^2V_3\dilp02,\qquad (d\neq3,\quad \Delta_+=d-1).
\eeq
Applied to $d=3$ (again for $\Delta_-=1$), this equation leads to the erroneous conclusion that $V_3=0$. However, as we have seen by now a number of times, this simply reflects the fact that the asymptotic expansions we have assumed do not in general apply to this case, unless extra logarithmic terms are included (see e.g. eqs. (2.18) and (2.19) in \cite{Anabalon:2015xvl}). However, both the potential for Theory I given in \eqref{potential} and the potential for Theory II in \eqref{pot2} have $V_3=0$, and so no logarithmic terms arise in the asymptotic expansions for the specific models we consider. This equation is then trivially satisfied, leaving $\vf_{(1)}$ undetermined. As we discuss in the main body of the paper, for Dirichlet boundary conditions $\vf_{(1)}$ corresponds to the {\em vev} of the dual scalar operator, while for Neumann and mixed boundary conditions it is related to the arbitrary source of the dual operator.   

Taking a further derivative with respect to $z$ and setting $z=0$ gives the next order equation
\begin{align}
2(\Delta_--\Delta_++2)\dil2&=-\Box_{(0)}\dil0-\Delta_-\tr(\gi0\g2)\dil0
+\frac12\ell^2V_3\dilp02\delta_{\Delta_-,2}
\nonumber\\
&\qquad
+\lp\ell^2V_3\dil0\dil1+\frac16\ell^2V_4\dilp03\rp\delta_{\Delta_-,1}
.
\end{align}
This equation determines $\dil2$ as long as $\Delta_--\Delta_++2\neq 0$,
\eq
\hskip-.3cm\dil2=
\left\{\begin{array}{l@{\quad}l}
\displaystyle
\frac{1}{2(d-2)}\Box_{(0)}\dil0, & (\Delta_-=0)\\[1em]
\displaystyle
\frac1{2(d-4)}\lp\Box_{(0)}\dil0+\tr(\gi0\g2)\dil0-\ell^2V_3\dil0\dil1-\frac16\ell^2V_4\dilp03\rp,
& (\Delta_-=1)\\
\displaystyle
\frac1{2(d-6)}\lp\Box_{(0)}\dil0+2\tr(\gi0\g2)\dil0-\frac12\ell^2V_3\dilp02\rp,
& (\Delta_-=2)\\[1em]
\displaystyle
-\frac1{2(\Delta_--\Delta_++2)}
\lp\Box_{(0)}\dil0+\Delta_-\tr(\gi0\g2)\dil0\rp,
& (\Delta_-\neq1,2)
\end{array}
\right.
\eeq
Using the previous results, the $\Delta_-=1$ case can be further simplified to
\eq
\dil2=
\left\{\begin{array}{l@{\;\,}l}
\displaystyle
-\frac12\Box_{(0)}\dil0-\frac12\tr(\gi0\g2)\dil0+\frac1{12}\ell^2V_4\dilp03,
& (d=3)\\[1em]
\displaystyle
\frac1{2(d-4)}\lp\Box_{(0)}\dil0+\tr(\gi0\g2)\dil0+\lp\frac{\ell^4V_3^2}{2(d-3)}
-\frac16\ell^2V_4\rp\dilp03\rp,
& (d\neq3)
\end{array}
\right.
\eeq
We stress again that these expressions for $\dil2$ only apply to odd $d$. 

\paragraph{Boundary counterterms} Finally, we can use the Fefferman-Graham expansions in order to determine the covariant local boundary counterterms. We will do this explicitly only for the case $d=3$, $\Delta_-=1$. Inserting the asymptotic expansions \eqref{FG} in the regularized action \eqref{Sreg} for the generic model \eqref{genAction} and performing the integration over the radial coordinate we obtain
\eq
S_{\textsf{reg}}=\int_{z=\ep}d^3x\!\sqrt{-\g0}\Big(\frac{4\ell^2}{\ep^3}-\frac{\ell^2}{\ep}\lp\dilp02+2\tr(\gi0\g2)\rp+\mc O(\ep^0)\Big).\label{div}
\eeq
To determine the boundary counterterms we need to invert the Fefferman-Graham expansions~\eqref{FG} in order to express the divergent part \eqref{div} of the regularized action in terms of the covariant induced fields on the radial cutoff $\gamma_{ij}(\e,x^i)$, $\phi(\ep,x^i)$, $\psi_I(\ep,x^i)$ and $A_i(\ep,x^j)$. A bit of algebra gives
\begin{align}
\gij0&=\frac{\ep^2}{\ell^2}\left[
\gamma_{ij}+\ell^2\lp R_{ij}[\gamma]-\frac14R[\gamma]\gamma_{ij}\right)
+\frac18\phi^2\ga_{ij}-\frac{\ell^2}2\lp\p_i\psi_I\p_j\psi_I-\frac14\ga^{kl}\p_k\psi_I\p_l\psi_I\ga_{ij}\rp
\right]+\mc O(\ep^3),
\nonumber\\
\gij2&=-\lp R_{ij}[\gamma]-\frac14R[\gamma]\gamma_{ij}\right)
-\frac{\phi^2}{8\ell^2}\ga_{ij}
+\frac{1}2\lp\p_i\psi_I\p_j\psi_I-\frac14\ga^{kl}\p_k\psi_I\p_l\psi_I\ga_{ij}\rp
+\mc O(\ep^2),
\nonumber\\
\ax0&=\psi_I-\frac12\ell^2\Box_\gamma\psi_I+\mc O(\ep^3),
\qquad
\dil0=\frac1\ep\phi+\mc O(\ep),
\qquad
A^{(0)}_i=A_i+\mc O(\ep),
\nonumber\\
\sqrt{-\g0}&=\frac{\ep^3}{\ell^3}\sqrt{-\gamma}\left[
1+\frac18\ep^2\lp R_{(0)}-\frac1{2}g^{ij}_{(0)}\p_i\ax0\p_j\ax0+\frac3{2}\dilp02\rp
+\mc O(\ep^3)\right].
\end{align}
Substituting these expressions in \eqref{div} we finally get
\eq
S_{\textsf{reg}}=\int_{z=\ep}\!d^3x\sqrt{-\ga}\left(
\frac4{\ell}+\ell R[\ga]+\frac1{2\ell}\phi^2-\frac\ell2\ga^{ij}\p_i\psi_I\p_j\psi_I
\right)+\textsf{finite},
\eeq
leading to the counterterms \eqref{counterterms}.


\section{Quantum effective potential}
\label{effpot}

The aim of this appendix is to provide some details of the calculation of the quantum effective potential $V\sbtx{QFT}(\s)$ for the {\em vev} $\s=\<\co_{\Delta_-}\>=\vf\sub{0}$ of the scalar operator $\co_{\Delta_-}$, as defined in \eqref{effaction}. Although the procedure we describe allows one in principle to compute the effective potential exactly, in practice we are only able to compute it perturbatively in the vicinity of a given vacuum. However, this suffices for the purpose of examining the stability of the black hole solutions we discuss in the main body of the article.   

\paragraph{Non relativistic flows} The main ingredient in the calculation of the quantum effective potential is writing the black brane solutions in terms of first order flow equations, governed by a fake superpotential \cite{Freedman:2003ax}. For solutions of the generic action \eqref{genAction} that are spatially homogeneous and isotropic, but break Lorentz invariance, such first order flow equations were obtained in \cite{Lindgren:2015lia}. These equations can be straightforwardly generalized to include non-zero axionic charge.

In order to obtain the first order flow equations it is convenient to parameterize the general background we are interested in as\footnote{As in the main body of this article we take $x,y$ to be dimensionless, which leads to the extra factor of $\ell^2$ in the spatial part of the metric.} 
\bal\label{Bans-new}
	&ds^2=du^2+e^{2a(u)}\left(-b(u)dt^2+\ell^2(dx^2+dy^2)\right), \NO\\
	&A=c(u)\tx dt+\frac{q\sbtx{m}}{2}(x\tx dy-y\tx dx), \NO\\
	&\f=\f(u),
	\qquad
	\j_I=p x^I,
\eal
where $q\sbtx{m}$ is the constant magnetic charge density, $p$ is the isotropic axion charge density, and we have introduced the canonical radial coordinate $u$. The field strength of the Maxwell field on such backgrounds is given by
\be\label{FB}
F=\tx dA=\dot{c}\;\tx du\wedge \tx dt+q\sbtx{m} \tx dx\wedge \tx dy,
\ee
where a dot $\dot{}$ indicates differentiation with respect to $u$. It may be interesting to note that the ansatz \eqref{Bans-new} can describe not only planar black hole solutions such as the ones we are studying in this article, but also non-relativistic zero temperature flows \cite{Lindgren:2015lia}. 

Generalizing the result of \cite{Lindgren:2015lia} to non-zero $p$, it can be shown that any solution of the first order flow equations  
\be\label{RGeqs-new}
\dot a=-\frac{1}{2}W,\qquad
\frac{\dot b}{b}=-\pa_aW,\qquad
\dot \f=2\pa_\f W,\qquad
\dot \j_I=0,\qquad
\dot c=\frac{q\sbtx{e}}{\ell^2}Z^{-1}e^{-a}b^{1/2},
\ee
where $q\sbtx{e}$ is the electric charge density, automatically solves the second order field equations provided, the superpotential $W(a,\f)$ satisfies the partial differential equation  
\be\label{superpotential-new}
2(\pa_\f W)^2-\frac{1}{2}\left(3+\pa_a\right)W^2=V(\f)+\frac{p^2}{\ell^2}e^{-2a}+\frac{1}{2\ell^4}\(q\sbtx{m}^2Z(\f) + q\sbtx{e}^2Z^{-1}(\f)\)e^{-4a}.
\ee
Notice that setting all the charges to zero and taking the superpotential to be a function of $\f$ only leads to the standard flow equations for Poincar\'e domain wall solutions.    

The key reason why this description of the solutions in terms of a fake superpotential is useful for computing the quantum effective potential is that it allows us to determine the renormalized on-shell action as a function of an {\em arbitrary} scalar {\em vev} $\vf\sub{0}$, since the quantum effective potential can be directly related to the superpotential. Extremizing the quantum effective potential with respect to the {\em vev} $\s=\vf\sub{0}$ then determines the extremum $\s_*$, corresponding to a smooth solution of the second order equations. In contrast, evaluating the on-shell action using the solution of the second order field equations yields only the value corresponding to the extremum of the quantum effective potential at a specific $\vf\sub{0}=\s_*$. 

The regularized on-shell action is related to the superpotential by \cite{Lindgren:2015lia}
\be
S\sbtx{reg}=-2\int d^3x \(\sqrt{-\gamma}\;W(a,\f)-\frac{q\sbtx{e}}{2}c-\ell^2\cv_0/2\),
\ee
where $\cv_0$ is a constant that does not depend on $\vf\sub{0}$. This constant can be computed, if desired, by evaluating the on-shell action of the bald solution of the theory. From the expression \eqref{effaction-e} for the effective action and the counterterms \eqref{counterterms} we obtain the general formula for the quantum effective potential on Minkowski space (with metric $g\sub{0}=\diag(-1,\ell^2,\ell^2)$)  
\be\label{effpot-formula}
\cv\sbtx{QFT}(\s)=\cv_0+\lim_{u\to\infty}\[-e^{3a}\(2W(a,\f)+\frac4{\ell}+\frac1{2\ell}\phi^2- \frac{p^2}{\ell} e^{-2a}\)+\frac{q\sbtx{e}}{\ell^2}c\]+\cf(\s).
\ee
Solving directly the partial differential equation \eqref{superpotential-new} that determines the superpotential is not an easy task, but it is also not necessary for our purposes, since we do not need to determine new solutions of the field equations, but rather to examine the stability of known solutions. This can be achieved by computing the effective potential in the vicinity of the extrema corresponding to the known background solutions.  

\paragraph{Solving the superpotential equation around a known solution} Any solution of the field equations of the form \eqref{Bans-new} provides a curve $Y(a,\f)=0$ on the configuration space $(a,\f)$. In the vicinity of the solution corresponding to such a curve, the superpotential $W(a,\f)$ can be expressed in the form of a Taylor expansion in $Y$ as 
\be\label{sup-exp}
W(a,\f)=\sum_{n=0}^\infty W_n(X)Y^n,
\ee
for some suitable function $X(a,\f)$. In order for $Y=0$ to correspond to a solution of the field equations, as required by the hypothesis, it must satisfy $\dot Y=\co(Y)$, as a consequence of the equations of motion, or equivalently of the flow equations \eqref{RGeqs-new}. This leads to the condition
\be\label{condition}
2\pa_\f Y\(W'_0(X)\pa_\f X +W_1(X)\pa_\f Y \)-\frac12 W_0(X)\pa_a Y=\co(Y).
\ee
Depending on the form of $Y(a,\f)$, this constraint restricts the form of $X(a,\f)$, as well as $W_1(X)$. 

We are only going to discuss the effective potential around the hairy solutions of either Theory I or II, and so we focus on hairy solutions in the rest of this appendix. Any hairy solution can be described by a curve of the form
\be\label{Y}
Y(a,\f)=e^{-2a}-Y_o(\f)=0,
\ee
for some non-trivial function $Y_o(\f)$. Setting $X=\f$ and taking into account the transformation of the partial derivatives according to
\be
\pa_a\to (\pa_a Y)\pa_Y=-2(Y+Y_o)\pa_Y,\qquad
\pa_\f \to \pa_X+(\pa_\f Y)\pa_Y=\pa_X-Y'_o\pa_Y,
\ee
the constraint \eqref{condition} and the first two orders in the expansion of \eqref{superpotential-new} lead to the three equations 
\bsub
\bal
&2Y_o'(W_0'-Y_o'W_1)=W_0Y_o,\\
&2(W_0'-Y_o'W_1)^2-\frac32 W_0^2+2Y_oW_0W_1=V(\f)+\frac{p^2}{\ell^2}Y_o+\frac{1}{2\ell^4}\(q\sbtx{m}^2Z + q\sbtx{e}^2Z^{-1}\)Y_o^2,\\
&4W_1'(W_0'-Y_o'W_1)-W_0W_1+2Y_oW_1^2=\frac{p^2}{\ell^2}+\frac{1}{\ell^4}\(q\sbtx{m}^2Z + q\sbtx{e}^2Z^{-1}\)Y_o.
\eal
\esub
Notice that the coefficient $W_2$ has dropped out of the last of these equations due to the first equation, which follows from the constraint \eqref{condition}. These three equations form a set of coupled non-linear system that determine the functions $W_0(\f)$, $W_1(\f)$ and $Y_o(\f)$. 

In order to solve these equations we write them in the form
\bsub\label{system0}
\bal
&W_1=\frac{2W_0'Y_o'-W_0Y_o}{2Y_o'^2},\\
&\frac{Y_o}{Y_o'}(W_0^2)'-\frac12\(\frac{Y_o^2}{Y_o'^2}+3\)W_0^2=V+\frac{p^2}{\ell^2}Y_o+\frac{1}{2\ell^4}\(q\sbtx{m}^2Z + q\sbtx{e}^2Z^{-1}\)Y_o^2,\\
&-\frac{Y_o^2}{Y_o'^2}(W_0^2)'+\frac12\(3\frac{Y_o}{Y_o'}+\frac{Y_o^3}{Y_o'^3}-\(\frac{Y_o^2}{Y_o'^2}\)'\)W_0^2=-V'-\frac{1}{2\ell^4}\(q\sbtx{m}^2Z' + q\sbtx{e}^2(Z^{-1})'\)Y_o^2.
\eal
\esub
The last two equations comprise an algebraic linear system for $W_0^2$ and $(W_0^2)'$ which determines 
\bsub
\label{system}
\bal
&W_0^2=\frac{-1}{\ck\ck'}\[\ck\Big(V+\frac{p^2}{\ell^2}Y_o+\frac{1}{2\ell^4}\(q\sbtx{m}^2Z + q\sbtx{e}^2Z^{-1}\)Y_o^2\Big)-V'-\frac{1}{2\ell^4}\(q\sbtx{m}^2Z'+q\sbtx{e}^2(Z^{-1})'\)Y_o^2\],\\
&(W_0^2)'=\frac{\ck^2+3-2\ck'}{2\ck}W_0^2+\frac{1}{\ck^2}\(V'+\frac{1}{2\ell^4}\(q\sbtx{m}^2Z' + q\sbtx{e}^2(Z^{-1})'\)Y_o^2\),
\eal
\esub
where
\be
\ck\equiv Y_o/Y_o'.
\ee

Taking the derivative of the first equation in \eqref{system} with respect to $\f$ and equating it with the second leads to a decoupled third order non-linear equation for $Y_o$. However, in the present context we do not need to solve this equation for $Y_o(\f)$ since we can simply read it off using the known background solution of the field equations. By construction, the $Y_o(\f)$ read off the background solution automatically solves the non-linear equation for $Y_o(\f)$ obtained from \eqref{system}. Given the function $Y_o(\f)$, the equations \eqref{system0} and \eqref{system} allow us to obtain $W_0(\f)$ and $W_1(\f)$ algebraically in terms of $Y_o$ and its derivatives. Moreover, the flow equations \eqref{RGeqs-new} imply that the background solution from which we read off $Y_o(\f)$ satisfies the first order equations
\be
\dot a=-\frac12W_0,\qquad \frac{\dot b}{b}=2Y_oW_1,\qquad 
\dot\f = W_0\t,
\ee
or equivalently
\be
e^{-2a}=Y_o(\f),\qquad b=W_0^2\exp\Big(-\int^\f d\bar\f\;\ck(\bar\f)\Big),\qquad \int^\f \frac{d\bar\f}{W_0(\bar\f)\ck(\bar\f)}=u.
\ee
This information is sufficient to obtain the effective potential to quadratic order in $\s-\s_*$ around the extremum $\s=\s_*$, corresponding to the specific background from which $Y_o(\f)$ is read off.

\paragraph{Hairy solution of Theory I} From \eqref{hairySol} we easily see that the hairy solution of Theory I corresponds to the curve  $e^{-2a}=Y_o(\f)$, where
\be\label{YoI}
Y_o(\f)=\frac{\ell^2}{v^2}\(\frac{\sinh\(\frac{\f}{2\sqrt{3}}\)}{1-\sqrt{\a}\tanh\(\frac{\f}{2\sqrt{3}}\)}\)^2,
\ee
and recall that the parameter $v$ is related to all the charge densities through  
\eqref{constraint}. Equations \eqref{system0} and \eqref{system} then determine $W_0(\f)$ and $W_1(\f)$ algebraically, but we will not write the explicit expressions here. However, inserting the asymptotic form of $W_0(\f)$ and $W_1(\f)$ and $Y_o(\f)$ in the constraint \eqref{condition} determines that to leading order asymptotically, i.e. as $\f\to0$, 
\be
\dot Y\sim -\frac{2}{\ell} Y,
\ee
and hence $Y\sim e^{-2u/\ell}$. Since also $Y_o(\f)\sim \f^2\sim e^{-2u/\ell}$, it follows that $Y$ is sourced by the deviation of the scalar VEV $\s=\vf\sub{0}$ from its value at the extremum (c.f. \eqref{vevs-I})
\be
\s_*=\frac{2\sqrt{3}v}{\ell^2},
\ee
that is
\be
Y=e^{-2A}-Y_o(\f)\sim \frac{1}{\ell^2}\(\frac{1}{\s^2}-\frac{1}{\s_*^2}\)\f^2.
\ee

Inserting these results in the general expression \eqref{effpot-formula} for the effective potential and using the boundary condition for the scalar specified in \eqref{bc} and \eqref{th-I}, we finally obtain 
\bal\label{effpotI}
\cv\sbtx{QFT}(\s)&=\cv_0+\frac{\m q\sbtx{e}}{\ell^2}\NO\\
&+\lim_{u\to\infty}e^{3u/\ell}\[-\Big(2W_0(\f)+2W_1(\f)Y+\co(Y^2)+\frac{4}{\ell}+\frac{1}{2\ell}\f^2-\frac{p^2}{\ell}(Y_o(\f)+Y)\Big)+\frac{\sqrt{\a}}{6\ell\sqrt{3}}\f^3\]\NO\\
&=\cv_0+\frac{\m q\sbtx{e}}{\ell^2}+\frac{\ve}{2\ell^2\s_*^3}\(\s^3-3\s_*^2\s\)\NO\\
&=\cv_0+\frac{\m q\sbtx{e}}{\ell^2}+\frac{\ve}{2\ell^2\s_*^3} \(-2\s^3_*+3\s_*(\s-\s_*)^2+(\s-\s_\ast)^3\),
\eal
where $\ve$ is the energy density given in \eqref{epDyon}. It follows that $\s=\s_*$ is a stable local extremum of the effective potential provided $\ve>0$, i.e. $v>0$.

\paragraph{Electric solutions of Theory II} For the magnetically charged solutions of Theory II given in \eqref{mag-sol-II} the function $Y_o(\f)$ takes the form 
\be\label{YoIIe}
Y_o(\f)=\frac{\ell^2}{v\sbtx{e}^2}\(e^{-\frac12\sqrt{\frac{2-\x}{\x}}\;\f}-e^{\frac12\sqrt{\frac{\x}{2-\x}}\;\f}\)^2.
\ee
Moreover, the scalar {\em vev} at the extremum takes the value (see \eqref{vevsIIe})
\be
\s_*=-\sqrt{\x(2-\x)}\;v\sbtx{e}/\ell^2.
\ee

Using these results, as well as the boundary condition for the scalar specified in \eqref{bc} and \eqref{th-II-m}, we find that the effective potential \eqref{effpot-formula} in this case is 
\bal\label{effpotIIe}
\cv\sbtx{QFT}(\s)&=\cv_0+\frac{\m q\sbtx{e}}{\ell^2}+\lim_{u\to\infty}e^{\frac{3u}{\ell}}\[-\Big(2W_0(\f)+2W_1(\f)Y+\co(Y^2)+\frac{4}{\ell}+\frac{1}{2\ell}\f^2-\frac{p^2}{\ell}(Y_o(\f)+Y)\Big)\right.\NO\\
&\left.\hskip4.5in-\frac{(1-\x)}{6\ell\sqrt{\x(2-\x)}}\f^3\]\NO\\
&=\cv_0+\frac{\m q\sbtx{e}}{\ell^2}+\frac{\ve}{2\ell^2\s_*^3} \(\s^3-3\s_*^2\s\)\NO\\
&=\cv_0+\frac{\m q\sbtx{e}}{\ell^2}+\frac{\ve}{2\ell^2\s_*^3} \(-2\s^3_*+3\s_*(\s-\s_*)^2+(\s-\s_\ast)^3\),
\eal
where now $\ve$ is the energy given in \eqref{energyDensityIIe}. Hence, again these solutions are stable provided the energy density is positive definite.

\paragraph{Magnetic solutions of Theory II} For the magnetically charged solutions of Theory II given in \eqref{mag-sol-II} the function $Y_o(\f)$ takes the form 
\be\label{YoIIm}
Y_o(\f)=\frac{\ell^2}{v\sbtx{m}^2}\(e^{\frac12\sqrt{\frac{2-\x}{\x}}\;\f}-e^{-\frac12\sqrt{\frac{\x}{2-\x}}\;\f}\)^2.
\ee
Moreover, the scalar {\em vev} at the extremum takes the value (see eq. \eqref{vevsIIm})
\be
\s_*=-\sqrt{\x(2-\x)}\;v\sbtx{m}/\ell^2.
\ee

Using these results, as well as the boundary condition for the scalar specified in \eqref{bc} and \eqref{th-II-e}, we find that the effective potential \eqref{effpot-formula} in this case is 
\bal\label{effpotIIm}
\cv\sbtx{QFT}(\s)&=\cv_0+\lim_{u\to\infty}e^{\frac{3u}{\ell}}\[-\Big(2W_0(\f)+2W_1(\f)Y+\co(Y^2)+\frac{4}{\ell}+\frac{1}{2\ell}\f^2-\frac{p^2}{\ell}(Y_o(\f)+Y)\Big)\right.\NO\\
&\left.\hskip4.5in+\frac{(1-\x)}{6\ell\sqrt{\x(2-\x)}}\f^3\]\NO\\
&=\cv_0+\frac{\ve}{2\ell^2\s_*^3} \(\s^3-3\s_*^2\s\)\NO\\
&=\cv_0+\frac{\ve}{2\ell^2\s_*^3} \(-2\s^3_*+3\s_*(\s-\s_*)^2+(\s-\s_\ast)^3\),
\eal
where the energy density $\ve$ is given in \eqref{energyDensityIIm}.

\bibliographystyle{jhepcap}
\bibliography{refs}

\providecommand{\href}[2]{#2}\begingroup\raggedright\begin{thebibliography}{10}

\bibitem{Benincasa:2005iv}
P.~Benincasa, A.~Buchel, and A.~O. Starinets, {\it {Sound waves in strongly
  coupled non-conformal gauge theory plasma}},  {\em Nucl. Phys.} {\bf B733}
  (2006) 160--187, [\href{http://xxx.lanl.gov/abs/hep-th/0507026}{{\tt
  hep-th/0507026}}].

\bibitem{Finazzo:2014cna}
S.~I. Finazzo, R.~Rougemont, H.~Marrochio, and J.~Noronha, {\it {Hydrodynamic
  transport coefficients for the non-conformal quark-gluon plasma from
  holography}},  {\em JHEP} {\bf 02} (2015) 051,
  [\href{http://xxx.lanl.gov/abs/1412.2968}{{\tt 1412.2968}}].

\bibitem{Gursoy:2015nza}
U.~Gursoy, M.~Jarvinen, and G.~Policastro, {\it {Late time behavior of
  non-conformal plasmas}},  {\em JHEP} {\bf 01} (2016) 134,
  [\href{http://xxx.lanl.gov/abs/1507.08628}{{\tt 1507.08628}}].

\bibitem{Attems:2016ugt}
M.~Attems, J.~Casalderrey-Solana, D.~Mateos, I.~Papadimitriou,
  D.~Santos-Oliván, C.~F. Sopuerta, M.~Triana, and M.~Zilhão, {\it
  {Thermodynamics, transport and relaxation in non-conformal theories}},  {\em
  JHEP} {\bf 10} (2016) 155, [\href{http://xxx.lanl.gov/abs/1603.01254}{{\tt
  1603.01254}}].

\bibitem{Gubser:2008px}
S.~S. Gubser, {\it {Breaking an Abelian gauge symmetry near a black hole
  horizon}},  {\em Phys. Rev.} {\bf D78} (2008) 065034,
  [\href{http://xxx.lanl.gov/abs/0801.2977}{{\tt 0801.2977}}].

\bibitem{Hartnoll:2008vx}
S.~A. Hartnoll, C.~P. Herzog, and G.~T. Horowitz, {\it {Building a Holographic
  Superconductor}},  {\em Phys. Rev. Lett.} {\bf 101} (2008) 031601,
  [\href{http://xxx.lanl.gov/abs/0803.3295}{{\tt 0803.3295}}].

\bibitem{Charmousis:2010zz}
C.~Charmousis, B.~Gouteraux, B.~S. Kim, E.~Kiritsis, and R.~Meyer, {\it
  {Effective Holographic Theories for low-temperature condensed matter
  systems}},  {\em JHEP} {\bf 11} (2010) 151,
  [\href{http://xxx.lanl.gov/abs/1005.4690}{{\tt 1005.4690}}].

\bibitem{Huijse:2011ef}
L.~Huijse, S.~Sachdev, and B.~Swingle, {\it {Hidden Fermi surfaces in
  compressible states of gauge-gravity duality}},  {\em Phys. Rev.} {\bf B85}
  (2012) 035121, [\href{http://xxx.lanl.gov/abs/1112.0573}{{\tt 1112.0573}}].

\bibitem{Iizuka:2011hg}
N.~Iizuka, N.~Kundu, P.~Narayan, and S.~P. Trivedi, {\it {Holographic Fermi and
  Non-Fermi Liquids with Transitions in Dilaton Gravity}},  {\em JHEP} {\bf 01}
  (2012) 094, [\href{http://xxx.lanl.gov/abs/1105.1162}{{\tt 1105.1162}}].

\bibitem{Dong:2012se}
X.~Dong, S.~Harrison, S.~Kachru, G.~Torroba, and H.~Wang, {\it {Aspects of
  holography for theories with hyperscaling violation}},  {\em JHEP} {\bf 06}
  (2012) 041, [\href{http://xxx.lanl.gov/abs/1201.1905}{{\tt 1201.1905}}].

\bibitem{Iizuka:2012iv}
N.~Iizuka, S.~Kachru, N.~Kundu, P.~Narayan, N.~Sircar, and S.~P. Trivedi, {\it
  {Bianchi Attractors: A Classification of Extremal Black Brane Geometries}},
  {\em JHEP} {\bf 07} (2012) 193,
  [\href{http://xxx.lanl.gov/abs/1201.4861}{{\tt 1201.4861}}].

\bibitem{Gouteraux:2012yr}
B.~Gouteraux and E.~Kiritsis, {\it {Quantum critical lines in holographic
  phases with (un)broken symmetry}},  {\em JHEP} {\bf 04} (2013) 053,
  [\href{http://xxx.lanl.gov/abs/1212.2625}{{\tt 1212.2625}}].

\bibitem{Bardoux:2012aw}
Y.~Bardoux, M.~M. Caldarelli, and C.~Charmousis, {\it {Shaping black holes with
  free fields}},  {\em JHEP} {\bf 05} (2012) 054,
  [\href{http://xxx.lanl.gov/abs/1202.4458}{{\tt 1202.4458}}].

\bibitem{Donos:2013eha}
A.~Donos and J.~P. Gauntlett, {\it {Holographic Q-lattices}},  {\em JHEP} {\bf
  04} (2014) 040, [\href{http://xxx.lanl.gov/abs/1311.3292}{{\tt 1311.3292}}].

\bibitem{Andrade:2013gsa}
T.~Andrade and B.~Withers, {\it {A simple holographic model of momentum
  relaxation}},  {\em JHEP} {\bf 05} (2014) 101,
  [\href{http://xxx.lanl.gov/abs/1311.5157}{{\tt 1311.5157}}].

\bibitem{Azeyanagi:2009pr}
T.~Azeyanagi, W.~Li, and T.~Takayanagi, {\it {On String Theory Duals of
  Lifshitz-like Fixed Points}},  {\em JHEP} {\bf 06} (2009) 084,
  [\href{http://xxx.lanl.gov/abs/0905.0688}{{\tt 0905.0688}}].

\bibitem{Mateos:2011ix}
D.~Mateos and D.~Trancanelli, {\it {The anisotropic N=4 super Yang-Mills plasma
  and its instabilities}},  {\em Phys. Rev. Lett.} {\bf 107} (2011) 101601,
  [\href{http://xxx.lanl.gov/abs/1105.3472}{{\tt 1105.3472}}].

\bibitem{Jain:2014vka}
S.~Jain, N.~Kundu, K.~Sen, A.~Sinha, and S.~P. Trivedi, {\it {A Strongly
  Coupled Anisotropic Fluid From Dilaton Driven Holography}},  {\em JHEP} {\bf
  01} (2015) 005, [\href{http://xxx.lanl.gov/abs/1406.4874}{{\tt 1406.4874}}].

\bibitem{Donos:2016zpf}
A.~Donos, J.~P. Gauntlett, and O.~Sosa-Rodriguez, {\it {Anisotropic plasmas
  from axion and dilaton deformations}},  {\em JHEP} {\bf 11} (2016) 002,
  [\href{http://xxx.lanl.gov/abs/1608.02970}{{\tt 1608.02970}}].

\bibitem{Bardoux:2012tr}
Y.~Bardoux, M.~M. Caldarelli, and C.~Charmousis, {\it {Conformally coupled
  scalar black holes admit a flat horizon due to axionic charge}},  {\em JHEP}
  {\bf 09} (2012) 008, [\href{http://xxx.lanl.gov/abs/1205.4025}{{\tt
  1205.4025}}].

\bibitem{Gouteraux:2014hca}
B.~Goutéraux, {\it {Charge transport in holography with momentum
  dissipation}},  {\em JHEP} {\bf 04} (2014) 181,
  [\href{http://xxx.lanl.gov/abs/1401.5436}{{\tt 1401.5436}}].

\bibitem{Coleman:1991ku}
S.~R. Coleman, J.~Preskill, and F.~Wilczek, {\it {Quantum hair on black
  holes}},  {\em Nucl. Phys.} {\bf B378} (1992) 175--246,
  [\href{http://xxx.lanl.gov/abs/hep-th/9201059}{{\tt hep-th/9201059}}].

\bibitem{Park:2016slj}
C.~Park, {\it {On black hole thermodynamics with a momentum relaxation}},
  \href{http://xxx.lanl.gov/abs/1606.07340}{{\tt 1606.07340}}.

\bibitem{Papadimitriou:2007sj}
I.~Papadimitriou, {\it {Multi-Trace Deformations in AdS/CFT: Exploring the
  Vacuum Structure of the Deformed CFT}},  {\em JHEP} {\bf 05} (2007) 075,
  [\href{http://xxx.lanl.gov/abs/hep-th/0703152}{{\tt hep-th/0703152}}].

\bibitem{Papadimitriou:2005ii}
I.~Papadimitriou and K.~Skenderis, {\it {Thermodynamics of asymptotically
  locally AdS spacetimes}},  {\em JHEP} {\bf 08} (2005) 004,
  [\href{http://xxx.lanl.gov/abs/hep-th/0505190}{{\tt hep-th/0505190}}].

\bibitem{Hertog:2004ns}
T.~Hertog and G.~T. Horowitz, {\it {Designer gravity and field theory effective
  potentials}},  {\em Phys. Rev. Lett.} {\bf 94} (2005) 221301,
  [\href{http://xxx.lanl.gov/abs/hep-th/0412169}{{\tt hep-th/0412169}}].

\bibitem{MR2129341}
V.~P. Maslov, {\it Zeroth-Order Phase Transitions},  {\em Mat. Zametki} {\bf
  76} (2004), no.~5 748--761.

\bibitem{Ibanez:2008xq}
M.~Ibanez, J.~Links, G.~Sierra, and S.-Y. Zhao, {\it {Exactly solvable pairing
  model for superconductors with a p + ip - wave symmetry}},  {\em Phys. Rev.}
  {\bf B79} (2009) 180501, [\href{http://xxx.lanl.gov/abs/0810.0340}{{\tt
  0810.0340}}].

\bibitem{Altamirano:2013ane}
N.~Altamirano, D.~Kubiznak, and R.~B. Mann, {\it {Reentrant phase transitions
  in rotating anti–de Sitter black holes}},  {\em Phys. Rev.} {\bf D88}
  (2013), no.~10 101502, [\href{http://xxx.lanl.gov/abs/1306.5756}{{\tt
  1306.5756}}].

\bibitem{Wald:1984rg}
R.~M. Wald, {\em {General Relativity}}.
\newblock The University of Chicago Press (1984).

\bibitem{Bianchi:2001de}
M.~Bianchi, D.~Z. Freedman, and K.~Skenderis, {\it {How to go with an RG
  flow}},  {\em JHEP} {\bf 08} (2001) 041,
  [\href{http://xxx.lanl.gov/abs/hep-th/0105276}{{\tt hep-th/0105276}}].

\bibitem{deHaro:2006ymc}
S.~de~Haro, I.~Papadimitriou, and A.~C. Petkou, {\it {Conformally Coupled
  Scalars, Instantons and Vacuum Instability in AdS(4)}},  {\em Phys. Rev.
  Lett.} {\bf 98} (2007) 231601,
  [\href{http://xxx.lanl.gov/abs/hep-th/0611315}{{\tt hep-th/0611315}}].

\bibitem{Martinez:2004nb}
C.~Martinez, R.~Troncoso, and J.~Zanelli, {\it {Exact black hole solution with
  a minimally coupled scalar field}},  {\em Phys. Rev.} {\bf D70} (2004)
  084035, [\href{http://xxx.lanl.gov/abs/hep-th/0406111}{{\tt
  hep-th/0406111}}].

\bibitem{Papadimitriou:2006dr}
I.~Papadimitriou, {\it {Non-Supersymmetric Membrane Flows from Fake
  Supergravity and Multi-Trace Deformations}},  {\em JHEP} {\bf 02} (2007) 008,
  [\href{http://xxx.lanl.gov/abs/hep-th/0606038}{{\tt hep-th/0606038}}].

\bibitem{Breitenlohner:1982bm}
P.~Breitenlohner and D.~Z. Freedman, {\it {Positive Energy in anti-De Sitter
  Backgrounds and Gauged Extended Supergravity}},  {\em Phys. Lett.} {\bf B115}
  (1982) 197--201.

\bibitem{Balasubramanian:1998sn}
V.~Balasubramanian, P.~Kraus, and A.~E. Lawrence, {\it {Bulk versus boundary
  dynamics in anti-de Sitter space-time}},  {\em Phys. Rev.} {\bf D59} (1999)
  046003, [\href{http://xxx.lanl.gov/abs/hep-th/9805171}{{\tt
  hep-th/9805171}}].

\bibitem{Klebanov:1999tb}
I.~R. Klebanov and E.~Witten, {\it {AdS / CFT correspondence and symmetry
  breaking}},  {\em Nucl. Phys.} {\bf B556} (1999) 89--114,
  [\href{http://xxx.lanl.gov/abs/hep-th/9905104}{{\tt hep-th/9905104}}].

\bibitem{Papadimitriou:2011qb}
I.~Papadimitriou, {\it {Holographic Renormalization of general dilaton-axion
  gravity}},  {\em JHEP} {\bf 08} (2011) 119,
  [\href{http://xxx.lanl.gov/abs/1106.4826}{{\tt 1106.4826}}].

\bibitem{Lindgren:2015lia}
J.~Lindgren, I.~Papadimitriou, A.~Taliotis, and J.~Vanhoof, {\it {Holographic
  Hall conductivities from dyonic backgrounds}},  {\em JHEP} {\bf 07} (2015)
  094, [\href{http://xxx.lanl.gov/abs/1505.04131}{{\tt 1505.04131}}].

\bibitem{Henningson:1998gx}
M.~Henningson and K.~Skenderis, {\it {The Holographic Weyl anomaly}},  {\em
  JHEP} {\bf 07} (1998) 023,
  [\href{http://xxx.lanl.gov/abs/hep-th/9806087}{{\tt hep-th/9806087}}].

\bibitem{Henningson:1998ey}
M.~Henningson and K.~Skenderis, {\it {Holography and the Weyl anomaly}},  {\em
  Fortsch. Phys.} {\bf 48} (2000) 125--128,
  [\href{http://xxx.lanl.gov/abs/hep-th/9812032}{{\tt hep-th/9812032}}].

\bibitem{deHaro:2000vlm}
S.~de~Haro, S.~N. Solodukhin, and K.~Skenderis, {\it {Holographic
  reconstruction of space-time and renormalization in the AdS / CFT
  correspondence}},  {\em Commun. Math. Phys.} {\bf 217} (2001) 595--622,
  [\href{http://xxx.lanl.gov/abs/hep-th/0002230}{{\tt hep-th/0002230}}].

\bibitem{Bianchi:2001kw}
M.~Bianchi, D.~Z. Freedman, and K.~Skenderis, {\it {Holographic
  renormalization}},  {\em Nucl. Phys.} {\bf B631} (2002) 159--194,
  [\href{http://xxx.lanl.gov/abs/hep-th/0112119}{{\tt hep-th/0112119}}].

\bibitem{Skenderis:2002wp}
K.~Skenderis, {\it {Lecture notes on holographic renormalization}},  {\em
  Class. Quant. Grav.} {\bf 19} (2002) 5849--5876,
  [\href{http://xxx.lanl.gov/abs/hep-th/0209067}{{\tt hep-th/0209067}}].

\bibitem{Witten:2001ua}
E.~Witten, {\it {Multitrace operators, boundary conditions, and AdS / CFT
  correspondence}},  \href{http://xxx.lanl.gov/abs/hep-th/0112258}{{\tt
  hep-th/0112258}}.

\bibitem{Berkooz:2002ug}
M.~Berkooz, A.~Sever, and A.~Shomer, {\it {'Double trace' deformations,
  boundary conditions and space-time singularities}},  {\em JHEP} {\bf 05}
  (2002) 034, [\href{http://xxx.lanl.gov/abs/hep-th/0112264}{{\tt
  hep-th/0112264}}].

\bibitem{Mueck:2002gm}
W.~Mueck, {\it {An Improved correspondence formula for AdS / CFT with
  multitrace operators}},  {\em Phys. Lett.} {\bf B531} (2002) 301--304,
  [\href{http://xxx.lanl.gov/abs/hep-th/0201100}{{\tt hep-th/0201100}}].

\bibitem{Davison:2014lua}
R.~A. Davison and B.~Goutéraux, {\it {Momentum dissipation and effective
  theories of coherent and incoherent transport}},  {\em JHEP} {\bf 01} (2015)
  039, [\href{http://xxx.lanl.gov/abs/1411.1062}{{\tt 1411.1062}}].

\bibitem{Castro:2010fd}
A.~Castro, A.~Maloney, and A.~Strominger, {\it {Hidden Conformal Symmetry of
  the Kerr Black Hole}},  {\em Phys. Rev.} {\bf D82} (2010) 024008,
  [\href{http://xxx.lanl.gov/abs/1004.0996}{{\tt 1004.0996}}].

\bibitem{Bertini:2011ga}
S.~Bertini, S.~L. Cacciatori, and D.~Klemm, {\it {Conformal structure of the
  Schwarzschild black hole}},  {\em Phys. Rev.} {\bf D85} (2012) 064018,
  [\href{http://xxx.lanl.gov/abs/1106.0999}{{\tt 1106.0999}}].

\bibitem{thermo2}
I.~Papadimitriou and K.~Skenderis, {\em {AdS back hole thermodynamics: General
  boundary conditions for scalars and p-form fields. {\rm In preparation}}}.

\bibitem{Gibbons:1976ue}
G.~W. Gibbons and S.~W. Hawking, {\it {Action Integrals and Partition Functions
  in Quantum Gravity}},  {\em Phys. Rev.} {\bf D15} (1977) 2752--2756.

\bibitem{Skenderis:1999mm}
K.~Skenderis and P.~K. Townsend, {\it {Gravitational stability and
  renormalization group flow}},  {\em Phys. Lett.} {\bf B468} (1999) 46--51,
  [\href{http://xxx.lanl.gov/abs/hep-th/9909070}{{\tt hep-th/9909070}}].

\bibitem{Freedman:2003ax}
D.~Z. Freedman, C.~Nunez, M.~Schnabl, and K.~Skenderis, {\it {Fake supergravity
  and domain wall stability}},  {\em Phys. Rev.} {\bf D69} (2004) 104027,
  [\href{http://xxx.lanl.gov/abs/hep-th/0312055}{{\tt hep-th/0312055}}].

\bibitem{Papadimitriou:2004rz}
I.~Papadimitriou and K.~Skenderis, {\it {Correlation functions in holographic
  RG flows}},  {\em JHEP} {\bf 10} (2004) 075,
  [\href{http://xxx.lanl.gov/abs/hep-th/0407071}{{\tt hep-th/0407071}}].

\bibitem{Brill:1997mf}
D.~R. Brill, J.~Louko, and P.~Peldan, {\it {Thermodynamics of (3+1)-dimensional
  black holes with toroidal or higher genus horizons}},  {\em Phys. Rev.} {\bf
  D56} (1997) 3600--3610, [\href{http://xxx.lanl.gov/abs/gr-qc/9705012}{{\tt
  gr-qc/9705012}}].

\bibitem{Caldarelli:1998hg}
M.~M. Caldarelli and D.~Klemm, {\it {Supersymmetry of Anti-de Sitter black
  holes}},  {\em Nucl. Phys.} {\bf B545} (1999) 434--460,
  [\href{http://xxx.lanl.gov/abs/hep-th/9808097}{{\tt hep-th/9808097}}].

\bibitem{Vanzo:1997gw}
L.~Vanzo, {\it {Black holes with unusual topology}},  {\em Phys. Rev.} {\bf
  D56} (1997) 6475--6483, [\href{http://xxx.lanl.gov/abs/gr-qc/9705004}{{\tt
  gr-qc/9705004}}].

\bibitem{Emparan:1999gf}
R.~Emparan, {\it {AdS / CFT duals of topological black holes and the entropy of
  zero energy states}},  {\em JHEP} {\bf 06} (1999) 036,
  [\href{http://xxx.lanl.gov/abs/hep-th/9906040}{{\tt hep-th/9906040}}].

\bibitem{Anabalon:2013sra}
A.~Anabalón and D.~Astefanesei, {\it {On attractor mechanism of $AdS_{4}$
  black holes}},  {\em Phys. Lett.} {\bf B727} (2013) 568--572,
  [\href{http://xxx.lanl.gov/abs/1309.5863}{{\tt 1309.5863}}].

\bibitem{Papadimitriou:2016yit}
I.~Papadimitriou, {\it {Lectures on Holographic Renormalization}},  {\em
  Springer Proc. Phys.} {\bf 176} (2016) 131--181.

\bibitem{Anabalon:2015xvl}
A.~Anabalon, D.~Astefanesei, D.~Choque, and C.~Martinez, {\it {Trace Anomaly
  and Counterterms in Designer Gravity}},  {\em JHEP} {\bf 03} (2016) 117,
  [\href{http://xxx.lanl.gov/abs/1511.08759}{{\tt 1511.08759}}].

\end{thebibliography}\endgroup

\end{document}